\documentclass[trackchanges,twocolumn,twocolappendix]{aastex701}
\usepackage{graphicx}	% Including figure files
\usepackage{amsmath}	% Advanced maths commands
\usepackage{gensymb}
\usepackage{nicefrac}
\usepackage{array}
\usepackage{tikz}
\usepackage{xcolor}
\usepackage[percent]{overpic}
\begin{document}

\title{Are most detected tidal disruption events partial?}

\author[orcid=0009-0004-2166-8461,sname='Sharma']{Megha Sharma}
\affiliation{School of Physics and Astronomy, Monash University, VIC, 3800, Australia}
\email[show]{megha.sharma@monash.edu}  

\author[orcid=0000-0002-4716-4235]{Daniel J.\ Price} 
\affiliation{School of Physics and Astronomy, Monash University, VIC, 3800, Australia}
\affiliation{IPAG, Univ. Grenoble Alpes, CNRS, 38000 Grenoble, France}
\email{daniel.price@monash.edu}

\author[orcid=0000-0002-3684-1325]{Alexander Heger}
\affiliation{School of Physics and Astronomy, Monash University, VIC, 3800, Australia}
\email{alexander.heger@monash.edu}

\author[orcid=0000-0002-4449-9152]{Katie Auchettl}
\affiliation{School of Physics, University of Melbourne, VIC 3010, Australia}
\affiliation{Department of Astronomy and Astrophysics, University of California, Santa Cruz, CA 95064, USA}
\email{katie.auchettl@unimelb.edu.au}

\begin{abstract}
During a tidal disruption event (TDE), a star loses mass due to the tidal gravitational forces of the black hole.  In a partial tidal disruption event, a stellar remnant is left behind.  Several dozen TDEs have been detected so far, including repeating partial events.  We use the \textsc{Phantom} smoothed particle hydrodynamics code to model the disruption of a $1\,\mathrm{M}_\odot$ star around a $10^6\,\mathrm{M}_\odot$ black hole for impact parameters resulting in $\leq 50\%$ mass loss. We only consider zero energy orbits.  Our simulations show that the mass fallback rate can exceed the Eddington limit for $\beta \geq 0.8$, allowing debris to obscure the accretion disc by forming a reprocessing layer, similar to full TDEs.  The mass fallback rate is shallower than $t^{-5/3}$, tracking closer to $t^{-9/4}$. Assuming thermal emission from the debris, that shock heating is trapped, that electron scattering dominates the opacity, and a color correction ($f_{\mathrm{col}}$) of $1.7$, we find temperatures of $\mathord\sim10^4\;\mathrm{K}$, optical bolometric luminosities of $\mathord\sim10^{42-44}\;\mathrm{erg\;s^{-1}}$ and blackbody radii ranging from $10$--$100\,\mathrm{au}$ for our simulations. We compare our values with observations and find support for the previous argument that some TDEs classified as full disruptions might actually be partial. Moreover, our results explain the detected optical/UV TDEs. We also find that our zero energy partial TDEs have properties similar to the repeating partial TDEs such as \textit{ASSASN-19dj}, \textit{ASSASN-14ko}, \textit{ASSASN-18ul}, \textit{ASSASN-22ci},
\textit{AT2020vdq}
and \textit{AT2022dbl}. In the $\beta=0.8$, isentropic simulation where radiation is assumed to escape, we find X-ray luminosities of $\sim 10^{44-45}\;\mathrm{erg\;s^{-1}}$ and radii lower than the inner most stable circular orbit.

\end{abstract}



\section{Introduction} 
 Most galaxies larger than $\approx10^{5-6}\,\mathrm{M}_{\odot}$ host a supermassive black hole (SMBH; \citealt{Kormendy2013}).  Whereas a fraction of these SMBHs tend to be active (i.e., are hosted by an active galactic nucleus; AGN) and easily detectable via luminous multi-wavelength emission, there is a population of quiescent or weakly accreting black holes whose accretion occurs at a slower rate and require other mechanisms to probe their properties \citep{Matthee2024}.  A tidal disruption event (TDE) provides a direct probe of these dormant SMBHs and occurs when a star passes so close to the black hole that the tidal forces exceed the star's self-gravity and is ripped apart \citep{Hills1975,Lacy1982,Rees1988}.  Since their theoretical prediction \citep{Hills1975}, about $100$ TDEs have been observed to date \citep{Gezari2021,Hammerstein2023}, and upcoming surveys such as Vera C.\ Rubin Observatory's Legacy Survey of Space and Time (LSST; \citealt{Bricman2023}), La Silla Schmidt Southern Survey (LS4; \citealt{Miller2025}), Asteroid Terrestrial-impact Last Alert System (ATLAS; \citealt{Tonry2018}), All-Sky Automated Survey for Supernovae (ASASSN; \citealt{Kochanek2017}) and Wide Field Survey Telescope (WFST; \citealt{Wang2023}) are expected to increase this number substantially.

After the disruption, some stellar debris remains bound to the black hole, forming a (compact) accretion disc.  The flare is thought to follow the mass fallback rate which is given by \citep{Rees1988,Phinney1989}
\begin{align}
\label{eq:mdot}
    \frac{\mathrm{d}M}{\mathrm{d}t} &= \frac{\mathrm{d}M}{\mathrm{d}\varepsilon} \frac{\mathrm{d}\varepsilon}{\mathrm{d}t}= \frac{\mathrm{d}M}{\mathrm{d}\varepsilon} t^{-5/3}\;,
\end{align}
where $\varepsilon$ is the specific orbital energy. Early studies assumed a uniform distribution for  $\frac{\mathrm{d}M}{\mathrm{d}\varepsilon}$ and a prompt accretion due to efficient circularisation of the material, with peak luminosity in soft X-ray/UV bands \citep{Rees1988,Evans1989,Ulmer1999}.  But most TDEs are detected in optical and UV \citep{vanVelzen2011}, and display a wide range of decay rates such as shallower, or exponential declines in UV/optical, and some even show re-brightening in X-ray and radio, and a range of photospheric radii, and fainter than expected X-rays \citep[e.g.,][]{Auchettl2017,Hinkle2020, VanVelzen2021,Cendes2022, Hammerstein2023}. Possible explanations include the presence of a reprocessing layer formed from optically thick outflows which mask the X-ray emissions \citep{Loeb1997,Ulmer1999}, stream-stream collisions during circularisation \citep{Dai2015,Piran2015,Jiang2016,Bennerot2017}, or a quasi-spherical pressure-supported envelope \citep{Coughlin2014,Metzger2022}.
 
Due to computational challenges several simulations focus on simulating the disruption of polytropic and MESA stellar models in grid based hydrodynamics codes \citep[e.g.,][]{Guillochon2013,Lodato2009,Law-Smith2020,Ryu2020} and Smoothed Particle Hydrodynamics (SPH) codes \citep[e.g.,][]{Golightly2019,Bandopadhyay2026} to determine the mass fallback rates, but without actually following the fallback of material onto the black hole and the disc formation process. Recent simulations have focused on the disc formation process in these events and demonstrate debris stream self-intersection due to apsidal precession \citep[e.g.,][]{Hayasaki2013,LiptaiPrice2019,Bonnerot2020,Andalman2022}. None of these works focus on producing the synthetic lightcurves from the simulations.  \citet{Steinberg2024} evolved a simulation for up to $62$ days and found that the shocks near pericentre power the light curve and stream–disc shocks result in outflows reproducing the UV/optical luminosities. \citet{Fitz2024} and \citet{Price2024} performed simulations of these events for up to an year around a $10^6\;\mathrm{M}_\odot$ Schwarzschild black hole using the \textsc{Phantom} SPH code \citep{Price2018,Liptai2019}. They used optical depth of the surrounding material to predict the
optical appearance, and found the formation of a reprocessing layer in their work.
 
 A caveat is that all recent simulations modelling the entire event \citep{Fitz2024,Price2024,Steinberg2024} assume full TDEs where the star is completely ripped apart due to the tidal forces of the black holes. The theoretical boundary at which this takes place is termed as the \emph{tidal radius}, $r_\mathrm{t}$ given by
 \begin{equation}
    r_\mathrm{t} = r_* \left(\frac{M_\bullet}{M_*}\right)^{\nicefrac{1}{3}}\;,
\end{equation}
 and $r_*$ is radius of star, $M_\bullet$ is mass of black hole, and $M_*$ is mass of the star \citep{Rees1988}. During a partial TDE the star survives the encounter with some stripping of material and the formation of a remnant \citep{Guillochon2013}.  The penetration factor ($\beta \equiv r_\mathrm{t}/r_\mathrm{p}$) is used to determine if the pericentre of the orbit of a star is within the tidal radius.  But the boundary of transition from a partial to a full TDE remains unclear due to several factors that affect the outcome such as the star’s structure and general relativistic effects  \citep{Guillochon2013,Law-Smith2020,Ryu2020,Sharma2024}. \emph{Mass stripping} begins for $\beta \sim 0.6$ for typical moderate-mass main-sequence stars ($\geq1-2\;\mathrm{M}_\odot$;  \citealt[][]{Nixon2022,Mockler2022, Sharma2024}).

Most of the TDEs are partial due to the larger cross section \citep{Stone2016,Zhong2022}. \citet{Bortolas2023} argued that the rate of partial TDEs, would be $\sim 10$ times larger than the full TDEs. For partial TDEs mass fallback is steeper than $t^{-5/3}$, with faster post-peak decay rate of $t^{-9/4}$ \citep{Guillochon2013,Coughlin2019,Ryu2020,Bandopadhyay2026}. \citet{Chen2021} suggested that no outflows would form due to efficient radiative diffusion. The lightcurves would have multiple peaks if partial disruption occurs for stars on bound orbits.  But the decay rates also depend on the stellar structure \citep{Lodato2009,Law-Smith2020}. Therefore, it is hard to differentiate between a full and a partial TDE event except for when the curve has multiple peaks. Furthermore, a new class of periodic events is emerging which could be repeating partial TDEs \citep{Hinkle2024}. 

In this paper, we extend previous work of \citet{Fitz2024,Price2024} to partial TDEs on zero-energy (``parabolic'') orbits. We test if partial TDEs result in bolometric light curves, radii and temperatures that are consistent with or different from that of a full TDE. We also compare with periodic events, but we note that none of our models are on similar orbits to the observed periodic events. With our results, we place constraints on whether partial TDEs dominate the population of events detected by observations. Section~\ref{sec:methods} describes the method used, followed by results in Section~\ref{sec:results}.  We discuss in Section~\ref{sec:discussion}, and conclude in Section~\ref{sec:conclusion}.

\section{Methods}
\label{sec:methods}
We perform simulations using the General Relativistic Smoothed Particle Hydrodynamics \citep{Lucy1977,Gingold1977,Monaghan1992,Price2012} code \textsc{Phantom v2025.0.0}\footnote{\url{https://github.com/danieljprice/phantom/}} \citep{Price2018,Liptai2019}.  A significant fraction of $1\;\mathrm{M}_\odot$ stars in the host galaxies of TDEs would be on the main-sequence \citep{Mockler2022}. Hence, we generate a $1\,\mathrm{M}_\odot$, middle-aged main-sequence (MAMS) stellar model ($t = 4.53\;\mathrm{Gyr}$) using the stellar-evolution code \textsc{Kepler}\footnote{\url{https://doi.org/10.5281/zenodo.10780856}} \citep{Weaver1978,Sharma2024}. 

Initial trajectories are computed using \textsc{Phantom-Geo}\footnote{\url{https://github.com/phantomSPH/phantom-geodesic/}} \citep{Liptai2019}, starting at $r_0 = 10\mathord,000\,r_\mathrm{t}$. The initial position and velocity are calculated for a Keplerian $E=0$ orbit in $x-y$ plane. The initial velocity is determined using $v_0 = \sqrt{2 G M_\bullet/r_0}$.  This is a valid assumption as the starting distance of the geodesic is far enough for the Keplerian orbital setup to be valid due to negligible effect of the black hole. We then use the geodesic obtained to determine the positions and velocities at $10\,r_\mathrm{t}$.
 This ensures that all trajectories follow zero-energy orbits.  The velocities and positions at $10\; r_\mathrm{t}$ are used as the starting points of the centre of mass of the star for different impact parameters, $\beta\equiv r_\mathrm{t}/r_\mathrm{p}$ in \textsc{Phantom}. As we are interested in understanding the properties of partial TDEs, we explore the properties of these events by assuming an impact parameter ranging from $0.6 - 1.6$. This corresponds to a range of mass loss $\leq 50\%$, where $r_\mathrm{p}$ is the Newtonian pericentre.  We used same method as \citet{Sharma2024} to determine what constitutes as a remnant post-disruption based on the energy of the particles with respect to the maximum density particle (centre of the remnant).

We model the star initially using $10^6$ SPH particles, adopting an adiabatic equation of state ($\gamma= \nicefrac53$).  We first relax the stellar model in the Minkowski metric which ensures that black hole's tidal forces do not affect the star.  We perform the relaxation until the kinetic to potential energy ratio falls below $\leq 10^{-7}$, ensuring hydrostatic equilibrium, following the procedure described in Appendix~C of \citet{Lau2022}.  We then place this relaxed star on the desired orbit around a $10^6\;\mathrm{M}_\odot$ black hole in the Schwarzschild metric. Our choice of a $10^6\;\mathrm{M}_\odot$ black hole is motivated by the fact that the rate of TDEs decreases around more massive black holes \citep{VanVelzen2018}.  We delete SPH particles inside the last stable circular orbit ($6\,G\,M_\bullet/c^{2}$) to avoid small timesteps similar to \citet{Price2024} as material accretes onto the black hole.  This does not remove any of the remnant produced in our simulations.  We use high-resolution shock-capturing, with shock capturing parameters  $\alpha=1$ and $\beta=2$.  We also use the default value of $0.1$ for shock conductivity in \textsc{Phantom} \citep{Price2018}.  Our primary set of simulations assume that all energy is trapped (adiabatic approximation), and not radiated instantaneously, making the flow inherently super-Eddington, appropriate for the accretion associated with TDEs close to peak brightness \citep[e.g.,][]{Dai2018}.  For low $\beta$ models ($\beta < 1.0$) where accretion becomes sub-Eddington, consistent with that expected from a partial TDE, we also performed isentropic simulations of the same setup above where we assume shock heating to be immediately radiated away.  We use $K=P\,\rho^{-\gamma}$ as the energy variable, where $P$ is pressure, and $\rho$ is the rest frame mass density (see \citealt{Liptai2019} for details). 

Since the surviving stellar core is the main computational cost due to small timestep near the maximum density particle in the simulation of partial TDEs, we accelerated the computation by replacing the remnant with a softened point-mass  (using Plummer softening with $\epsilon = 20\;\mathrm{R}_\odot$). We do this at $4$ days since the beginning of the simulation in each model.  We performed tests with a smaller softening radius ($\epsilon = 5\;\mathrm{R}_\odot$) and without replacing the core, but found negligible differences. \citet{Sharma2024} found that the size of the remnants can increase by a factor of $100$ in tidal disruption events, which makes this choice reasonable, ensuring our calculations evolve faster.
 
 % of $5\;\mathrm{R}_\odot$, and $20\;\mathrm{R}_\odot$. We also performed the simulation with no point mass replacement for  (all simulations start with $10^4$ particles). The evolution of simulations is similar with or without replacement of the remnant, and the size of softening radius does not significantly alter the results. 
 
Because only a fraction of the stellar mass is removed from the star, we use \textsc{Phantom}'s particle splitting module, \textsc{splitpart}, to increase the resolution in the debris stream post-disruption (see \citealt{Nealon2025} for details, although we only use a global split of all particles in this paper). After replacing the stellar core, we repeatedly split the SPH particles until each simulation contained a minimum of $10^6$ particles in the remaining bound and unbound debris, maintaining comparable particle numbers across all $\beta$ values, albeit with different resultant SPH particle masses in each simulation.  Resolution tests show that our results converge for up to $\sim100$ days post-disruption (pericentre passage time is $\sim 7.0\;\mathrm{hours}$; see Appendix~\ref{app:res_study}) across different $\beta$ values. Hence, we only show the results up to $100$ days since the start of the simulations for our models.  Table~\ref{tab:all_sims} lists the mass of the remnant, followed by the number of particles after splitting, and mass per SPH particle after splitting.  We performed the splitting in stages with each step splitting into no more than $13$ particles, and evolving the simulation for $7.6$ hours between splits to allow the simulation to relax in between the splits. This corresponds to the maximum time used to write full snapshots for our simulations.

We test if splitting results in different result compared with a simulation without particle splitting. Comparison of a $\beta=1.6$ model generated using \textsc{Splitpart} with a simulation with no  particle splitting shows no significant morphological differences (see Appendix~\ref{app:splitting_vs_nosplitting}).  

% For the $\beta=0.6$ we first add the default number of children ($13$) twice, then add $6$ more children. For $\beta=1.0$, we first add $6$ particles followed by $9$. For $\beta=1.0$, we add $2$ particles followed by $8$. For $\beta=1.2$, we add $2$ particles followed by $4$. For $\beta=1.4$ we add $4$ particles and for $\beta=1.6$ we add $2$.

 % We increase the resolution of our simulations by adding particles in our point mass replaced star simulations at $4$ days. Our $\beta=1.6$ model has $\sim500\mathord,000$ particles after point mass remnant replacement post disruption. Due to limitation on computational resources, maximum number of particles to feasibly simulate is $\sim1\mathord,000\mathord,000$. Hence, we add particles to our other simulations such that we have $\sim1\mathord,000\mathord,000$ particles in the stream post-disruption for all of our models. 
 
\subsection{Lightcurve calculation}
To create synthetic lightcurves, we post-processed the simulations by  discretizing an image plane of $5\times10^{16}\;\mathrm{cm}$ by $5\times10^{16}\;\mathrm{cm}$ into $1024\times1024$ pixels similar to \citet{Fitz2024} and \citet{Price2024}. Each pixel samples $128$ frequencies from $10^8\,\mathrm{Hz}$ to $10^{22}\;\mathrm{Hz}$. 
We compute the lightcurves by solving the following equation of radiative transfer 
\begin{equation}
    \frac{\mathrm{d}I_\nu}{\mathrm{d}\tau} = B_\nu - I_\nu \; ,
\end{equation}
where $I_\nu$ is the intensity, $B_\nu = 2 h \nu^3c^{-2}\exp[(h\nu/k_\mathrm{B}T)-1]^{-1}$ is the Planck function, and $\tau$ is optical depth at each frequency for each particle.  $T$ is absolute temperature, $\nu$ is frequency, $k_\mathrm{B}$ is Boltzmann constant, $h$ is Planck constant, and $c$ is speed of light.  We consider the radiation and gas to be in local thermal equilibrium, and determine the temperature by solving the implicit equation 
\begin{equation}
\label{eq:temp_calc}
    f(T) = \frac{3}{2} \frac{k_\mathrm{B}T}{\mu m_\mathrm{H}} + \frac{a T^4}{\rho} - u = 0\;,
\end{equation}
where $\mu = 0.6$ is the mean molecular weight, and $u$ is the internal energy.  We use the Newton-Raphson method to solve for the temperature from the internal energy, which, in turn, is obtained using $u = K \rho^{\gamma-1} /(\gamma-1)$ where $K$ is our evolved entropy variable \citep{Liptai2019}, $\gamma$ is the adiabatic index and $\rho$ is the density.  Optical depth is given by  
\begin{equation}
    \mathrm{d}\tau = \frac{\nu_0}{\nu} \kappa \rho\, \mathrm{d} z'=\gamma\, (1-\beta_z) \kappa \rho\, \mathrm{d}z'_0\;,
\end{equation}
where  $\beta = v_z/c$, $\gamma = 1/\sqrt{1-v^2/c^2}$, and $\nu_0/\nu = \gamma\, (1-\beta_z)$ accounts for the optical depth change caused by the moving photosphere \citep{Price2024}.  We only use electron opacity, $\kappa$ for our calculations, with a value of $\sim 0.34\;\mathrm{cm^2\; g^{-1}}$ for ionised gas. 

%We neglect recombination of material as the gas cools with time. 

To calculate the specific flux, we integrate the specific intensity ($I_\nu$) over the emitting plane, given by
\begin{equation}
    F_\nu = \int I_\nu\, \mathrm{d}x\, \mathrm{d}y\;.
\end{equation}
We compute the synthetic spectral energy distributions ($\nu F_\nu$ vs energy) by integrating over all pixels in the image.  We assume that each particle in the debris emits like a blackbody. This results in a lower blackbody temperature than in reality. The contribution of electron and inverse Compton scattering would result in a higher temperature \citep{Li2005}. To account for this a color ($f_{\mathrm{col}}$) correction factor can be used, e.g., $f_{\mathrm{col}}$ has been used in comparing disc models with observed TDEs \citep[e.g.][]{Berger2026,Chakrobatory2026,Yao2026}. Recently, \citet{Berger2026} used a $f_{\mathrm{col}}=2.4$ for their disc modelling to determine the mass of the black hole using a TDE. They found a lower $f_{\mathrm{col}}$ results in lower black hole mass. For our work, we use a $f_{\mathrm{col}}$ of $1.7$ \citep{Shimura1995}. A higher value of $f_{\mathrm{col}}$ such as $2.4$ results in higher calculated bolometric temperatures and luminosities for our simulations. It also has an effect on the synthetic spectral energy distributions, translating them to higher energies.

Inferred bolometric luminosities ($L_\mathrm{bb}$) are computed by integrating the blackbody spectrum fitted for the optical band in the wavelength range of $367.6–901.0\; \mathrm{nm}$, which covers the commonly used $g$, $r$, and $i$ optical wavelengths used to discover TDEs. The blackbody temperature ($T_\mathrm{bb}$) is calculated from fitting a single-temperature blackbody to the optical blackbody fit, and the blackbody radius $R_\mathrm{bb}$ is given by solving $L_\mathrm{bb} = 4 \pi R_\mathrm{bb}^2 \sigma T^4_\mathrm{bb}$. 

We compare our inferred blackbody radii with observational data obtained from \citet{Neustadt2020}, \citet{VanVelzen2021} which includes full TDEs. Sources presented in \citet{Hinkle2024} are classified as repeating TDEs where events have recurrence times ranging from days to years. These include \textit{ASSASN-19dj}, \textit{ASSASN-14ko}, \textit{ASSASN-18ul}, \textit{ASSASN-22ci}, \textit{AT2020vdq}. We additionally include \textit{AT2022dbl} \citep{Makrygianni2025}.  To visualise these repeating events, we phase-fold the data for the sample when comparing with our simulations. We note that for partial TDEs the stars would need to be on elliptical orbits, but for this work we have only considered zero-energy orbits.

\begin{figure*}
    \centering
    % The 'percent' option makes the grid 1-100 for easy positioning
    \begin{overpic}[width=\textwidth,keepaspectratio]{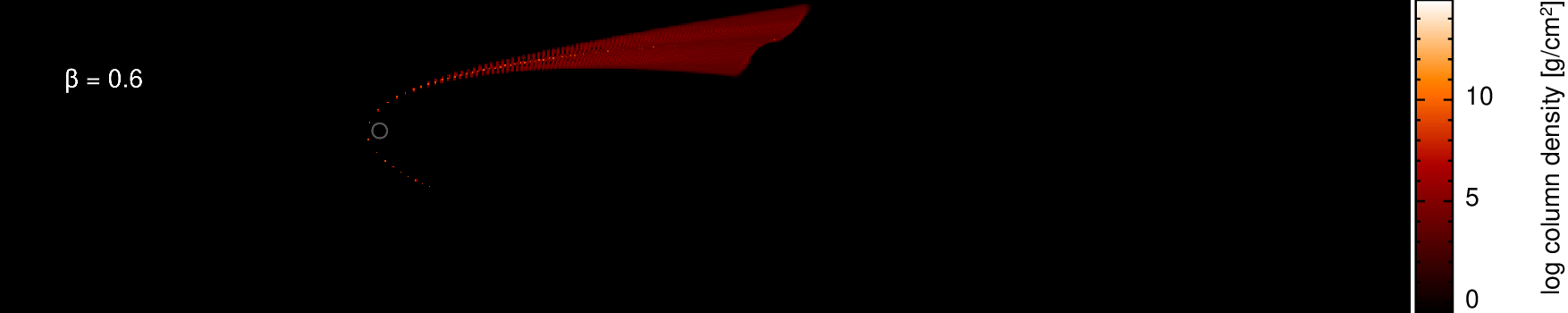}
        % \put(x,y) where x and y are % of the main image width/height
        \put(76,8){%
    \setlength{\fboxrule}{1pt} % Set border thickness
    \setlength{\fboxsep}{0pt}  % Set gap to zero
    \fcolorbox{white}{white}{\includegraphics[width=0.08\textwidth]{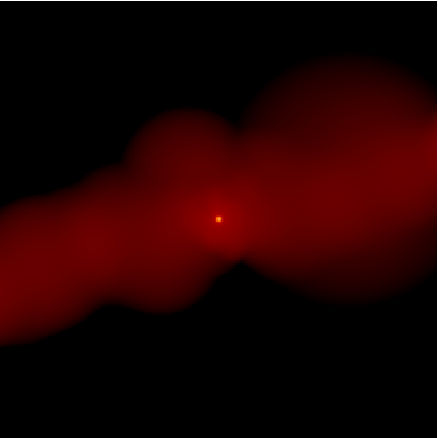}}}
    \end{overpic}
    
    \vspace{0.2cm}
        \begin{overpic}[width=\textwidth,keepaspectratio]{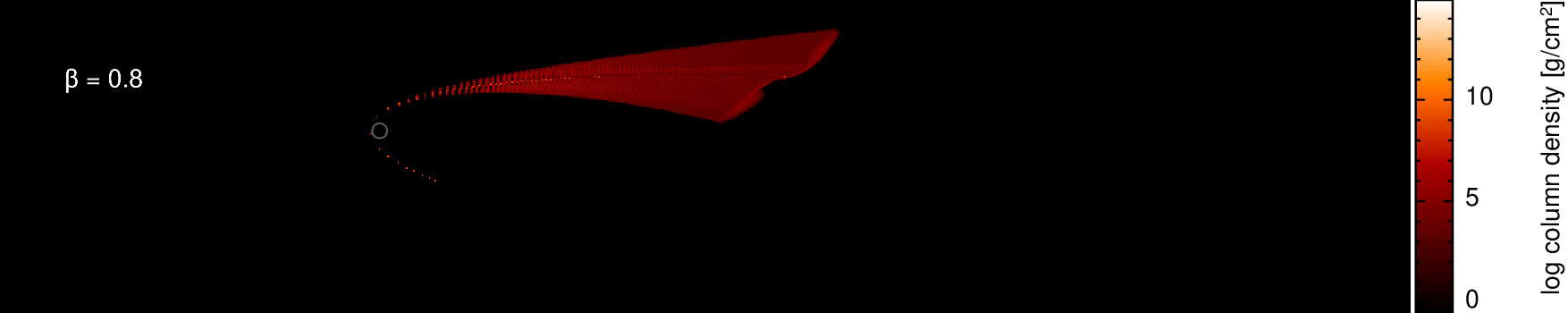}
        % \put(x,y) where x and y are % of the main image width/height
        \put(76,8){%
    \setlength{\fboxrule}{1pt} % Set border thickness
    \setlength{\fboxsep}{0pt}  % Set gap to zero
    \fcolorbox{white}{white}{\includegraphics[width=0.08\textwidth]{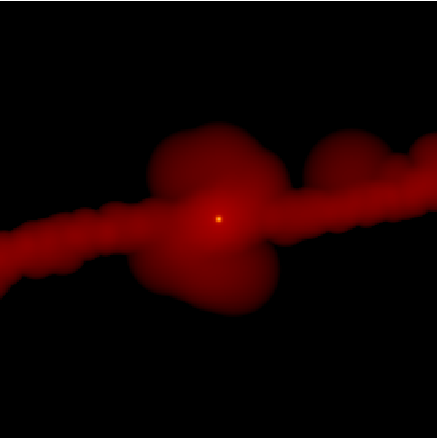}}}
    \end{overpic}
    
        \vspace{0.2cm}
        \begin{overpic}[width=\textwidth,keepaspectratio]{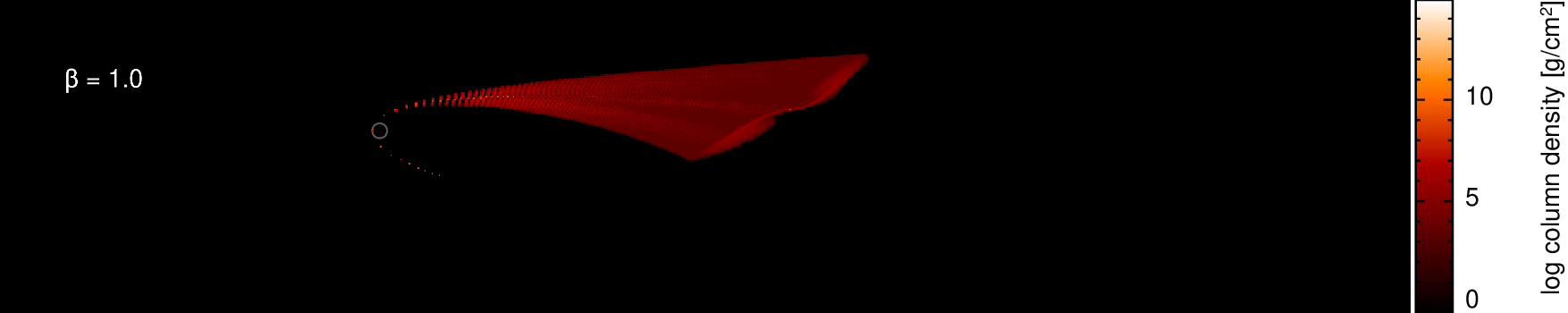}
        % \put(x,y) where x and y are % of the main image width/height
        \put(76,8){%
    \setlength{\fboxrule}{1pt} % Set border thickness
    \setlength{\fboxsep}{0pt}  % Set gap to zero
    \fcolorbox{white}{white}{\includegraphics[width=0.08\textwidth]{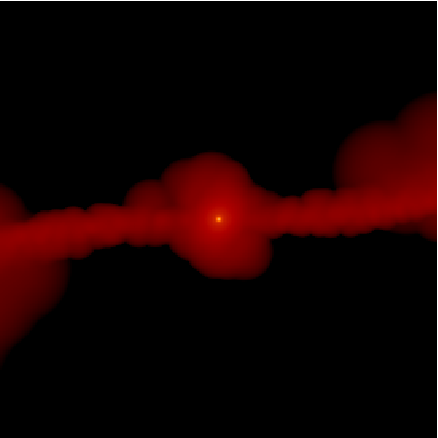}}}
    \end{overpic}
    
    \vspace{0.2cm}
        \begin{overpic}[width=\textwidth,keepaspectratio]{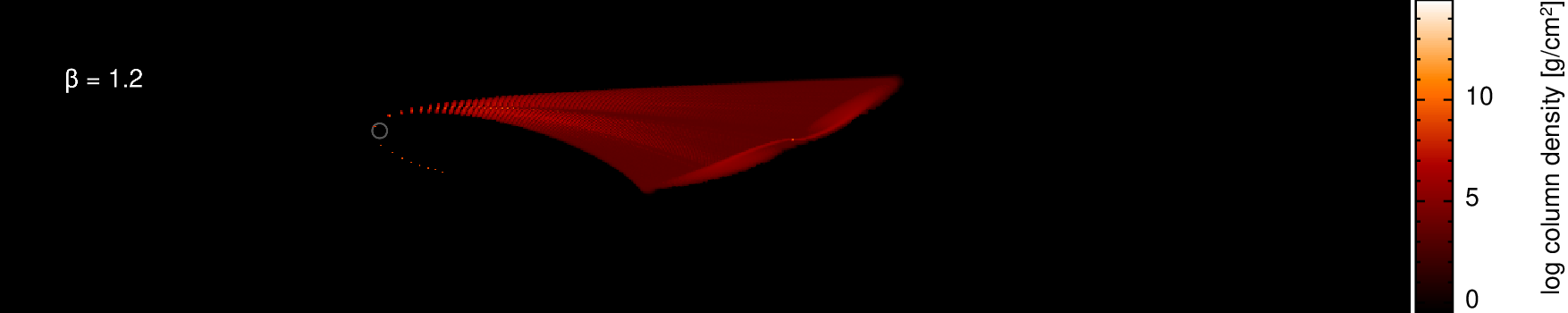}
        % \put(x,y) where x and y are % of the main image width/height
        \put(76,8){%
    \setlength{\fboxrule}{1pt} % Set border thickness
    \setlength{\fboxsep}{0pt}  % Set gap to zero
    \fcolorbox{white}{white}{\includegraphics[width=0.08\textwidth]{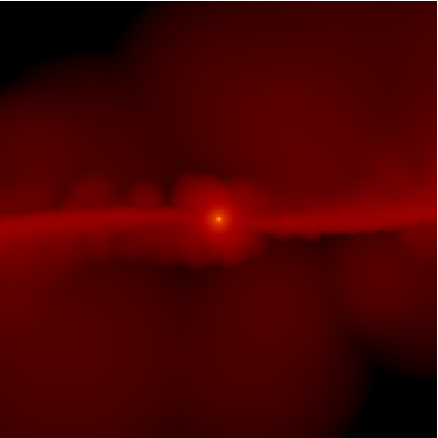}}}
    \end{overpic}
    
    \vspace{0.2cm}
        \begin{overpic}[width=\textwidth,keepaspectratio]{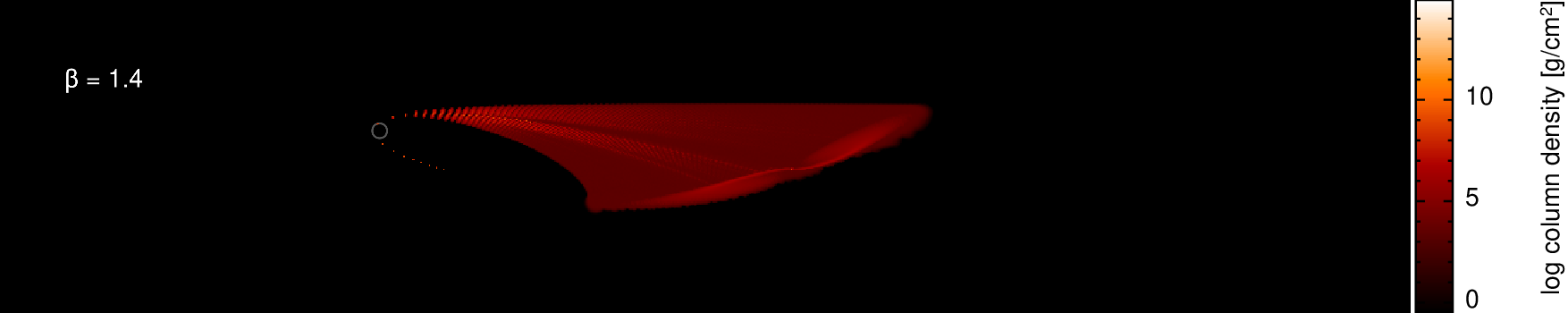}
        % \put(x,y) where x and y are % of the main image width/height
        \put(76,8){%
    \setlength{\fboxrule}{1pt} % Set border thickness
    \setlength{\fboxsep}{0pt}  % Set gap to zero
    \fcolorbox{white}{white}{\includegraphics[width=0.08\textwidth]{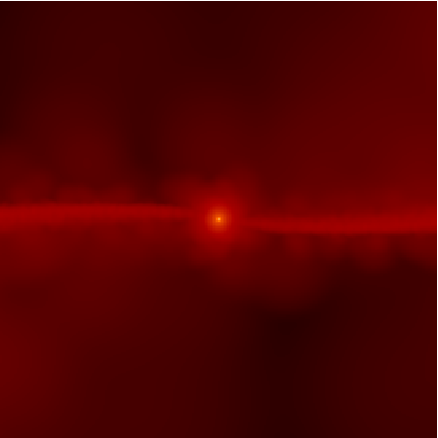}}}
    \end{overpic}
    
    \vspace{0.2cm}
        \begin{overpic}[width=\textwidth,keepaspectratio]{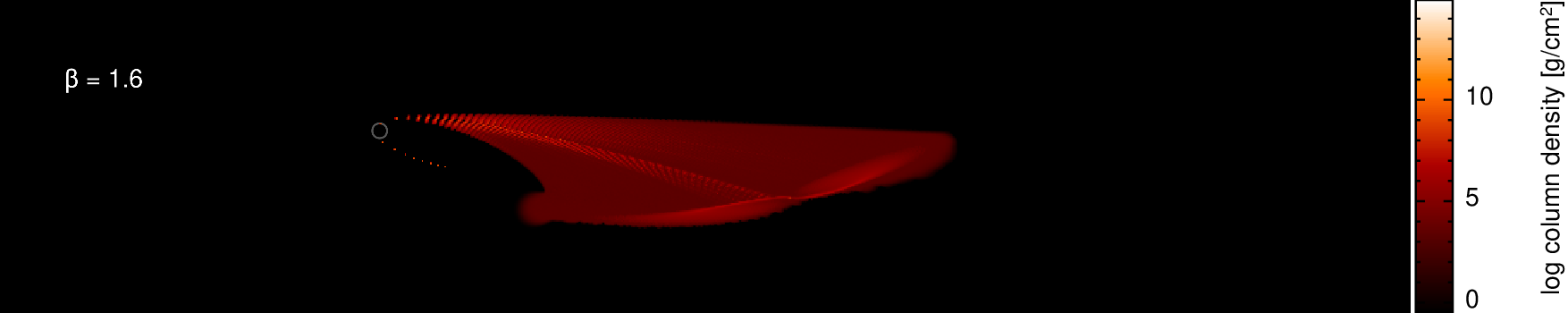}
        % \put(x,y) where x and y are % of the main image width/height
        \put(76,8){%
    \setlength{\fboxrule}{1pt} % Set border thickness
    \setlength{\fboxsep}{0pt}  % Set gap to zero
    \fcolorbox{white}{white}{\includegraphics[width=0.08\textwidth]{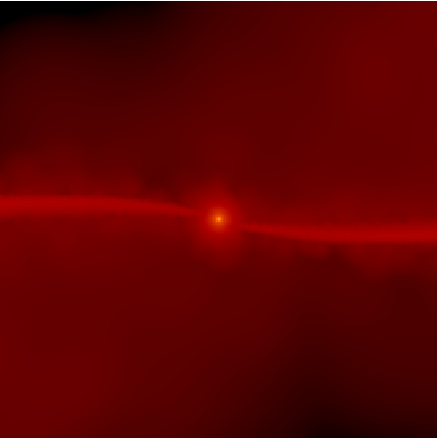}}}
    \end{overpic}
\caption{Debris evolution over the initial $4$ days of the simulation since $t=0$. Pericentre passage occurs around $6.8-7.6\;\mathrm{hours}$ in all models. Each panel shows different ratios of the tidal radius to the pericentre distance ($\beta$; top to bottom), showing the column density perpendicular to the orbital
plane ($y$-$x$). The grey circle corresponds to the tidal radius around the $10^{6}M_{\odot}$ black hole. Snapshots are shown super-imposed every $0.96$ hours. The star loses more mass
as it approaches the SMBH at a closer distance (i.e., higher $\beta$). Each panel is $88\,\mathrm{au}\times19.6\,\mathrm{au}$, respectively. Each inset of $1\;\mathrm{au}\times1\;\mathrm{au}$ focuses on the remnant wrt to the maximum density particle in the remnant at $4$ days. }
    \label{fig:partial_disruption}
    
\end{figure*}

\begin{figure*}
    \centering
    % The 'percent' option makes the grid 1-100 for easy positioning
    \begin{overpic}[width=\textwidth,keepaspectratio]{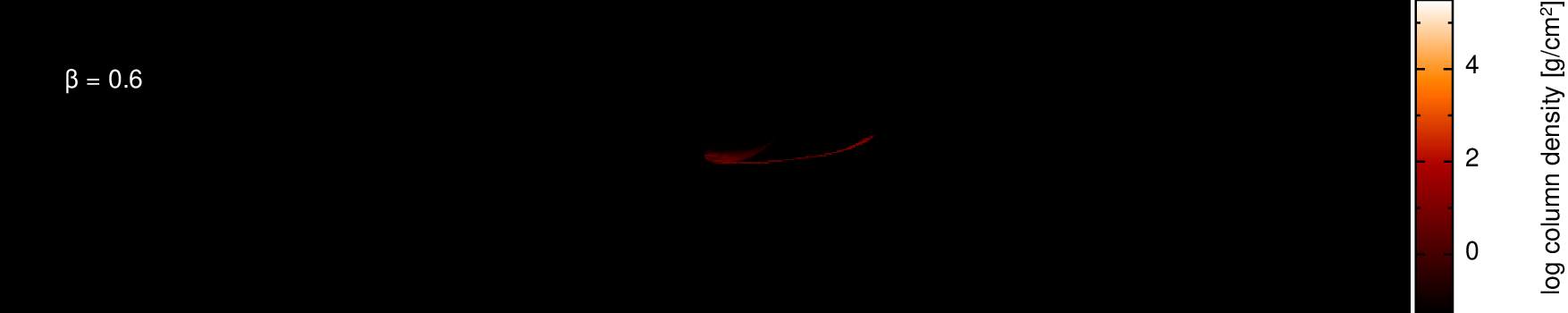}
        % \put(x,y) where x and y are % of the main image width/height
        \put(76,8){%
    \setlength{\fboxrule}{1pt} % Set border thickness
    \setlength{\fboxsep}{0pt}  % Set gap to zero
    \fcolorbox{white}{white}{\includegraphics[width=0.08\textwidth]{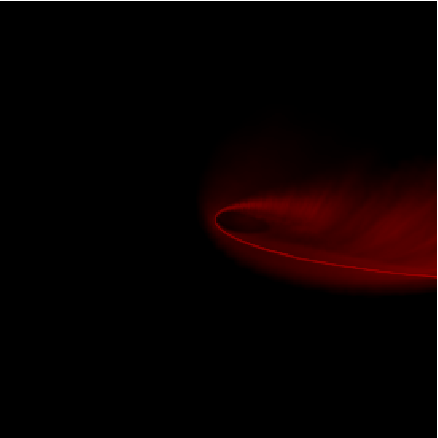}}}
    \end{overpic}
    
    \vspace{0.2cm}
        \begin{overpic}[width=\textwidth,keepaspectratio]{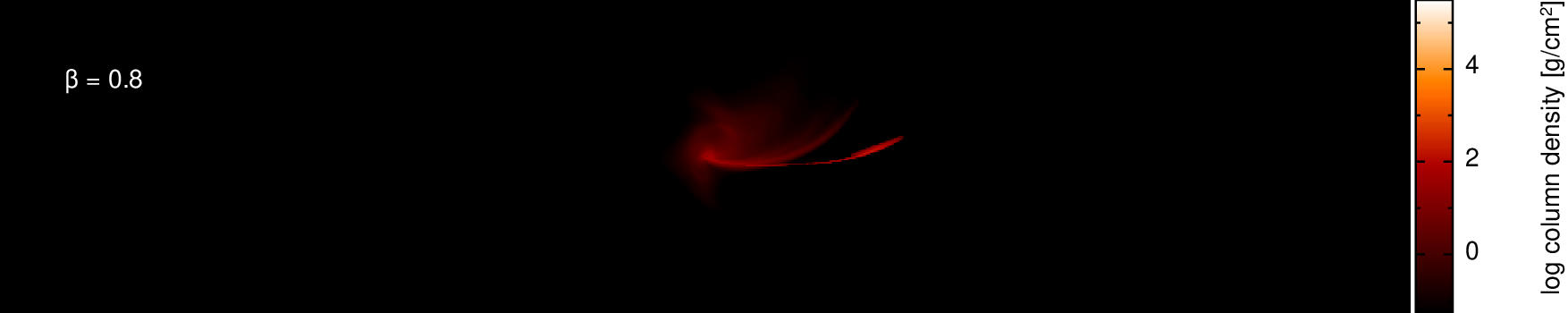}
        % \put(x,y) where x and y are % of the main image width/height
        \put(76,8){%
    \setlength{\fboxrule}{1pt} % Set border thickness
    \setlength{\fboxsep}{0pt}  % Set gap to zero
    \fcolorbox{white}{white}{\includegraphics[width=0.08\textwidth]{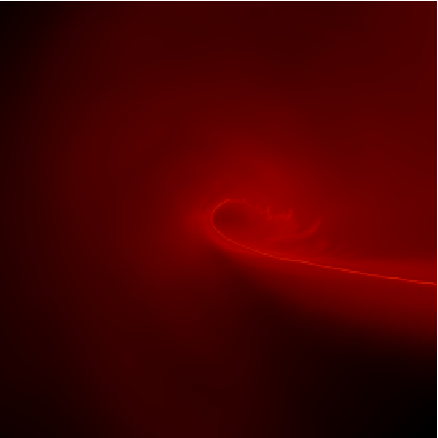}}}
    \end{overpic}
    
        \vspace{0.2cm}
        \begin{overpic}[width=\textwidth,keepaspectratio]{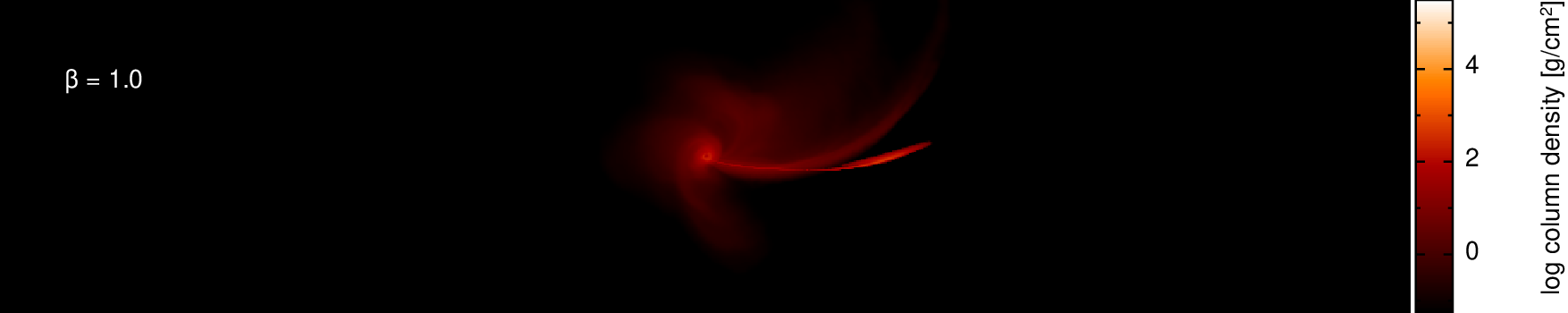}
        % \put(x,y) where x and y are % of the main image width/height
        \put(76,8){%
    \setlength{\fboxrule}{1pt} % Set border thickness
    \setlength{\fboxsep}{0pt}  % Set gap to zero
    \fcolorbox{white}{white}{\includegraphics[width=0.08\textwidth]{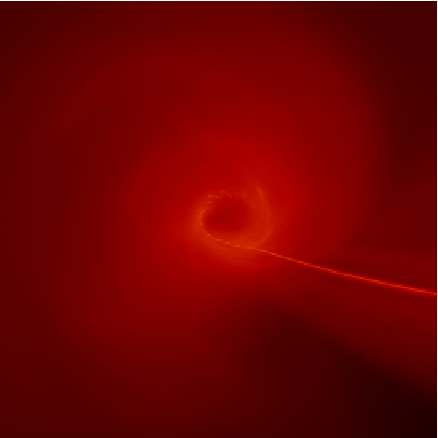}}}
    \end{overpic}
    
    \vspace{0.2cm}
        \begin{overpic}[width=\textwidth,keepaspectratio]{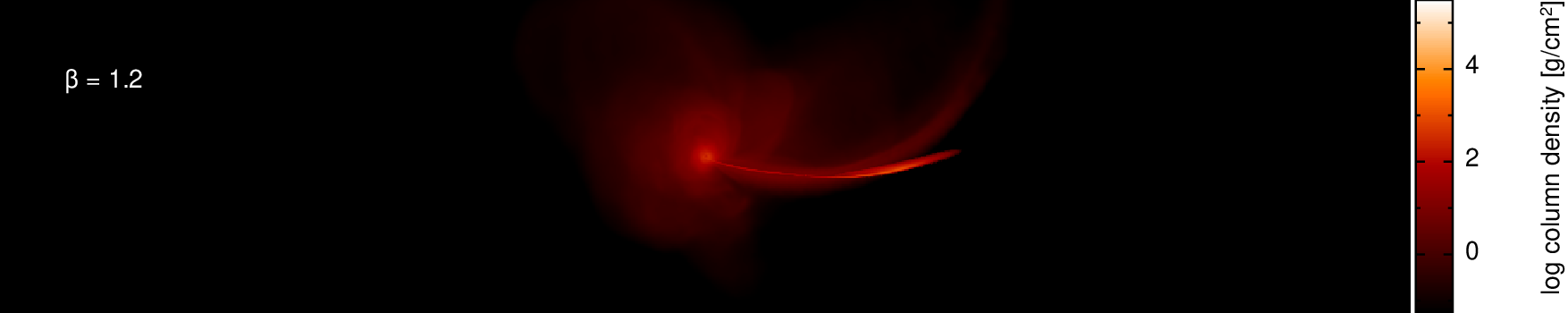}
        % \put(x,y) where x and y are % of the main image width/height
        \put(76,8){%
    \setlength{\fboxrule}{1pt} % Set border thickness
    \setlength{\fboxsep}{0pt}  % Set gap to zero
    \fcolorbox{white}{white}{\includegraphics[width=0.08\textwidth]{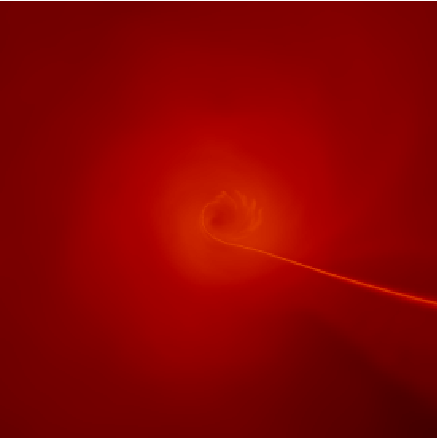}}}
    \end{overpic}
    
    \vspace{0.2cm}
        \begin{overpic}[width=\textwidth,keepaspectratio]{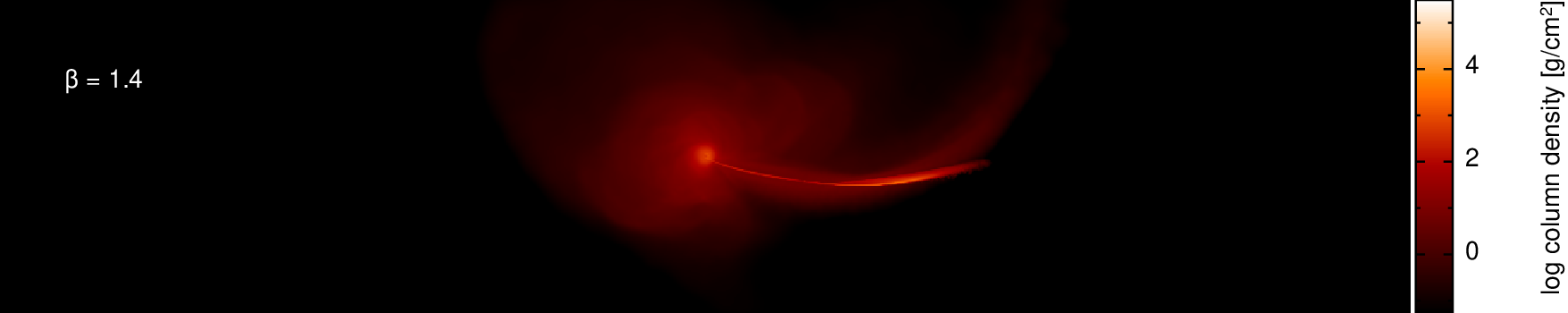}
        % \put(x,y) where x and y are % of the main image width/height
        \put(76,8){%
    \setlength{\fboxrule}{1pt} % Set border thickness
    \setlength{\fboxsep}{0pt}  % Set gap to zero
    \fcolorbox{white}{white}{\includegraphics[width=0.08\textwidth]{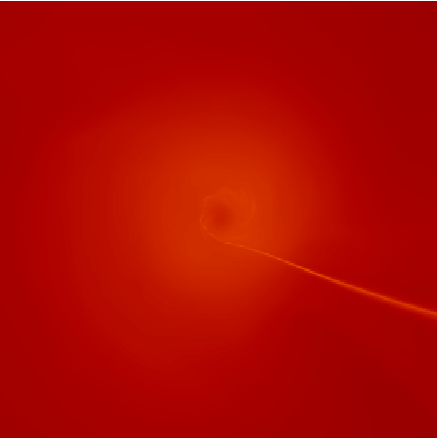}}}
    \end{overpic}
    
    \vspace{0.2cm}
        \begin{overpic}[width=\textwidth,keepaspectratio]{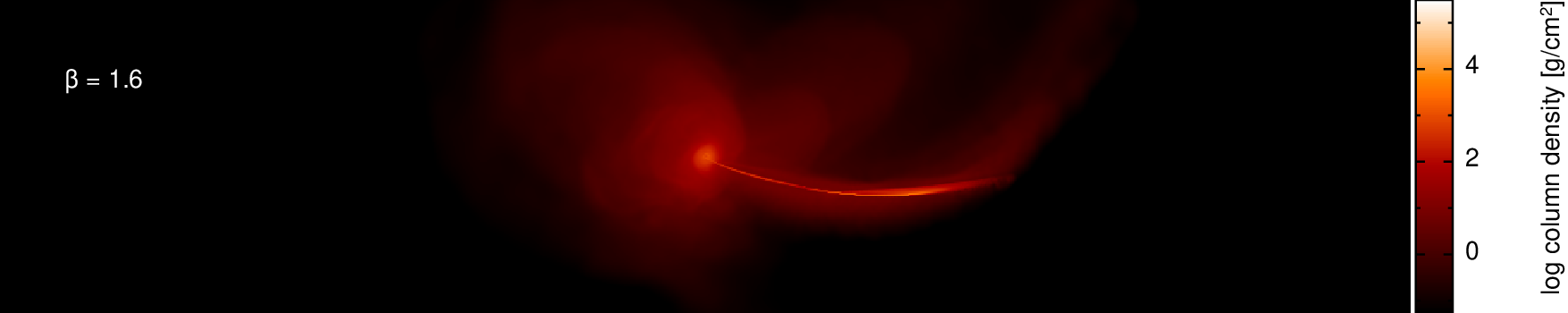}
        % \put(x,y) where x and y are % of the main image width/height
        \put(76,8){%
    \setlength{\fboxrule}{1pt} % Set border thickness
    \setlength{\fboxsep}{0pt}  % Set gap to zero
    \fcolorbox{white}{white}{\includegraphics[width=0.08\textwidth]{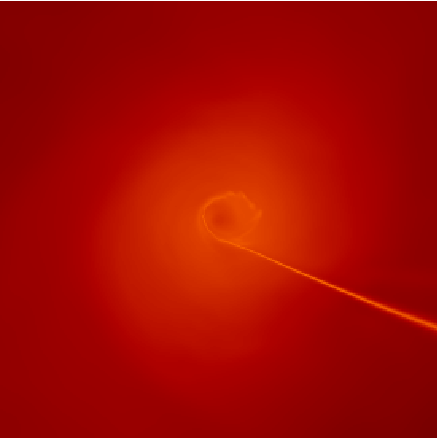}}}
    \end{overpic}

    \caption{Debris evolution, showing column density perpendicular to the orbital plane at $t=95$ days post-disruption for different ratios of the tidal radius to pericentre distance ($\beta$; top to bottom). The size of the reprocessing layer changes with the amount of material available in fallback material with low $\beta$ disruptions producing a small/low mass reprocessing layer compared to high $\beta$ disruptions. Each panel is $3\mathord,150\;\mathrm{au}\times700\;\mathrm{au}$, respectively. Each inset shows the $100\;\mathrm{au} \times 100\;\mathrm{au}$ zoomed region around the black hole.}
    \label{fig:longterm}
\end{figure*}

% \begin{figure*}
%     \centering
% 	\includegraphics[width=\textwidth,keepaspectratio]{beta06.eps}
% 	\includegraphics[width=\textwidth,keepaspectratio]{beta08.eps}
% 	\includegraphics[width=\textwidth,keepaspectratio]{beta10.eps}
% 	\includegraphics[width=\textwidth,keepaspectratio]{beta12.eps}
% 	\includegraphics[width=\textwidth,keepaspectratio]{beta14.eps}
%     \includegraphics[width=\textwidth,keepaspectratio]{beta16.eps}
    
% \end{figure*}

\begin{figure*}
      \begin{overpic}[width=\textwidth,keepaspectratio]{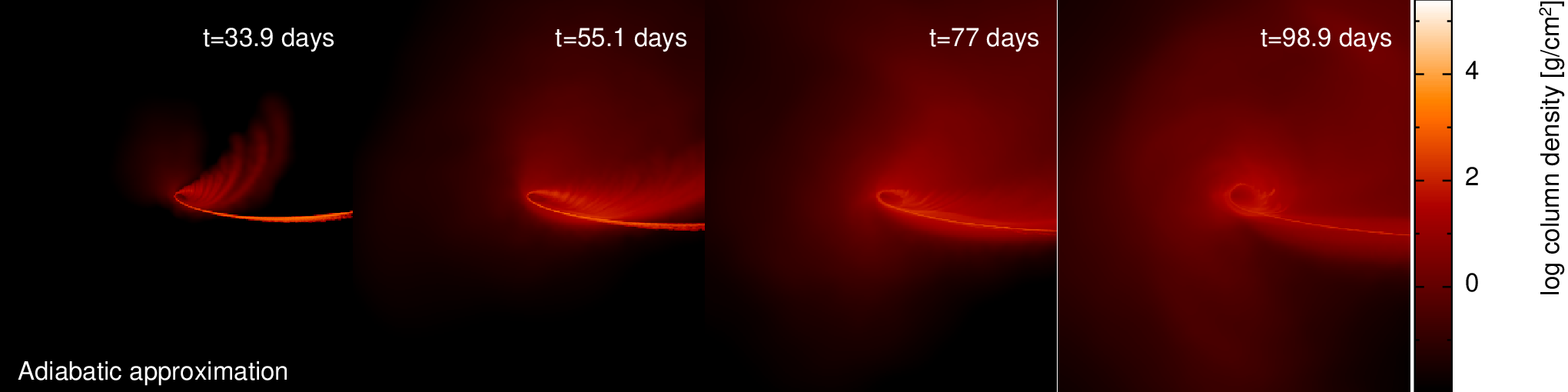}
    
    % First Inset Image
    \put(68,16){%
        \setlength{\fboxrule}{1pt}
        \setlength{\fboxsep}{0pt}
        \fcolorbox{white}{white}{\includegraphics[width=0.07\textwidth]{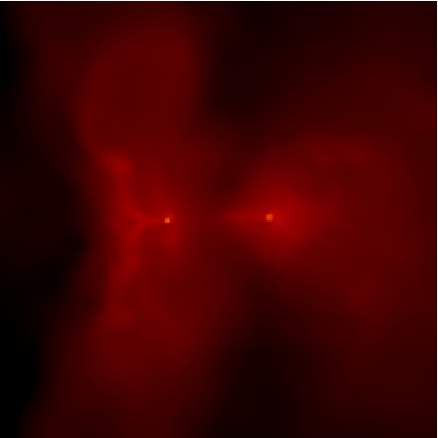}}
    }

    % Second Inset Image
    \put(68,1){%
        \setlength{\fboxrule}{1pt}
        \setlength{\fboxsep}{0pt}
        \fcolorbox{white}{white}{\includegraphics[width=0.07\textwidth]{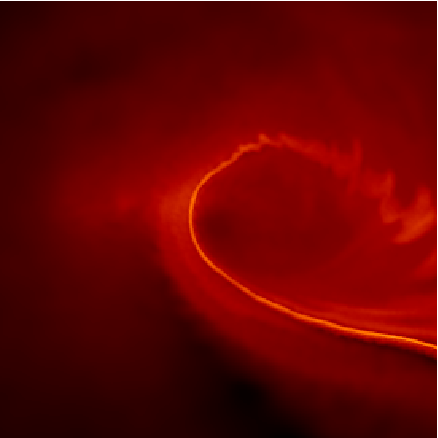}}
    }

    % The Text Label (Independent of the insets)
    % Adjust these coordinates to move the text freely
    \put(70,7){\color{white} $y-x$}

    \put(70,22){\color{white} $x-z$}

\end{overpic}
    % \end{overpic}
    
    \vspace{0.04cm}
        
        \begin{overpic}[width=\textwidth,keepaspectratio]{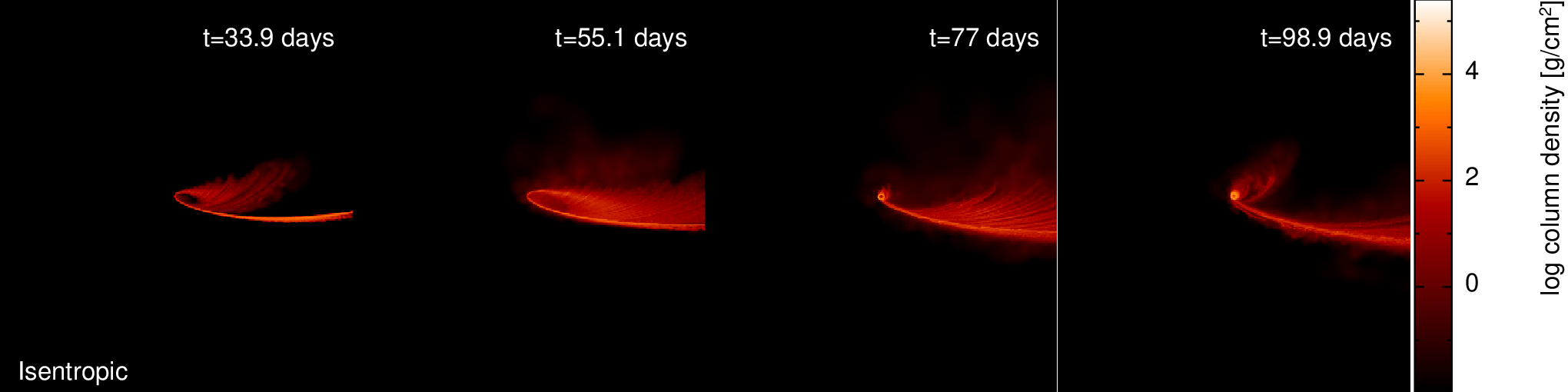}
    
    % --- UPPER INSET (at y=16) ---
    \put(68,16){%
        \setlength{\fboxrule}{1pt}
        \setlength{\fboxsep}{0pt}
        \fcolorbox{white}{white}{\includegraphics[width=0.07\textwidth]{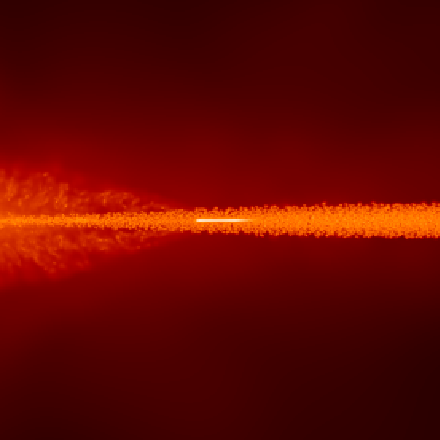}}
    }

    % --- LOWER INSET (at y=1) ---
    \put(68,1){%
        \setlength{\fboxrule}{1pt}
        \setlength{\fboxsep}{0pt}
        \fcolorbox{white}{white}{\includegraphics[width=0.07\textwidth]{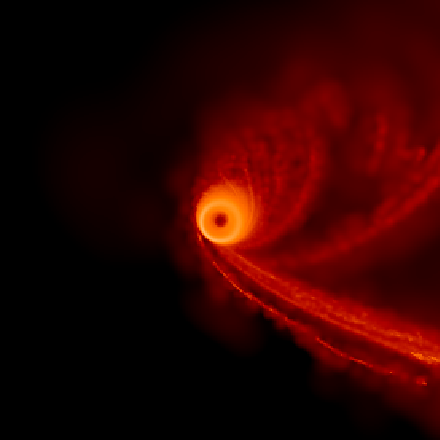}}
    }
 \put(70,7){\color{white}$y-x$}

    \put(70,22){\color{white}$x-z$}

\end{overpic}
    % \centering
    % \includegraphics[width=\textwidth]{beta08_adia.eps}
    %     \includegraphics[width=\textwidth]{beta08_isen.eps}
      \caption{Long term evolution of the $\beta=0.8$ simulation, using adiabatic (top row) and isentropic (bottom row) approximations for the heating and cooling. The material returns to the pericentre around $t=26$ days. Pericentre occurs at $t\sim 7.9\; \mathrm{hours}$. Stream-stream collisions at apocentre result in the formation of a disc and (in the adiabatic case; top row) outflow structure around the black hole. With fast cooling (bottom row), there is no overflow structure around the black hole, but a disc with more defined structure is formed as seen at $t=77$ days.  Each tile is $160\;\mathrm{au} \times 180 \;\mathrm{au}$. The insets on the disc at $98.9$ days are $30\;\mathrm{au}\times30\;\mathrm{au}$ and made when $y,z=0$, respectively, with density scales shown in column density scale. }
    \label{fig:beta08_long}
\end{figure*}

\begin{figure}
    \centering
    \includegraphics[width=\columnwidth]{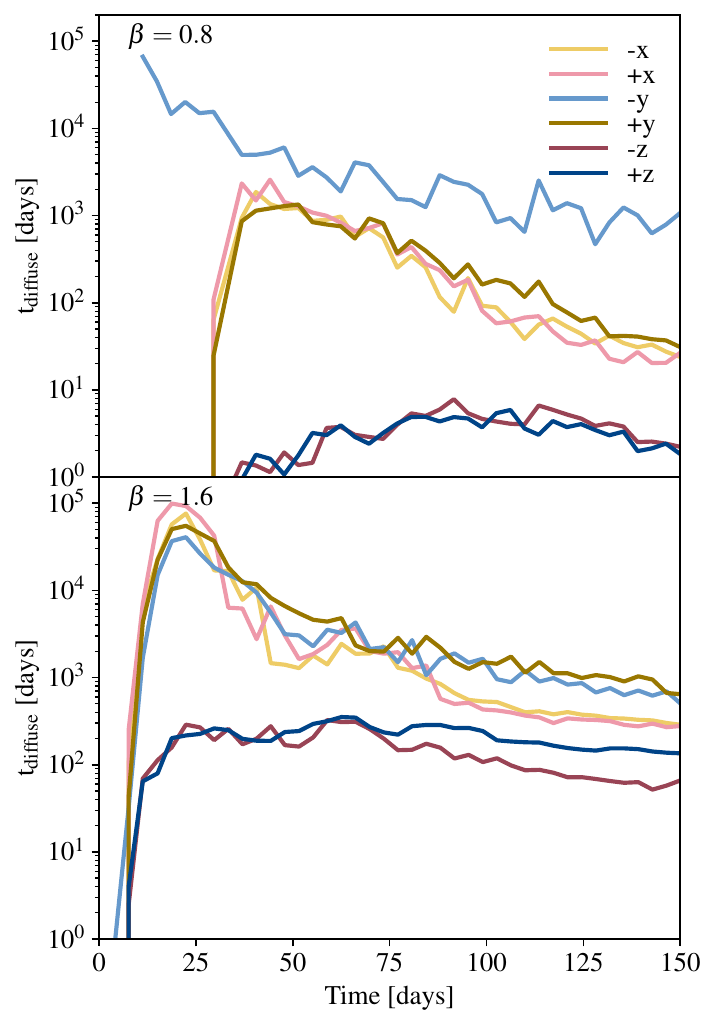}
      \caption{$t_\mathrm{diffuse}$ as function of time for $\beta=0.8$ (top panel) and $\beta=1.6$ (bottom panel) for adiabatic simulations.  The photon diffusion time is $\sim 1$ day for $\beta=0.8$, whereas for $\beta=1.6$, its about $100$ days.}
    \label{fig:diffuse_time}
\end{figure}

\begin{figure}
    \centering
    \includegraphics[width=\columnwidth]{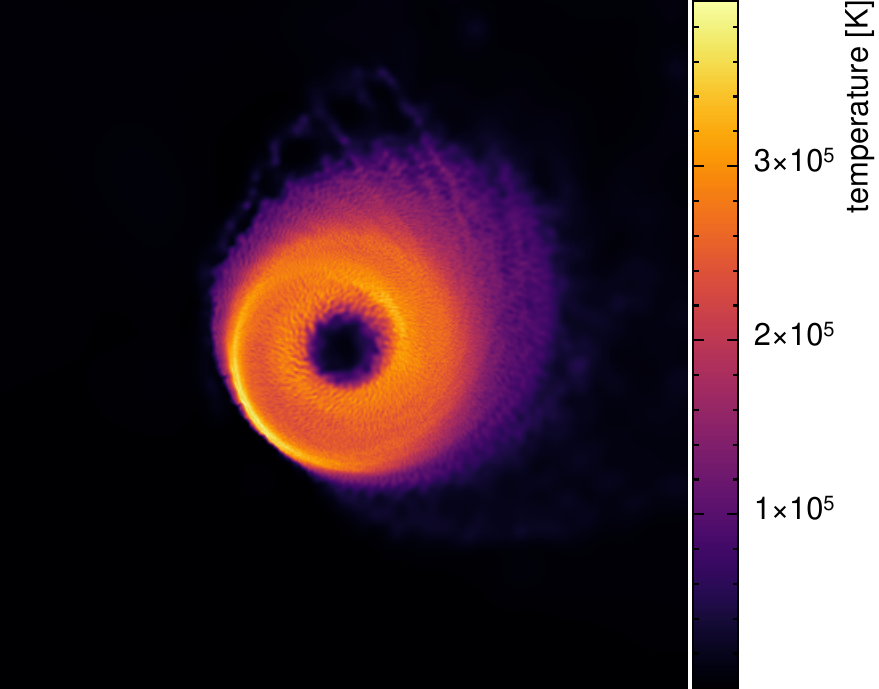}
      \caption{Temperature rendered $z=0$ slice in $y-x$ plane for $\beta=0.8$ isentropic simulation at $98.9\;\mathrm{days}$. Panel is $8\;\mathrm{au}\times8\;\mathrm{au}$. The peak temperature is $\sim4\times10^5\;\mathrm{K}$, implying hard X-ray emission.} 
    \label{fig:temp_isen}
\end{figure}

\begin{figure}
    \centering
    \includegraphics[width=\columnwidth]{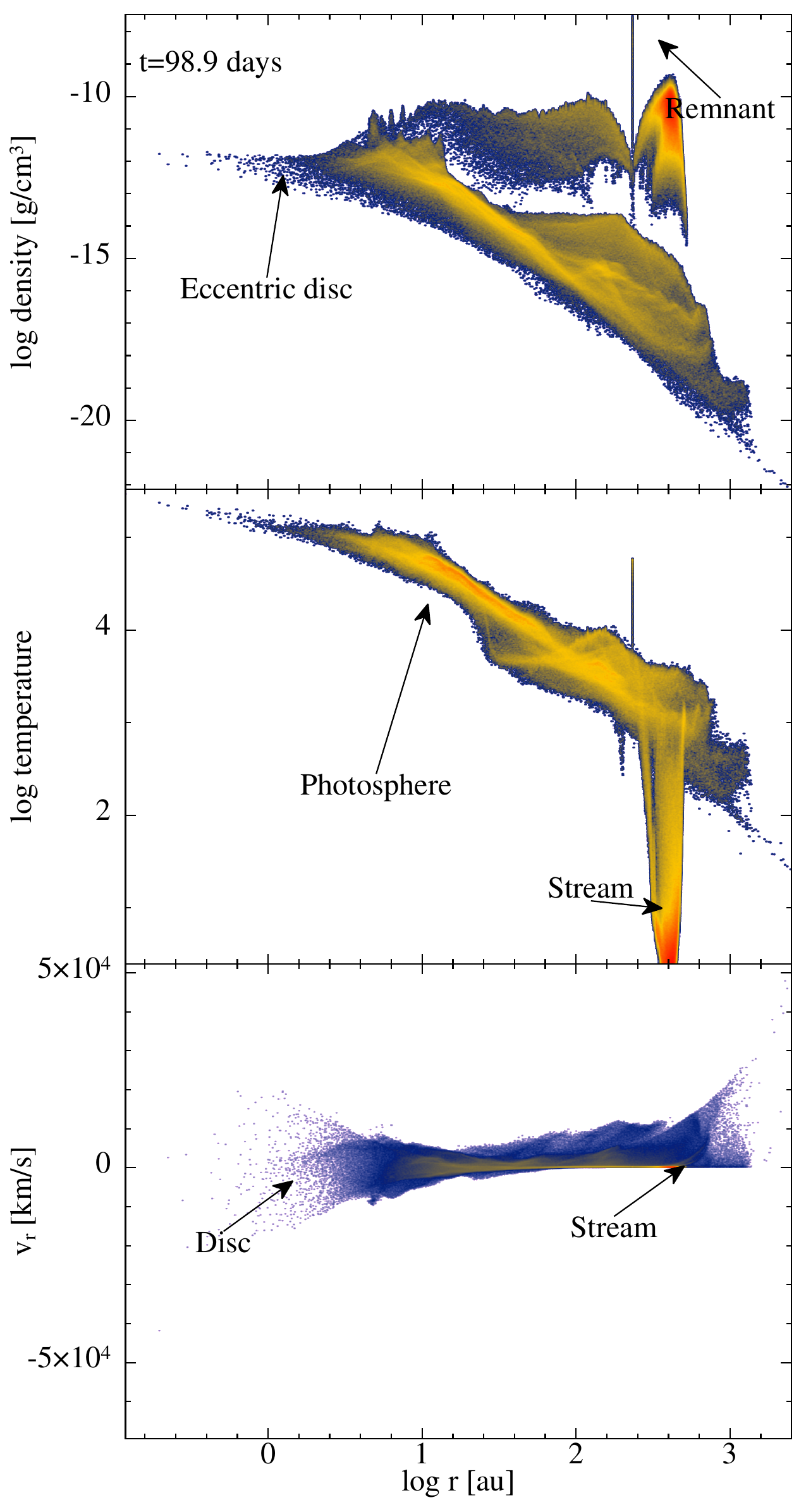}
      \caption{Spatial distribution of density (top), temperature (middle) and radial velocity (bottom) for $\beta=1.0$ disruption at $98.9$ days. Colours represent the density of points (mass per pixel) on the plot, with orange being a high density of points and blue being low. The x-axis represents the location of particles with respect to to the black hole placed at origin. The blackbody radius of this model is $32\;\mathrm{au}$. The stream is cold but high density, whereas the outflowing material has temperature $\sim10^4 \;\mathrm{K}$ but with densities $\lesssim 10^{-13}\mathrm{g}\,\mathrm{cm}^{-3}$. As the time progresses, the density of material close to the black hole decreases. }
    \label{fig:radial_profile}
\end{figure}

\begin{figure*}
    \centering
    \includegraphics[width=\columnwidth]{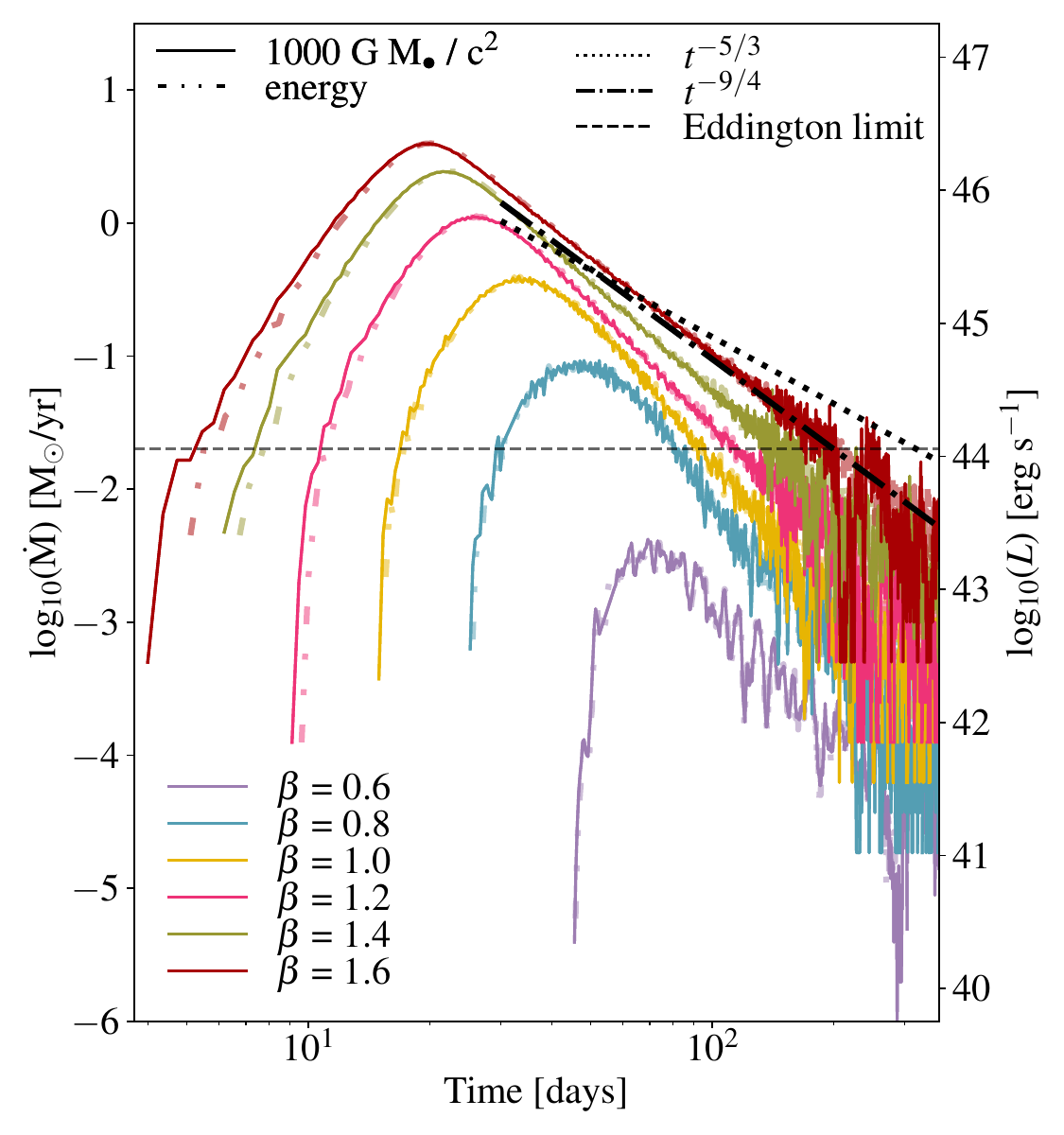}
    \includegraphics[width=\columnwidth]{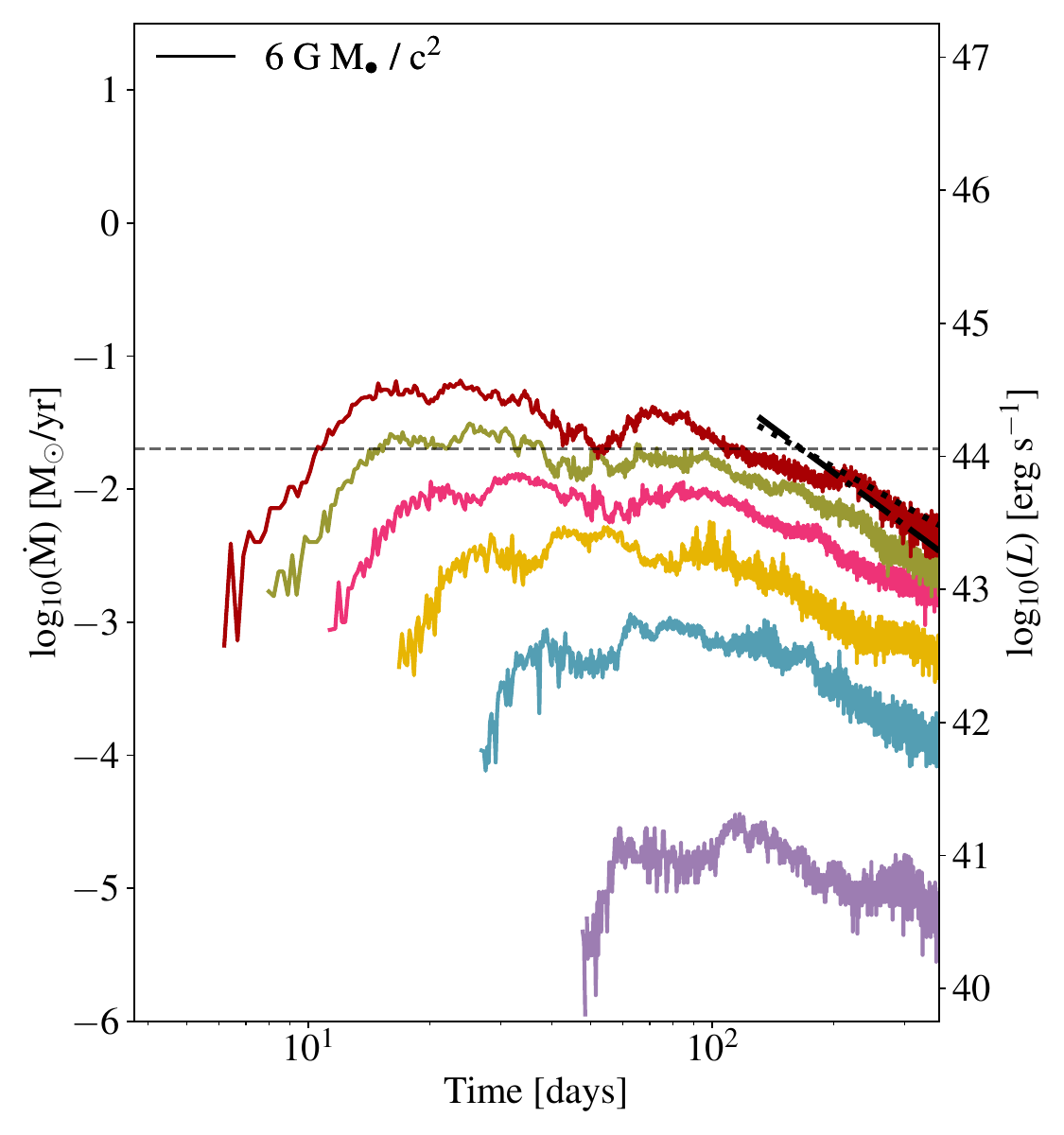}
      \caption{Mass fallback rate as a function of time for all our models determined using material that enters $1\mathord,000\; G M_\bullet /c^2$ (left panel, solid lines) and $6\;G M_\bullet/c^2$ (right panel, solid lines), and calculated using the Newtonian energy to determine period at $\sim4$ days post-disruption (left panel, dash-dot-dot line). Solid and dash-dot-dot curves in the left panel match within a few percent, with higher $\beta$ showing slight differences at the start of the curve due to the GR effects on particles as the simulation progresses. Dotted line shows the theoretically predicted $t^{-5/3}$ slope associated with fallback for a full disruption, which none of our simulation follow. Dashed-dotted line shows the $t^{-9/4}$ power law predicted for partial TDEs, which matches with our models. Dashed line represents the Eddington limit for a $10^{6}M_{\odot}$ black hole, assuming an accretion efficiency, $\eta=0.1$.  }
    \label{fig:mdot}
\end{figure*}

% \begin{figure}
%     \centering
%     \includegraphics[width=\columnwidth]{mdot_all_6.pdf}
%        \caption{\textbf{Mass accretion rate based on particles crossing $6G M_\bullet/c^2$ as function of time for all our models. We assume an accretion efficiency of $\eta=0.1$. The accretion rate is lower than mass fallback rate.} }
%     \label{fig:mdot_acc}
% \end{figure}

\begin{figure*}
    \centering
    \includegraphics[width=\textwidth]{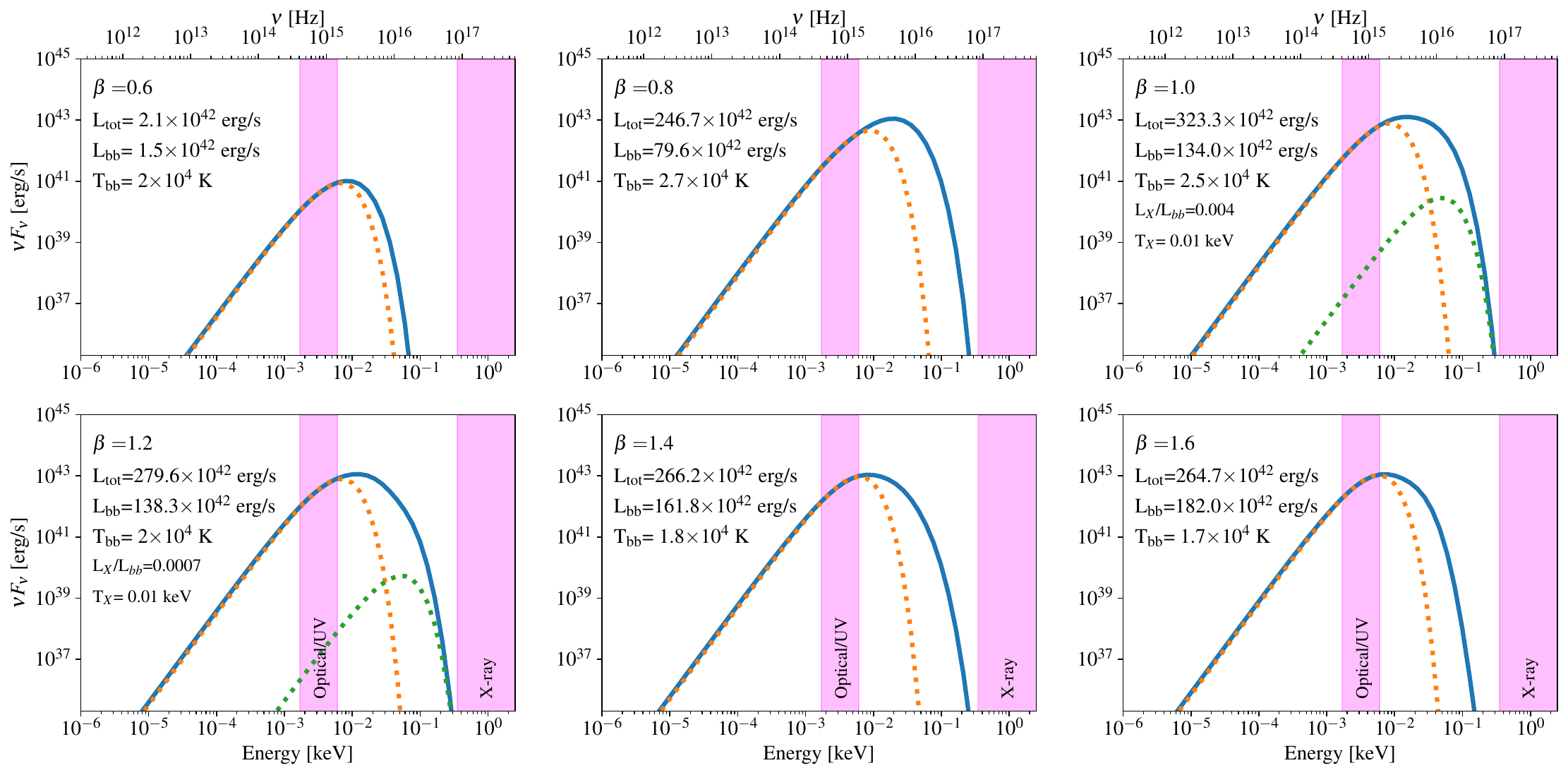}
      \caption{Synthetic spectral energy distributions computed for all models at $95$ days (blue line).  The blue line represents the observer in $z$ axis and orange shows single temperature blackbody fit to the optical band. The green line represents the blackbody fit to the X-ray band.} The total bolometric and inferred optical bolometric luminosity, and blackbody temperature for each model are listed in each panel. 
    \label{fig:sdf}
\end{figure*}
\begin{figure}
    \centering
    \includegraphics[width=\columnwidth]{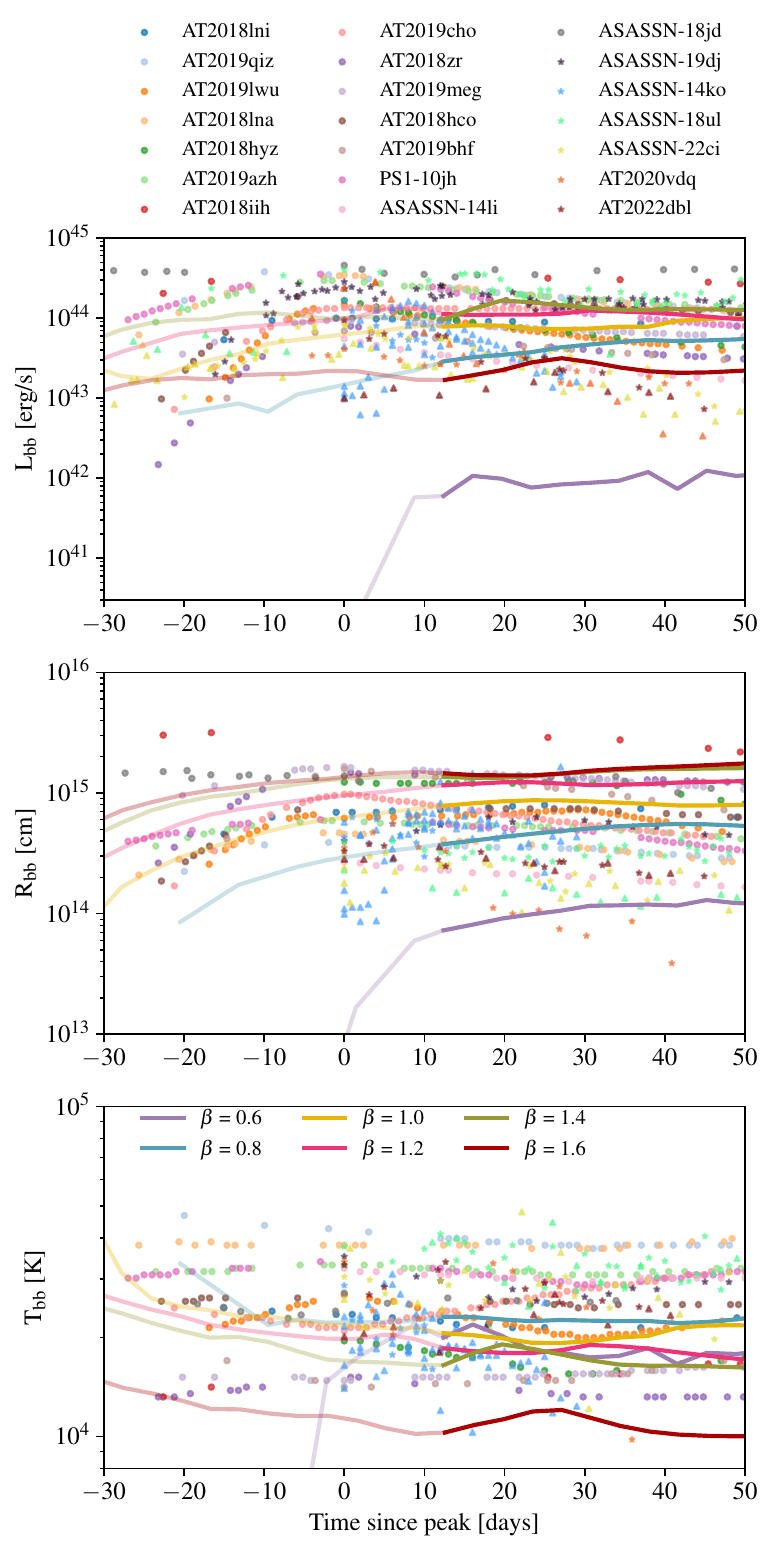}
    \caption{Evolution of bolometric (top panel) luminosity derived from the area under the orange curve in Figure~\ref{fig:sdf}, radius (middle panel) and temperature (bottom panel) with time for adiabatic simulations. Circle markers represent the observed full TDEs from \citet{Neustadt2020} and \citet{VanVelzen2021}. First flare of a repeating partial TDEs \citep{Hinkle2024,Makrygianni2025} is shown with a star marker and subsequent ones are shown with triangle.  Solid lines show simulation results for a range of $\beta$, with translucent part representing where the photosphere is unresolved in our post-processing (\citealt{Price2024}). The blackbody radii, luminosities and temperatures {\bf fall within the range of the observed population}.} 
    \label{fig:bolometric_radius}
\end{figure}
\section{Results}
\label{sec:results}
Figure~\ref{fig:partial_disruption} shows snapshots of column density integrated perpendicular to the orbital plane from the \textsc{Phantom} simulations. This figure shows all $\beta$ values considered in this work covering the first $4$ days of evolution post-disruption. We find that as the pericentre of the orbit is brought closer (increasing $\beta$), the black hole strips more material from the star and alters its orbit. Table~\ref{tab:all_sims} lists the Newtonian pericentre along with the the general relativistic pericentre obtained from the \textsc{Phantom-Geo} code. Compared to the Newtonian case, applying general relativity results in a closer approach to the black hole consistent with previous work \citep[][]{Sharma2024,Price2024}. Column $4$ lists the mass of the remnant in each simulation.  In simulations with $\beta=0.6$, the star loses about $1\%$ of its mass, compared with $\approx50\%$ for $\beta=1.6$. All models have pericentre times ranging from $\sim8.3-7.0\;\mathrm{hours}$, for longer times for lower $\beta$ value models.  

\begin{table*}
\caption{Properties of \textsc{Phantom} simulations. Column 1 lists the penetration factor, Column 2 lists the Newtonian pericentre distance, followed by Column 3 which gives the general relativistic pericentre. Column 4 lists the mass of the remnant at $\sim4$ days post-disruption. Columns 5 and 6 list the number of particles in the simulations post point mass replacement and particle splitting as well as the resultant SPH particle mass. Columns 7, 8 and 9 correspond to remnant properties such as eccentricity, period and escape velocity all calculated at $4\;\mathrm{days}$ post simulation start. All models have pericentre times ranging from $8.3 - 7.0\;\mathrm{hours}$.  
\label{tab:all_sims}}
    \centering
    % \begin{tabular}{rrrrp{1.7cm}p{1.85cm}}
    % \begin{tabular}{rrrrp{1.7cm}rrrr>{\raggedright\arraybackslash}p{1.7cm}>{\raggedright\arraybackslash}p{1.85cm}}
    % \begin{tabular}{rrrrrrrr>{\raggedright\arraybackslash}p{1.7cm}>{\raggedright\arraybackslash}p{1.85cm}}
    \begin{tabular}{rrrrrrrrll}
    \hline
    \hline
    $\beta$ & $r_\mathrm{p;N}$ &  $r_\mathrm{p;GR}$ & $M_\mathrm{rem}$ & $N_{\rm part}$ after splitting & mass per particle & $1-e$ & $P$ & $v_\infty$\\
   & ($\mathrm{R}_\odot$) & ($\mathrm{R}_\odot$) & ($\mathrm{M}_\odot$)& & ($\mathrm{M}_\odot$) & & ($\mathrm{years}$) & ($\mathrm{km \;s^{-1}}$) \\
    \hline
       0.6 & 167.87 & 163.52 & 0.99 & 1,008,930 & $9.86\times10^{-12}$ &$7.1\times10^{-5}$  & 1,148& ---\\
       0.8 & 125.90 & 121.51 & 0.98 & 1,021,464 & $1.85\times10^{-8}$  & $3.2\times10^{-5}$ & 2,398 & ---\\
       1.0 & 100.72 & 96.29 & 0.93 & 1,076,800 & $6.25\times10^{-8}$  & $5.5\times10^{-5}$ & 785 & ---\\
       1.2 & 83.94 & 79.47 & 0.84 & 1,296,363 & $1.25\times10^{-7}$ & $2.3\times10^{-5}$ & 2,241 & ---\\
       1.4 & 71.94 & 67.42 & 0.68 & 1,263,959 & $2.50\times10^{-7}$ & ---& --- &  257\\
       1.6 & 62.95 & 58.38 & 0.48  & 1,030,399 & $5.00\times10^{-7}$& --- & --- & 500\\
       \hline
    \end{tabular}
\end{table*}
On a $E=0$ orbit about half of the material has $E<0$ and other half follows $E>0$. When the disruption takes place, a large fraction of material stays in the form of a remnant, but the stripped material is bound and unbound, with about half of the stripped material ending up bound to the black hole. At $t=4$ days, we calculated the Newtonian energy of the remnant with respect to the black hole to determine if it is bound or unbound. We find that the remnant stars are bound to the black hole on highly eccentric orbits for $\beta \leq 1.2$, whereas for higher $\beta$ the remnant star becomes unbound. Using this Newtonian energy, we determined $v_\infty$ (the velocity at $r=\infty$). We list the eccentricity and period of the bound stars in columns $7$ and $8$, and velocity of unbound remnant in column $9$.  None of the stellar remnants in our simulations have velocities high enough to escape the nucleus of their host Galaxy ($\sqrt{2G M_\bullet/r_\mathrm{t}}\sim 10^4\;\mathrm{km\;s^{-1}}$). For the case of the Galactic centre, then we expect that these stellar remnants could potentially contribute to the estimated $10\mathord,000-100\mathord,000$ remnants from TDEs \citep{Alexander2001,Manukian2013}. 

Figure~\ref{fig:longterm} shows the snapshots at $95$ days for the same simulations as in Figure~\ref{fig:partial_disruption}, but with the core of the stellar remnant replaced by a point-mass particle. The time when material starts to return to the pericentre varies with $\beta$, with smaller $\beta$ implying later accretion. For example, at $\beta=0.8$, material returns to the pericentre after around $30$ days, whereas for $\beta=1.6$, the time to return is shorter, at about $7$ days.  The simulation behaviour is similar to what \citet{Price2024} found for their full tidal disruption of a polytropic star around a $10^6\;\mathrm{M}_\odot$ black hole. After the material passes through the pericentre nozzle shock, the resultant
apsidal precession leads to the stream-stream collisions at apocentre (the outer shock), which in turn allows material to fall towards the black hole and drive kinetic outflows.  Only a small amount of material forms a Keplerian disc around the black hole, and most material ends up as part of the expanding cloud of material.  The size of this cloud varies between simulations, with higher $\beta$ calculations producing clouds roughly ten times larger than lower $\beta$ at around $100$ days. 

Figure~\ref{fig:beta08_long} shows the evolution of material in the $\beta=0.8$ simulation, where the star loses $\sim 2\%$ of its mass during the encounter. Top panel shows the adiabatic approximation where the energy is trapped or advected rather than radiated, and bottom panel shows the isentropic case where the energy is efficiently radiated. A disc with outflows forms around the black hole, where the disc dissipates as the simulation progresses in the adiabatic case. We only considered isentropic simulations of $\beta \leq 0.8$ and found that they do not power large scale outflows, due to the assumption of efficient radiative cooling as shown in the bottom panel of Figure~\ref{fig:beta08_long}. The insets in the last panel show slices of the mass density in linear scale for the disc, edge-on and in-plane. The disc not an isolated circular disc even for our isentropic calculation. The angular momentum transport is mainly driven by stream induced shocks, rather than any slow viscous transport.
  
To evaluate whether isentropic or adiabatic modelling more accurately represents the system, we calculated the photon diffusion time, $t_\mathrm{diffuse}$ as 
\begin{equation}
    t_\mathrm{diffuse} \sim \frac{1}{c} \int \kappa r \rho \mathrm{d}r\;.
\end{equation}
We sum the contributions from each SPH particle intersecting a line of sight along each coordinate axis \citep{Price2024}.  The top panel of Figure~\ref{fig:diffuse_time} shows the $t_\mathrm{diffuse}$ as function of time for the adiabatic simulation with $\beta=0.8$. The $\pm z$, $t_\mathrm{diffuse}\sim 1$ day suggests that radiative cooling is efficient, making an isentropic model a more suitable approximation. Conversely, for $\beta=1.6$, in the adiabatic simulation $t_\mathrm{diffuse}\sim100$ days in $\pm z$ direction.  In this regime, energy is primarily transported outward by mechanical outflows before it can be radiated through the photosphere. In \citet{Price2024}, the $t_\mathrm{diffuse}\sim10$ days, whereas in our simulations it is slightly longer. Figure~\ref{fig:temp_isen} shows the temperatures in our isentropic, $\beta=0.8$ simulation (computed using Equation~\ref{eq:temp_calc}). We find temperatures $\sim 10^5\;\mathrm{K}$, which would result in soft X-ray emission. The disc is eccentric because material is falling back onto it.  

To understand the properties of the outflows, we plot in Figure~\ref{fig:radial_profile} the density (top), temperature (middle), and radial velocity (bottom) in the $\beta=1.0$ simulation at $98.9$ days (we chose this time for this plot as our simulations converge up to $100$ days post-disruption), with temperature calculated using Eq.~\ref{eq:temp_calc}.  The x-axis shows the radial distance from the black hole placed at origin. We note that as the material passes close to the black hole, heating from nozzle and pancake shocks spreads gas onto wider range of orbits around the black hole, although this spreading can be overestimated at low resolution \citep{Hu2025}. The stream has a density roughly six orders of magnitudes higher than the outflowing envelope ($\sim 10^{-16} \; \mathrm{g/cm^3}$). The outer layers of the remnant are visible, resulting from the smoothed point-mass potential. As the model evolves, some particles form an envelope around the point-mass remnant, which remain part of the stream.  The stream is cooler than the outflowing envelope, with the outflowing material having a temperature of approximately $10^4\;\mathrm{K}$, expected of UV/optical bands.  The outflowing material has velocity $\sim 10^4\;\mathrm{km\;s^{-1}}$, and the photosphere extends to $\sim 100\;\mathrm{au}$. This behaviour is similar to what was found by \citet{Price2024} for their full disruptions.  For other $\beta$ values at the same time, we find that the material's radial velocity becomes less spread as $\beta$ increases as material returns faster due to stronger gravitational effects of the black hole.  Our stream is two dex colder than \citet{Steinberg2024} due to them only considering the radiation term in their temperature calculation, whereas we consider both gas and radiation pressure terms (Equation~\ref{eq:temp_calc}), and find that our stream is gas pressure dominated.Furthermore, the temperature and density of the stream increases with $\beta$. In an isentropic simulation, a dense, hot accretion disc forms. 

To understand how much mass accretes onto the black hole, and thus the partial TDEs mass loss rate and luminosity evolution, we plot in Figure~\ref{fig:mdot} the mass fallback rate ($\dot{M}$) as a function of time for all the models over one year.  The right $y$-axis shows the luminosity ($L = \eta \dot{M}c^{2}$ assuming $\eta=0.1$). This is not the true luminosity but the maximum possible luminosity if all energy was radiated locally. In our adiabatic simulations, this is not the case as energy is carried out by kinetic outflows as discussed below.  We determine the mass fallback rates by tracking material that enters within a radius of $1\mathord,000\; G M_\bullet/c^2$. Any particle that enters this distance once is considered as part of the fallback rate and is ignored from the subsequent calculations.  This differs from the method used in \citet{Fitz2024}, who had used a radius of $150\;G M_\bullet/c^2$ and determined the slope. This is because in our partial TDE simulations, material can escape this smaller radius as the nozzle shock forms.  We compared our calculated mass fallback of material to the mass fallback determined by calculating the Newtonian energy and the Keplerian orbital periods at $4$ days since the beginning of the simulations (dash-dot-dot lines). We find a good match between our simulations and energy calculated predictions, but note that the mass fallback rates are steeper than the theoretically predicted $t^{-5/3}$ power-law for full TDEs, but similar to, if not slightly shallower than the $t^{-9/4}$ predicted from earlier partial TDE simulations \citep{Guillochon2013,Coughlin2019,Ryu2020}, and has also been observed in some optical/UV TDEs \citep[e.g.,][]{Charalampopoulos2023,Makrygianni2025}.  The slower decline of the accretion rate derived in our simulations compared to previous full TDE simulations is a result of the remnant affecting the energy of the bound material, and availability of less accretion material post-disruption \citep{Coughlin2019}.  Furthermore, for smaller $\beta$ values, we find that the accretion peaks at later time compared to larger $\beta$ values, with the mass fallback rate peaking $\approx80$ days for $\beta=0.6$ and $\approx20$ days for $\beta=1.6$.  \citet{Guillochon2013} found that smaller
values of $\beta$ reduce the debris energy spread, creating lower, more delayed peak fallback rates, which is consistent with our result. We note that effects such as magnetorotational instability (MRI) can increase the accretion rate due to turbulent stress \citep{Balbus1999}, but this work ignores any magnetic field effects.  

With the exception of $\beta=0.6$, the mass fallback rate at and around peak is super-Eddington, consistent with that expected from TDEs (i.e., if the debris  circularises efficiently forming a disc, the disc will be a geometrically thick, radiation pressure dominated slim disc). A significant fraction of the matter is expected to be blown off forming outflow. Despite this the accretion rate onto the black hole can be super-Eddington as shown in panel (b) of Figure~\ref{fig:mdot} for $\beta\geq1.4$ \citep[e.g.,][]{Dai2018}. However, at late times with $t >100$ days the mass accretion becomes sub-Eddington in nature as we expect a slim disc to transition into a thin disc. 

We now consider the frozen-in approach to derive the Eddington time ($t_\mathrm{Edd}$). The minimum time at which the bound material starts to return back to the pericentre, i.e., $t_\mathrm{min}$ after the disruption is given by \citep{Lodato2011}
\begin{equation}
    t_\mathrm{min} = 41 \left(\frac{M_\bullet}{10^6\mathrm{M}_\odot}\right)^{\!\nicefrac{1}{2}}\left(\frac{R_*}{\mathrm{R}_\odot} \right)^{\!\nicefrac{3}{2}} \left(\frac{M_*}{\mathrm{M}_\odot} \right)^{\!-1} \beta^{-3}\,\mathrm{days}\;.
\end{equation}
The frozen-in approximation does not hold true in the numerical simulations \citep{Lodato2009,Mockler2025}, and explains $4$ times lower values of $t_\mathrm{min}$ in our simulations. Assuming that the partially disrupted material follows mass fallback given by 
\begin{equation}
    \dot{M}_\mathrm{fb} = \dot{M}_\mathrm{peak} \left(\frac{t}{t_\mathrm{min}}\right)^{\!\nicefrac{-9}{4}}\;,
\end{equation}
and that all the bound stripped mass ($\Delta M$) would eventually fall onto the black hole, we can write 
\begin{align}
    \Delta M &=\int_{t_\mathrm{min}}^{\infty} \dot{M}_\mathrm{fb} \mathrm{d}t \;,
\end{align}
which gives 
\begin{equation}
    \dot{M}_\mathrm{peak} = \frac{5}{4} \frac{\Delta M}{t_\mathrm{min}}\;.
\end{equation}
Using Eddington mass accretion rate of
\begin{equation}
    \dot{M}_\mathrm{Edd} = 1.4 \times 10^{24} \left(\frac{M_\bullet}{10^6\, \mathrm{M}_\odot} \right)\left(\frac{0.1}{\eta} \right) \;\mathrm{g\;s^{-1}}\;,
\end{equation}
and equating this to $\dot{M}_\mathrm{fb}(t_\mathrm{Edd})$ one can find 
\begin{align}
t_\mathrm{Edd} &= 650\,\mathrm{days}\; \left(\frac{\Delta M}{\mathrm{M}_\odot}\right)^{\nicefrac{4}{9}} \left(\frac{M_\bullet}{10^6\, \mathrm{M}_\odot}\right)^{\nicefrac{-1}{6}} \left( \frac{M_*}{\mathrm{M}_\odot}\right)^{\nicefrac{5}{9}}  \nonumber \\ &\qquad\qquad\quad \left(\frac{R_*}{\mathrm{R}_\odot} \right)^{\!\nicefrac{5}{6}} \,\left( \frac{\eta}{0.1}\right)^{\!\nicefrac{4}{9}}\beta^{\nicefrac{-5}{3}}\;.
\label{eq:tedd}
\end{align}
 We test the Equations~\ref{eq:tedd} and find that our estimated Eddington times are within $30-10\%$ for $\eta=0.1$ compared to our simulations for $\beta \geq 0.8$.

To understand how mass accretion rate onto the black hole based on the particles crossing $6\;G M_\bullet/c^2$ compares with the mass fallback rate, we plot this as function a of time in the right panel of Figure~\ref{fig:mdot}. Only $\beta=1.6$ and $\beta=1.4$ actually accrete material at a super-Eddington rate. Since this really measures material that enters the last stable orbit it would not result in any observational signature, but rather represents the actual mass growth rate of the black hole. At late times ($t>130\;\mathrm{days}$), the mass growth rate follows close to both $t^{-5/3}$ and $t^{-9/4}$.

Figure~\ref{fig:sdf} shows the synthetic spectral energy distributions for all simulations at $95$ day post-disruption. We chose this time to show the results as we show calculations of the bolometric lightcurves, radii and temperature for up to $100$ days.  We observe soft X-ray emission in most of our simulations except $\beta=0.6$ (Appendix~\ref{app:spectral_evo}). For all $\beta$ values, the SEDs peak in the optical/UV band, consistent with that seen from the population of optical/UV discovered TDEs \citep[e.g.,][]{VanVelzen2020}. The fitted blackbody curves for optical and X-ray are given in \textit{orange} and \textit{green}, respectively, and the calculations shown in the following plots use this fit. 

Figure~\ref{fig:bolometric_radius} shows the optical luminosity (top) inferred blackbody radius (middle) and fitted blackbody temperature (lower) of our simulations compared with observed TDE events obtained from \citet{Neustadt2020,VanVelzen2021,Hinkle2024}.  Observational data are plotted in terms of time since peak bolometric luminosity, and the simulated data are shifted by $50$ days due to difficulty in determining the peak in our luminosity curves. The shift helps us to plot the curves closer to the peaks in the observational data. A different shift would result in the peak of our calculated values for radii matching with the observations at a different time.  The blackbody radius remains within $10^{14}-10^{15}\;\mathrm{cm}$ ($10-100\;\mathrm{au}$), closely matching both partial and full disruptions.  But unlike observations \citep[e.g.,][]{VanVelzen2020} where post-peak TDE blackbody radii decay slightly, ours stays flat similar to \citet{Fitz2024} and \citet{Price2024}. Moreover, the inferred blackbody radii increase with $\beta$.

For all simulations with $\beta > 0.6$, the simulated luminosities range between $10^{42}\;\mathrm{erg\,s^{-1}}$ and $10^{44}\;\mathrm{erg\,s^{-1}}$, and the temperatures are between $\sim 10^4-2\times 10^4 \;\mathrm{K}$. This {\bf comparable to} the bolometric luminosities and blackbody temperatures of population of observed full and partial TDEs which have luminosities between $10^{43}-3 \times 10^{44}\;\mathrm{erg\,s^{-1}}$ and temperatures between $\sim 10^4-5\times 10^4 \;\mathrm{K}$ \citep{Gezari2021}.  \textit{ASASSN-18ul} is an outlier \citep{Wevers2023}. \citet{Pasham20204} showed that the higher luminosity of this event is not a result of AGN activity. \citet{Wen2024} argued that the luminosity can be explained due to the presence of a secondary supermassive black hole.   Compared to observations which show a decrease in the luminosity post peak, the persistence of the disc and absence of radiative cooling in our simulations results in flat lightcurves.  We note that lower resolution simulations show declining lightcurves, but this is a numerical artefact caused by the lack of disc formation. 
 
As the blackbody radius is the most robustly determined quantity in our simulations, we compare it in detail with a sample of observed partial and full TDEs.  Our $\beta=0.6$ model matches the periodic event \textit{AT2020vdq} around $20-40$ days with $\beta=1.0$ lying closer to the peak. $\beta=0.8$ curve lies close to the periodic event  \textit{ASASSN-18ul} for the first flare, whereas $\beta=0.6$ is closer to the second flare. Periodic events like \textit{ASASSN-22ci} and \textit{AT2022dbl} lie close to $\beta=0.8$ at the start with the radius decreasing with time. $\beta=1.0$ is close to the periodic event \textit{ASSASN-14ko}. \textit{ASSASN-14ko} is an interesting case with $5$ flares which repeat every $\sim100$ days, with similar values. It also matches with the full TDE events such as \textit{AT2019azh}, \textit{AT2018lni} and \textit{AT2018hco}. $\beta=1.2$, $\beta=1.4$ and $\beta=1.6$ lie close to full TDEs such as \textit{ASASSN-18jd}, \textit{AT2018zr} and \textit{AT2019meg}. Overall, we propose that periodic TDEs may be partial TDEs, as currently suggested in the literature, and our results indicate that also some events currently classified as full TDEs could, instead, be partial events. This is consistent with shallower encounters and less complete disruptions found in the \textsc{Mosfit} fits of \citet{Mockler2019} and \citet{Nicholl2022} for non repeating TDEs (i.e., assumed full). 

\begin{figure}
    \centering
    \includegraphics[width=\columnwidth]{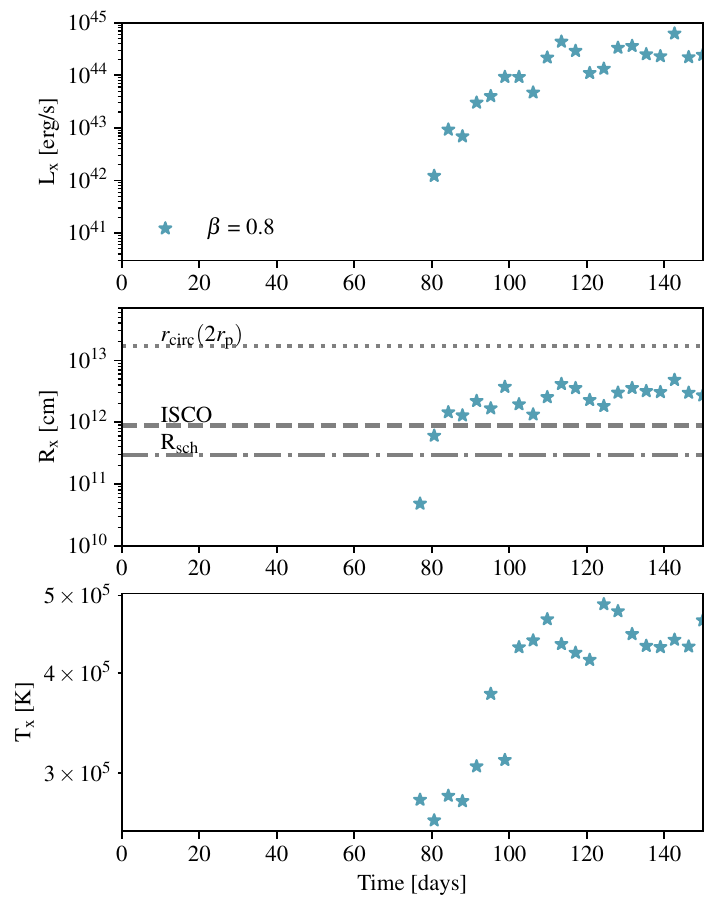}
    \caption{X-ray luminosity, radius and temperature of $\beta=0.8$ isentropic simulation. Dotted line corresponds to circularisation radius of $2\;r_\mathrm{p}$, dashed is the ISCO and dash-dotted is the Schwarzschild radius.}
    \label{fig:x_ray_lum}
\end{figure}

We also compare our isentropic simulations of $\beta=0.6$ and $\beta=0.8$ with observed X-ray emission. Our isentropic $\beta=0.6$ calculation has zero X-ray luminosity up to $170\;\mathrm{days}$ as the disc forms slower due to less material available. Hence, we ignore this simulation and only plot the X-ray luminosity, radius and temperature from the $\beta=0.8$ simulation in Figure~\ref{fig:x_ray_lum}. The blackbody fit radius lies closer to the Schwarzschild radius, and never exceeds the circularisation radius of $2\;r_\mathrm{p}$.
% Notably, $\beta=0.6$ simulation lies close to the observations of \textit{AT2020vdq} at later times, which is currently classified as a partial TDE. Our $\beta=0.8$ model matches \textit{AT2019qiz} at early times, and \textit{AT2019azh} at later times. Both these events are classified as full TDEs. Our $\beta=1.0$ lies close to classified partial TDE, \textit{ASASSN-18ul} at later times, and a full TDE \textit{AT2018lni} at early times. Our $\beta\ge1.2$ models match with full TDE events, \textit{AT2018zr}, \textit{AT2019meg}, and \textit{AT2018iih}. 

% \clearpage

\section{Discussion}
\label{sec:discussion}
We performed a suite of partial TDE simulations across a range of penetration factors ($\beta$) corresponding to both the partial and full disruption of a $1\;\mathrm{M}_\odot$ star around a $10^6\;\mathrm{M}_\odot$ black hole on zero energy orbits in the Schwarzschild metric. Such simulations allow us to probe the effect of different $\beta$ have on the mass accreted onto the black hole.  Our results demonstrate that even partial TDEs (i.e., where the mass loss is $<50\%$) can produce outflows with velocities of order $\sim 10^4 \;\mathrm{km \;s^{-1}}$, sufficient to obscure any underlying accretion disc (Figures~\ref{fig:longterm}-\ref{fig:beta08_long}). The outer layer of the photosphere has temperatures of about $10^4\;\mathrm{K}$ (Figure~\ref{fig:radial_profile}) which would correspond to the Balmer lines seen in optical spectra \citep{vanVelzen2011}. This morphology closely resembles that of full disruptions found in recent simulations \citep{Fitz2024,Price2024}.  The formation time and extent of the envelope depend on the impact parameter, with lower $\beta$ values resulting in longer fallback timescales and delayed envelope development (Figure~\ref{fig:beta08_long} and Figure~\ref{fig:mdot}).

%\textbf{add nozzle shock explanation}

Such an optically thick layer explains the prevalence of UV/optical detections of TDEs. Radiation pressure from Eddington-limited accretion slows the debris fallback, creating a quasi-spherical, outflowing envelope as long as the cooling time is longer than the dynamical timescale \citep{VanVelzen2020,Mockler2025}. This is what we see in our simulations, where stream-stream collisions at apocentre lead to accretion-powered outflows which form the `Eddington envelopes' that smother the black hole \citep{Price2024}.

Our simulations with mass loss less than $7\%$ i.e., $\beta < 0.8$ are sub-Eddington and others are super-Eddington for some time around the $10^{6}\;\mathrm{M}_{\odot}$ black hole.  Isentropic models show that efficient radiation would not form a reprocessing layer for $\beta=0.6$ and $\beta=0.8$ (Figure \ref{fig:beta08_long}). Hence, pure emission from the disruption would be observed instead of reprocessed lightcurve. The truth lies somewhere in the middle, as all our simulations which are super-Eddington would become sub-Eddington after some time (Figure~\ref{fig:mdot}).  Our isentropic result is similar to \citet{Chen2021} who proposed, based on their analytic model, that partial TDEs would not result in any outflows due to efficient radiative diffusion. Contrary to their result, we find that accretion rates of our partial TDEs with $\beta \geq 0.8$ are super-Eddington, and outflows can form a reprocessing layer. \citet{Hu2025} performed converged high-resolution simulations showing that stream fanning at the pericentre nozzle shock can be overestimated at low resolution, but is negligible in reality.  The converged result, however, is similar to that found by \citet{Price2024} where the stream continues at high Mach number through the nozzle shock, and subsequent stream-stream collisions at apocentre cause material to plunge towards the black hole and form an optically thick envelope. \citet{Kubli2025} found the same result but neither they nor \cite{Hu2025} were able to simulate to the resultant stream-stream collision and envelope formation directly.  \citet{Fitz2024} also found that stream-stream collisions lead to outflows and an optically thick reprocessing layer in their simulations of tidal disruptions of stars on initially bound orbits, without any resolution issues associated with the nozzle shock. 

The inferred blackbody radii obtained from our simulations agree well with TDEs classified as repeating partial \citep{Hinkle2024} and full events \citep{Neustadt2020,VanVelzen2021} as shown in Figure~\ref{fig:bolometric_radius}. A colour correction by a factor close to canonical value of $\sim 1.7$ on our temperature \citep{Shimura1995,Davis2006} makes both our temperatures and luminosities match better with observed values. \citet{Fitz2024,Price2024} had not considered such a correction to their simulations, but had argued that looking deeper into the photosphere would result in higher temperatures (i.e., considering lower opacity, see Figure~5 in \citealt{Fitz2024}).

% But our temperatures are about $2 \times$ lower than the observations. This is because we are not considering radiative transfer properly, and looking deeper into the photosphere would result in higher temperatures from our simulations (i.e., considering lower opacity, see Figure~5 in \citealt{Fitz2024}). This is likely because our assumed temperature does not take into account the difference between the last scattering surface and the thermalization surface in highly scattering-dominated flows. 

We note that repeating TDEs, \textit{AT2020vdq} and \textit{ASASSN-18ul} match with our $\beta=0.6$ adiabatic model, but as the accretion is not super-Eddington as shown in Figure~\ref{fig:mdot}, radiation would escape efficiently, and outflows would not form.   Our other $\beta$ models match with repeating partial events such as  \textit{AT2019azh}, \textit{ASSASN-22ci} and \textit{ASSASN-14ko}, \textit{AT2022dbl}, and full TDEs such as \textit{AT2018hco}, \textit{AT2019azh}, \textit{AT2018lni}, \textit{ASASSN-18jd}, \textit{AT2019meg} and \textit{AT2018zr} (Figure~\ref{fig:bolometric_radius}).  Our X-ray emission for $\beta=0.8$ is within a dex of the X-ray TDEs reported by \citet{Auchettl2017, Wevers2023} (Figure~\ref{fig:x_ray_lum}). The radius obtained from our simulations lies $\sim 0.5\;\mathrm{dex}$ of the Schwarzschild radius, with the first data point lying lower than the $R_\mathrm{sch}$. This offers an explanation for the current observations of X-ray TDEs with radii lower than the $R_\mathrm{sch}$ \citep{Gezari2021}. Our findings support the results from \textsc{Mosfit} \citep{Mockler2019} where low impact parameters fits can also match detections classified as full TDEs.  Our results suggest for the need of caution when classifying events as full TDEs from the upcoming surveys such as LSST \citep{Bricman2020} which would detect thousands of TDEs over $10$ years. Misclassification would have an effect on the determined black hole masses \citep{Mockler2019,Nicholl2022,Angus2026} as the masses from these fits may be systematically biased. But one also has to be careful about the effect of the mass of star, black hole and the impact parameter on the fits. 

Finally, we find that the remnants produced at $\beta \leq 1.2$ remain bound on highly eccentric orbits, implying the potential for recurrent partial disruptions. Although our simulations start on zero energy orbits, and have periods post-disruption that are much longer than those of currently detected repeating partial TDEs. More realistic eccentric ($e \sim 1$; \citealt{Zhong2014}) stellar orbits could yield shorter recurrence times, consistent with observations.

Our study has several limitations.  We employ simplified equations of state and neglect radiative cooling, recombination, and detailed opacities.  These factors likely affect the decline of bolometric luminosities as the accretion-induced shock heating is either entirely trapped or entirely radiated in our simulations. Furthermore, we replace the remnant with a softened point mass to accelerate computations and use particle splitting to achieve a high resolution.  Our simulations only converge up to $100$ days post-disruption (Appendix~\ref{app:res_study}).  
Future work is needed with higher resolution simulations to confirm this picture.  Radiation hydrodynamics also needs to be incorporated to more accurately model energy transport and outflow evolution.

\section{Conclusions}
\label{sec:conclusion}
In summary, we simulated the partial disruptions of $1\;\mathrm{M}_\odot$ Sun-like star around a $10^6\;\mathrm{M}_\odot$ black hole using the GRSPH code \textsc{Phantom}. Here we varied the ratio of tidal radius to pericentre distance,  $\beta$, between $0.6$ and $1.6$ such that during the disruption mass loss is $\approx 1 - 50\%$. From our simulations, we found that:
\begin{enumerate}
\item Partial TDEs can produce mass fallback rates that can exceed the Eddington limit for $\beta \geq 0.8$, in which case they exhibit morphologies similar to full TDEs both theoretically \citep{Fitz2024,Price2024} and observationally \citep[e.g.,][]{VanVelzen2021,Makrygianni2025}, powering optically thick outflows that form a reprocessing layer around the black hole. But these fallbacks eventually become sub-Eddington, with shorter transition time for lower $\beta$ values.  
\item The calculated mass fallback rates from energy distributions are steeper than $t^{-5/3}$ and follow close to the predicted $t^{-9/4}$ power-law \citep{Guillochon2013,Coughlin2019}.
\item The blackbody radii calculated from the synthetic lightcurves produced from our adiabatic simulations range from $10\,\mathrm{au}$--$100\,\mathrm{au}$ and increase with the impact parameter.  They approximately agree with the observations of full and repeating partial TDEs within $0.3$ dex. Luminosities agree with the observations within $\sim 0.2$ dex. Our work provides an explanation for the optical-ultraviolet TDEs \citep{VanVelzen2020,Makrygianni2025}. Our isentropic simulation results in radii close to the Schwarzschild radius, similar to observations \citep{Gezari2021}.
\item Our results suggest that most detected TDEs may be partial, supporting work presented in the literature \citep{Mockler2019,Nicholl2022,Angus2026}. Our results also match the observed repeating partial TDEs \citep{Hinkle2024}, but we only perform zero energy orbits in this paper. Further work with stars on eccentric orbits is required to model these events rigorously. 
\end{enumerate}

%% Please use the acknowledgment and contribution environments. This will 
%% be anonomyized when the "anonymous" style option is used. 

\begin{acknowledgments}
This work
was supported by resources awarded under Astronomy Australia Ltd’s ASTAC merit allocation scheme on the OzSTAR and Ngarrgu Tindebeek national
facilities at the Swinburne University of Technology, and gadi at the National Computing Initiative (NCI) by the Monash-NCI scheme.  OzSTAR receives funding from the Australian Government and the Victorian Government. AH was supported by the Australian Research Council (DP240101786, DP240103174). MS acknowledges support from Monash Graduate Scholarship and Monash International Tuition Scholarship. Parts of this research were supported by the Australian Research Council Discovery Early Career Researcher Award (DECRA) through project number DE230101069. DP thanks IPAG, CNRS and the University of Grenoble-Alpes for hospitality and support during his sabbatical.

\end{acknowledgments}
\section*{Data Availability}
Our simulation files and code used for plotting can be found at  \url{https://dx.doi.org/10.5281/zenodo.19696366}.

\software{ \textsc{Numpy} \citep{Harris2020}, \textsc{Scipy} \citep{Virtanen2020},
\textsc{Matplotlib} \citep{Hunter2007}, \textsc{Splash} \citep{Price2007}. \textsc{Kepler} \citep{Weaver1978}, \textsc{Phantom} \citep{Price2018}
          }


\appendix

\begin{figure}
    \centering
    \includegraphics[width=\columnwidth]{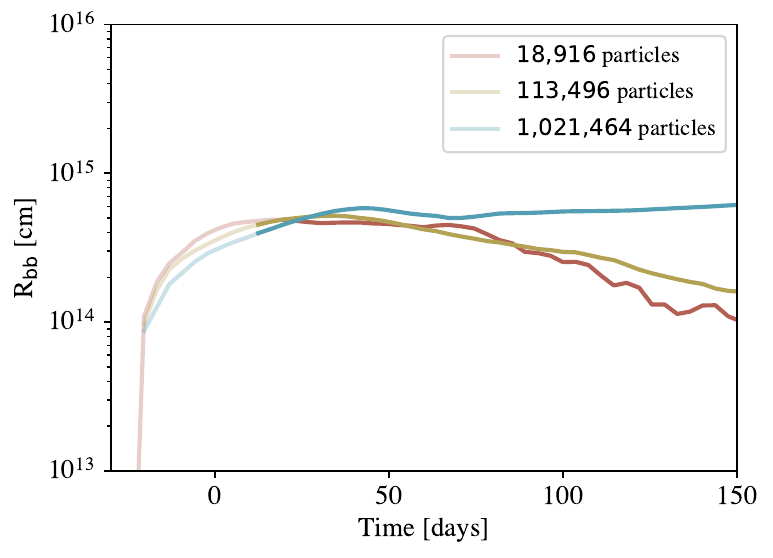}
    \caption{Blackbody radius as function of time for $\beta=0.8$ model. Brown, mustard, and teal-green show three resolutions of the simulation after we replace the remnant with a point mass potential at $4$ days, and use \textsc{Splitpart} to increase resolution. Our estimated values of $R_{\rm bb}$ are converged with numerical resolution up to $\sim100$ days.}
    \label{fig:res_08beta}
\end{figure}

% \section{Orbit positions and velocities}
% Table~\ref{tab:pos_vel_data} lists the position and velocity of each $\beta$ at $10\times r_\mathrm{p}$. All columns are listed in code units, where $M=G=c=1$. 

% \begin{table}
% \caption{Positions and velocities of the orbits around the $10^6\;\mathrm{M}_\odot$ BH, obtained from the \textsc{Geodesic} code. }
%     \centering
%     \begin{tabular}{rrr}
%     \hline
%     \hline
%     $\beta$ & $\vec{x}$ & $\vec{v}$ \\
%     \hline
%        0.6 & (-353.95, 316.57, 0.0) & (0.026, -0.059, 0.0) \\
%        0.8 & (-314.11, 356.14, 0.0) & (0.023, -0.060, 0.0) \\
%        1.0 & (-284.93, 379.88, 0.0) & (0.020, -0.061, 0.0) \\
%        1.2 & (-262.51, 395.71, 0.0) & (0.019, -0.062, 0.0) \\
%        1.4 & (-244.61, 407.02, 0.0) & (0.017, -0.062, 0.0) \\
%        1.6 & (-299.91, 415.50, 0.0) & (0.016, -0.063, 0.0) \\
%        \hline
%        \hline
%     \end{tabular}
%     \label{tab:pos_vel_data}
% \end{table}

% \section{Point mass replacement of the core}
% \label{app:core_replace}

% \begin{figure*}
%     \centering
%     \includegraphics[width=\linewidth]{cut_compare.pdf}
%     \caption{$10\mathord,000$ particle simulations of $\beta=1.6$ case at $22$ days since the beginning of the simulations. Left panel shows the model without any point mass potential, middle panel uses a point mass potential with softening length of $5\;\mathrm{R}_\odot$, and right panel uses $20\;\mathrm{R}_\odot$ softening radius. }
%     \label{fig:core_replace}
% \end{figure*}

\section{Resolution study}
\label{app:res_study}

\begin{figure*}
\label{fig:den}
    \centering
    \includegraphics[width=\textwidth]{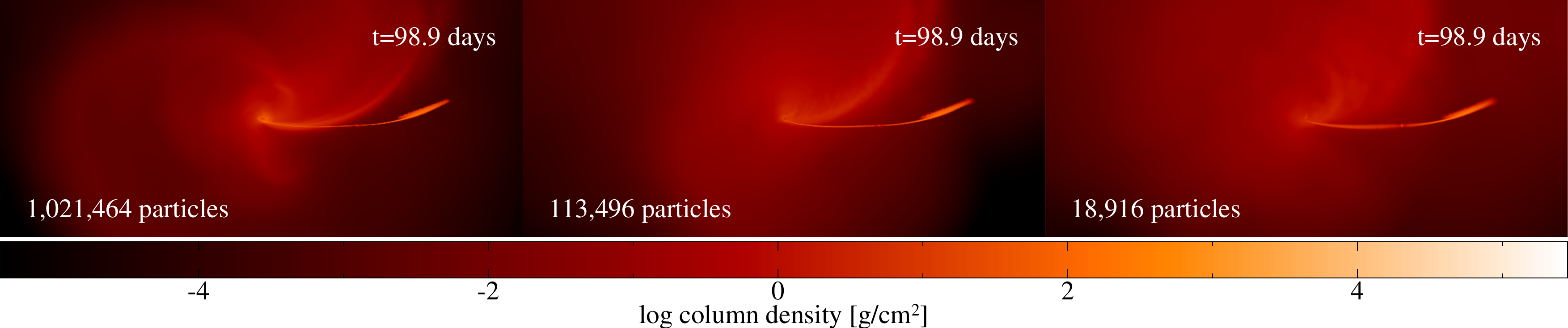}
    \caption{Column density of particles at $98.9\;\mathrm{days}$ since the start of the simulation. Pericentre approach takes place around $7.8\;\mathrm{hours}$ for $\beta=0.8$ model. Left, middle and right panels corresponds to $1\mathord,021\mathord,464$, $113\mathord,496$ and $18\mathord,916$ particles in our simulations. More particles results in disc formation even at $98\;\mathrm{days}$.  Each panel is $1\mathord,380\;\mathrm{au}\times 580\;\mathrm{au}$. }
    
\end{figure*}

Figure~\ref{fig:res_08beta} shows the blackbody radius as function of time at three different resolutions for $\beta=0.8$ simulations.  We only show one $\beta$ value for comparison, but we note that we found similar results for other $\beta$ values. The initial simulation has $1\mathord,000\mathord,000$ particles, and the star is disrupted by the black hole.  The star loses about $2\,\%$ of its mass during the interaction with the black hole.  We replace the core with a point mass potential, leaving $18\mathord,916$ particles in the bound and unbound streams. The brown line shows the result of long-term evolution of this model.  We then use \textsc{Splitpart} \citep{Price2018} to add more particles in the stream, resulting in $113\mathord,496$ particles.  We use \textsc{Splitpart} again to reach $1\mathord,021\mathord,464$ particles in the simulation (teal-blue line).  The brown and mustard lines converge, however, the teal-blue line matches for $\sim 100$ days, and then diverges.  This is due to higher resolution calculations resolving the formation of an eccentric accretion disc around the black hole as shown in Figure~\ref{fig:den} We see a similar behaviour in all of our simulations of any impact parameter, which is why we only show the results up to $100$ days in Figure~\ref{fig:bolometric_radius}. 

\begin{figure}
    \centering
    \includegraphics[width=\linewidth]{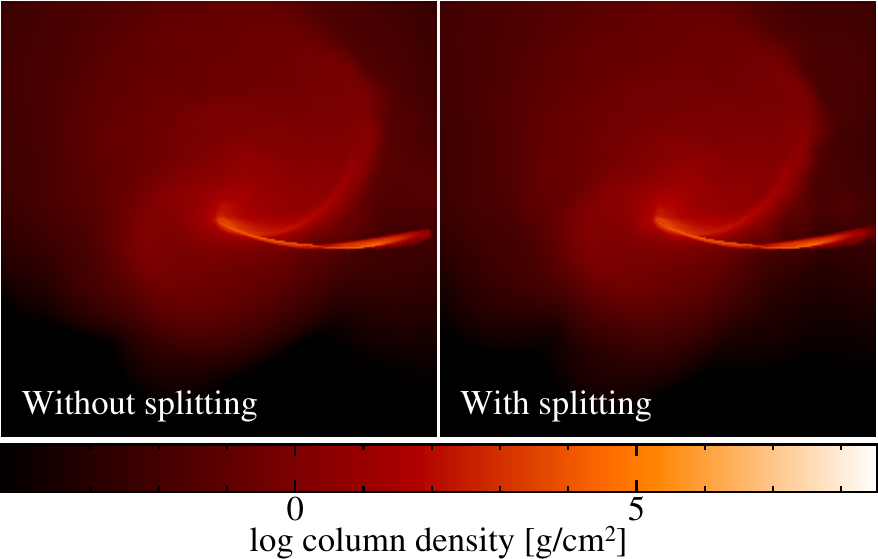}
    \caption{Our fiducial $\beta=1.6$ simulation (left panel; see Table~\ref{tab:all_sims}), compared to an equivalent $2\times10^6$ particle simulation (right panel) where we do not use particle splitting to achieve this resolution at $22\;\mathrm{days}$. Each panel is $180\;\mathrm{au} \times 180\;\mathrm{au}$. The stellar core has been replaced by a softened point mass particle in both cases. The gas column density appears similar between the models, showing the validity of our particle splitting method.}
    \label{fig:res_16_beta}
\end{figure}
\begin{figure}
    \centering
    \includegraphics[width=\columnwidth]{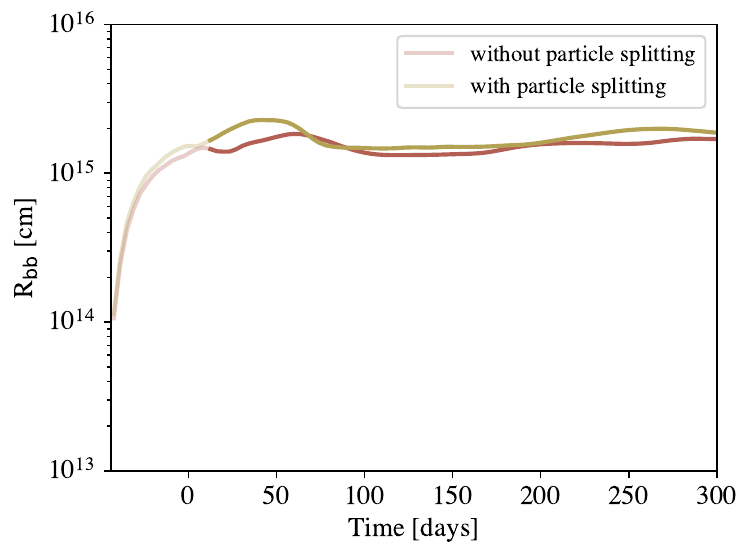}
    \caption{Blackbody radius as function of time for $\beta=1.6$ model. Brown and mustard lines show with and without particle splitting simulations. Our estimated values of $R_{\rm bb}$ are converged within $1\%$, with a maximum difference of $20\%$ around $50$ days. The computational run time is longer by $\sim 25\%$ for the the model where we use particle splitting.}
    \label{fig:res_16_rad}
\end{figure}

% Figure~\ref{fig:res_16_beta} shows the $\beta=1.6$ simulation at $\sim22.2$ days. Left panel shows simulation with $2\times10^6$ particles where we replace the remnant with a sink, resulting in $\sim10^6$ particles in the stream. Right panel shows the $\sim 10^6$ particles simulation, where we started with a $10^6$ particles, and had $\sim 5\times10^5$ particles after replacing remnant with a sink. Using \textsc{Splitpart} we have $10^6$ particles in the simulation. We can see that both simulations evolve in similar manner, with the size of the outflowing material consistent between them. This shows that the use of \textsc{Splitpart} is acceptable.  

\section{Effect of particle splitting}
\label{app:splitting_vs_nosplitting}

Figure~\ref{fig:res_16_beta} shows our $\beta=1.6$ simulation, compared with simulation without particle splitting at $22.2$ days. The returning stream appears identical in both simulations, showing that the particle splitting procedure does not significantly affect the outcome compared to a simulation performed at equivalently high resolution without splitting particles. Figure~\ref{fig:res_16_rad} shows the blackbody radius as function of time, showing that the models converge with each other and plateau.

% \section{Disc formation}
% Figure~\ref{fig:disc_beta08} shows the $\beta=0.8$ simulations for both adiabatic and isentropic cases in $x-z$ plane. The disc is spread in the adiabatic approximation scenario while is a thin disc in the isentropic case.

% \begin{figure}

%     \centering
%     \includegraphics[width=\columnwidth]{disc_beta08.eps}
%     \caption{$\beta=0.8$ simulation in $x-z$ plane for adiabatic and isentropic cases at $98.9$ days. Each panel is $240\;\mathrm{au}\times240\;\mathrm{au}$. The disc is more spread in the adiabatic case, showing the formation of a thin disc in isentropic simulation.}
%     \label{fig:disc_beta08}
% \end{figure}

\section{Spectral evolution}
\label{app:spectral_evo}
\begin{figure*}
    \centering
    \includegraphics[width=\linewidth]{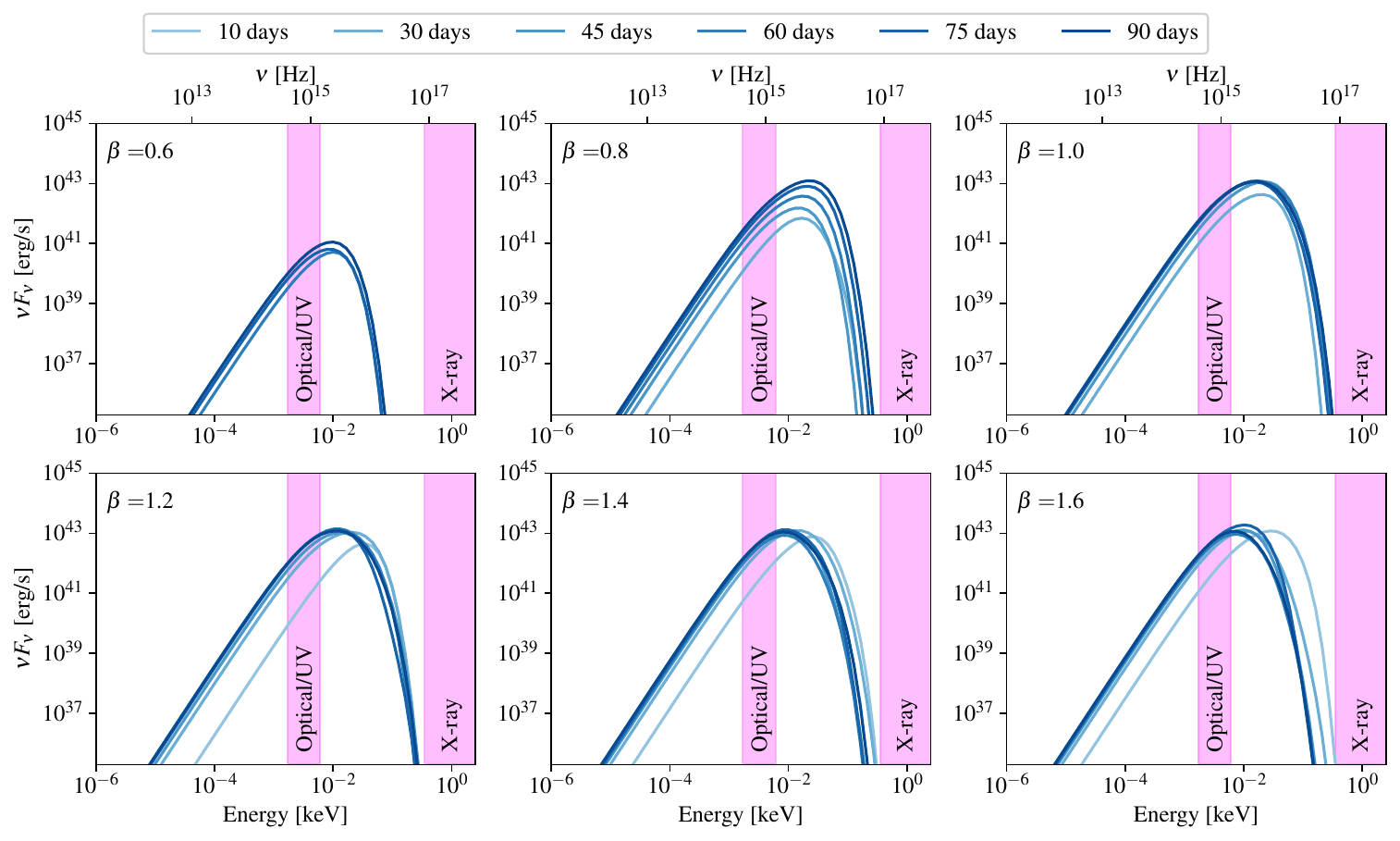}
    \caption{Spectral energy distributions at times ranging from $10-90$ days for all simulations. As time increases, the colors of the lines moves towards darker shades of blue. Except for $\beta=0.6$ simulation, other simulations show soft X-ray emission.}
    \label{fig:sdf_longterm}
\end{figure*}

Figure~\ref{fig:sdf_longterm} shows the synthetic spectral energy distributions for all models at different times. At lower times the spectra is closer to X-rays for our $\beta \geq 1.0$ simulations. We note that for lower $\beta$ simulations, the spectra gets closer to X-rays at later times ($\sim 60$ days).
%% For this sample we use BibTeX plus aasjournalv7.bst to generate the
\bibliographystyle{aasjournalv7}
\bibliography{sample701}{}

@ARTICLE{Shimura1995,
       author = {{Shimura}, Toshiya and {Takahara}, Fumio},
        title = "{On the Spectral Hardening Factor of the X-Ray Emission from Accretion Disks in Black Hole Candidates}",
      journal = {\apj},
         year = 1995,
        month = jun,
       volume = {445},
        pages = {780},
          doi = {10.1086/175740},
       adsurl = {https://ui.adsabs.harvard.edu/abs/1995ApJ...445..780S}
}

@ARTICLE{Davis2006,
       author = {{Davis}, Shane W. and {Hubeny}, Ivan},
        title = "{A Grid of Relativistic, Non-LTE Accretion Disk Models for Spectral Fitting of Black Hole Binaries}",
      journal = {\apjs},
         year = 2006,
        month = jun,
       volume = {164},
       number = {2},
        pages = {530-535},
          doi = {10.1086/503549},
archivePrefix = {arXiv},
       eprint = {astro-ph/0602499},
 primaryClass = {astro-ph},
       adsurl = {https://ui.adsabs.harvard.edu/abs/2006ApJS..164..530D}
}

@ARTICLE{Coughlin2014,
       author = {{Coughlin}, Eric R. and {Begelman}, Mitchell C.},
        title = "{Hyperaccretion during Tidal Disruption Events: Weakly Bound Debris Envelopes and Jets}",
      journal = {\apj},
         year = 2014,
        month = feb,
       volume = {781},
       number = {2},
          eid = {82},
        pages = {82},
          doi = {10.1088/0004-637X/781/2/82},
archivePrefix = {arXiv},
       eprint = {1312.5314},
 primaryClass = {astro-ph.HE},
       adsurl = {https://ui.adsabs.harvard.edu/abs/2014ApJ...781...82C}
}

@ARTICLE{Lucy1977,
       author = {{Lucy}, L.~B.},
        title = "{A numerical approach to the testing of the fission hypothesis.}",
      journal = {\aj},
         year = 1977,
        month = dec,
       volume = {82},
        pages = {1013-1024},
          doi = {10.1086/112164},
       adsurl = {https://ui.adsabs.harvard.edu/abs/1977AJ.....82.1013L}
}

@ARTICLE{Gingold1977,
       author = {{Gingold}, R.~A. and {Monaghan}, J.~J.},
        title = "{Smoothed particle hydrodynamics: theory and application to non-spherical stars.}",
      journal = {\mnras},
         year = 1977,
        month = nov,
       volume = {181},
        pages = {375-389},
          doi = {10.1093/mnras/181.3.375},
       adsurl = {https://ui.adsabs.harvard.edu/abs/1977MNRAS.181..375G}
}

@ARTICLE{Monaghan1992,
       author = {{Monaghan}, J.~J.},
        title = "{Smoothed particle hydrodynamics.}",
      journal = {\araa},
         year = 1992,
        month = jan,
       volume = {30},
        pages = {543-574},
          doi = {10.1146/annurev.aa.30.090192.002551},
       adsurl = {https://ui.adsabs.harvard.edu/abs/1992ARA&A..30..543M}
}

@ARTICLE{Nealon2025,
       author = {{Nealon}, Rebecca and {Price}, Daniel J.},
        title = "{Adaptive particle refinement for compressible smoothed particle hydrodynamics}",
      journal = {\pasa},
         year = 2025,
        month = jan,
       volume = {42},
          eid = {e016},
        pages = {e016},
          doi = {10.1017/pasa.2024.88},
archivePrefix = {arXiv},
       eprint = {2409.11470},
 primaryClass = {astro-ph.IM},
       adsurl = {https://ui.adsabs.harvard.edu/abs/2025PASA...42...16N}
}

@INPROCEEDINGS{Phinney1989,
       author = {{Phinney}, E.~S.},
        title = "{Manifestations of a Massive Black Hole in the Galactic Center}",
    booktitle = {The Center of the Galaxy},
         year = 1989,
       editor = {{Morris}, Mark},
       series = {IAU Symposium},
       volume = {136},
        month = jan,
        pages = {543},
       adsurl = {https://ui.adsabs.harvard.edu/abs/1989IAUS..136..543P}
}

@ARTICLE{Weaver1978,
       author = {{Weaver}, T.~A. and {Zimmerman}, G.~B. and {Woosley}, S.~E.},
        title = "{Presupernova evolution of massive stars.}",
      journal = {\apj},
         year = 1978,
        month = nov,
       volume = {225},
        pages = {1021-1029},
          doi = {10.1086/156569},
       adsurl = {https://ui.adsabs.harvard.edu/abs/1978ApJ...225.1021W}
}

@ARTICLE{Liptai2019,
       author = {{Liptai}, David and {Price}, Daniel J.},
        title = "{General relativistic smoothed particle hydrodynamics}",
      journal = {\mnras},
         year = 2019,
        month = may,
       volume = {485},
       number = {1},
        pages = {819-842},
          doi = {10.1093/mnras/stz111},
archivePrefix = {arXiv},
       eprint = {1901.08064},
 primaryClass = {astro-ph.IM},
       adsurl = {https://ui.adsabs.harvard.edu/abs/2019MNRAS.485..819L}
}

@ARTICLE{Price2018,
       author = {{Price}, Daniel J. and {Wurster}, James and {Tricco}, Terrence S. and {Nixon}, Chris and {Toupin}, St{\'e}ven and {Pettitt}, Alex and {Chan}, Conrad and {Mentiplay}, Daniel and {Laibe}, Guillaume and {Glover}, Simon and {Dobbs}, Clare and {Nealon}, Rebecca and {Liptai}, David and {Worpel}, Hauke and {Bonnerot}, Cl{\'e}ment and {Dipierro}, Giovanni and {Ballabio}, Giulia and {Ragusa}, Enrico and {Federrath}, Christoph and {Iaconi}, Roberto and {Reichardt}, Thomas and {Forgan}, Duncan and {Hutchison}, Mark and {Constantino}, Thomas and {Ayliffe}, Ben and {Hirsh}, Kieran and {Lodato}, Giuseppe},
        title = "{Phantom: A Smoothed Particle Hydrodynamics and Magnetohydrodynamics Code for Astrophysics}",
      journal = {\pasa},
         year = 2018,
        month = sep,
       volume = {35},
          eid = {e031},
        pages = {e031},
          doi = {10.1017/pasa.2018.25},
archivePrefix = {arXiv},
       eprint = {1702.03930},
 primaryClass = {astro-ph.IM},
       adsurl = {https://ui.adsabs.harvard.edu/abs/2018PASA...35...31P}
}

@ARTICLE{Lau2022,
       author = {{Lau}, Mike Y.~M. and {Hirai}, Ryosuke and {Gonz{\'a}lez-Bol{\'\i}var}, Miguel and {Price}, Daniel J. and {De Marco}, Orsola and {Mandel}, Ilya},
        title = "{Common envelopes in massive stars: towards the role of radiation pressure and recombination energy in ejecting red supergiant envelopes}",
      journal = {\mnras},
         year = 2022,
        month = jun,
       volume = {512},
       number = {4},
        pages = {5462-5480},
          doi = {10.1093/mnras/stac049},
archivePrefix = {arXiv},
       eprint = {2111.00923},
 primaryClass = {astro-ph.SR},
       adsurl = {https://ui.adsabs.harvard.edu/abs/2022MNRAS.512.5462L}
}

@ARTICLE{Sharma2024,
       author = {{Sharma}, Megha and {Price}, Daniel J. and {Heger}, Alexander},
        title = "{Partial tidal disruption events: the elixir of life}",
      journal = {\mnras},
         year = 2024,
        month = jul,
       volume = {532},
       number = {1},
        pages = {89-111},
          doi = {10.1093/mnras/stae1455},
archivePrefix = {arXiv},
       eprint = {2403.04211},
 primaryClass = {astro-ph.HE},
       adsurl = {https://ui.adsabs.harvard.edu/abs/2024MNRAS.532...89S}
}

@ARTICLE{Fitz2024,
       author = {{Hu}, Fangyi (Fitz) and {Price}, Daniel J. and {Mandel}, Ilya},
        title = "{Optical Appearance of Eccentric Tidal Disruption Events}",
      journal = {\apjl},
         year = 2024,
        month = mar,
       volume = {963},
       number = {1},
          eid = {L27},
        pages = {L27},
          doi = {10.3847/2041-8213/ad29ec},
archivePrefix = {arXiv},
       eprint = {2312.03210},
 primaryClass = {astro-ph.HE},
       adsurl = {https://ui.adsabs.harvard.edu/abs/2024ApJ...963L..27H}
}

@ARTICLE{Price2024,
       author = {{Price}, Daniel J. and {Liptai}, David and {Mandel}, Ilya and {Shepherd}, Joanna and {Lodato}, Giuseppe and {Levin}, Yuri},
        title = "{Eddington Envelopes: The Fate of Stars on Parabolic Orbits Tidally Disrupted by Supermassive Black Holes}",
      journal = {\apjl},
         year = 2024,
        month = aug,
       volume = {971},
       number = {2},
          eid = {L46},
        pages = {L46},
          doi = {10.3847/2041-8213/ad6862},
archivePrefix = {arXiv},
       eprint = {2404.09381},
 primaryClass = {astro-ph.HE},
       adsurl = {https://ui.adsabs.harvard.edu/abs/2024ApJ...971L..46P}
}

@ARTICLE{Rees1988,
       author = {{Rees}, Martin J.},
        title = "{Tidal disruption of stars by black holes of {}10$^{6}$-{}10$^{8}$ solar masses in nearby galaxies}",
      journal = {\nat},
         year = 1988,
        month = jun,
       volume = {333},
       number = {6173},
        pages = {523-528},
          doi = {10.1038/333523a0},
       adsurl = {https://ui.adsabs.harvard.edu/abs/1988Natur.333..523R}
}

@ARTICLE{Ulmer1999,
       author = {{Ulmer}, Andrew and {Paczynski}, Bohdan and {Goodman}, Jeremy},
        title = "{Tidal disruption Eddington envelopes around massive black holes}",
      journal = {\aap},
         year = 1998,
        month = may,
       volume = {333},
        pages = {379-384},
          doi = {10.48550/arXiv.astro-ph/9711199},
archivePrefix = {arXiv},
       eprint = {astro-ph/9711199},
 primaryClass = {astro-ph},
       adsurl = {https://ui.adsabs.harvard.edu/abs/1998A&A...333..379U}
}

@ARTICLE{Lodato2011,
       author = {{Lodato}, Giuseppe and {Rossi}, Elena M.},
        title = "{Multiband light curves of tidal disruption events}",
      journal = {\mnras},
         year = 2011,
        month = jan,
       volume = {410},
       number = {1},
        pages = {359-367},
          doi = {10.1111/j.1365-2966.2010.17448.x},
archivePrefix = {arXiv},
       eprint = {1008.4589},
 primaryClass = {astro-ph.CO},
       adsurl = {https://ui.adsabs.harvard.edu/abs/2011MNRAS.410..359L}
}

@ARTICLE{Lodato2009,
       author = {{Lodato}, G. and {King}, A.~R. and {Pringle}, J.~E.},
        title = "{Stellar disruption by a supermassive black hole: is the light curve really proportional to t$^{-5/3}$?}",
      journal = {\mnras},
         year = 2009,
        month = jan,
       volume = {392},
       number = {1},
        pages = {332-340},
          doi = {10.1111/j.1365-2966.2008.14049.x},
archivePrefix = {arXiv},
       eprint = {0810.1288},
 primaryClass = {astro-ph},
       adsurl = {https://ui.adsabs.harvard.edu/abs/2009MNRAS.392..332L}
}

@ARTICLE{Guillochon2013,
       author = {{Guillochon}, James and {Ramirez-Ruiz}, Enrico},
        title = "{Hydrodynamical Simulations to Determine the Feeding Rate of Black Holes by the Tidal Disruption of Stars: The Importance of the Impact Parameter and Stellar Structure}",
      journal = {\apj},
         year = 2013,
        month = apr,
       volume = {767},
       number = {1},
          eid = {25},
        pages = {25},
          doi = {10.1088/0004-637X/767/1/25},
archivePrefix = {arXiv},
       eprint = {1206.2350},
 primaryClass = {astro-ph.HE},
       adsurl = {https://ui.adsabs.harvard.edu/abs/2013ApJ...767...25G}
}

@ARTICLE{Steinberg2024,
       author = {{Steinberg}, Elad and {Stone}, Nicholas C.},
        title = "{Stream-disk shocks as the origins of peak light in tidal disruption events}",
      journal = {\nat},
         year = 2024,
        month = jan,
       volume = {625},
       number = {7995},
        pages = {463-467},
          doi = {10.1038/s41586-023-06875-y},
archivePrefix = {arXiv},
       eprint = {2206.10641},
 primaryClass = {astro-ph.HE},
       adsurl = {https://ui.adsabs.harvard.edu/abs/2024Natur.625..463S}
}

@ARTICLE{Gezari2021,
       author = {{Gezari}, Suvi},
        title = "{Tidal Disruption Events}",
      journal = {\araa},
         year = 2021,
        month = sep,
       volume = {59},
        pages = {21-58},
          doi = {10.1146/annurev-astro-111720-030029},
archivePrefix = {arXiv},
       eprint = {2104.14580},
 primaryClass = {astro-ph.HE},
       adsurl = {https://ui.adsabs.harvard.edu/abs/2021ARA&A..59...21G}
}

@ARTICLE{Bortolas2023,
       author = {{Bortolas}, Elisa and {Ryu}, Taeho and {Broggi}, Luca and {Sesana}, Alberto},
        title = "{Partial stellar tidal disruption events and their rates}",
      journal = {\mnras},
         year = 2023,
        month = sep,
       volume = {524},
       number = {2},
        pages = {3026-3038},
          doi = {10.1093/mnras/stad2024},
archivePrefix = {arXiv},
       eprint = {2303.03408},
 primaryClass = {astro-ph.HE},
       adsurl = {https://ui.adsabs.harvard.edu/abs/2023MNRAS.524.3026B}
}

@ARTICLE{Zhong2022,
       author = {{Zhong}, Shiyan and {Li}, Shuo and {Berczik}, Peter and {Spurzem}, Rainer},
        title = "{Revisit the Rate of Tidal Disruption Events: The Role of the Partial Tidal Disruption Event}",
      journal = {\apj},
         year = 2022,
        month = jul,
       volume = {933},
       number = {1},
          eid = {96},
        pages = {96},
          doi = {10.3847/1538-4357/ac71ad},
archivePrefix = {arXiv},
       eprint = {2205.09945},
 primaryClass = {astro-ph.GA},
       adsurl = {https://ui.adsabs.harvard.edu/abs/2022ApJ...933...96Z}
}

@ARTICLE{Stone2016,
       author = {{Stone}, Nicholas C. and {Metzger}, Brian D.},
        title = "{Rates of stellar tidal disruption as probes of the supermassive black hole mass function}",
      journal = {\mnras},
         year = 2016,
        month = jan,
       volume = {455},
       number = {1},
        pages = {859-883},
          doi = {10.1093/mnras/stv2281},
archivePrefix = {arXiv},
       eprint = {1410.7772},
 primaryClass = {astro-ph.HE},
       adsurl = {https://ui.adsabs.harvard.edu/abs/2016MNRAS.455..859S}
}

@ARTICLE{Hills1975,
       author = {{Hills}, J.~G.},
        title = "{Possible power source of Seyfert galaxies and QSOs}",
      journal = {\nat},
         year = 1975,
        month = mar,
       volume = {254},
       number = {5498},
        pages = {295-298},
          doi = {10.1038/254295a0},
       adsurl = {https://ui.adsabs.harvard.edu/abs/1975Natur.254..295H}
}

@ARTICLE{Chen2021,
       author = {{Chen}, Jin-Hong and {Shen}, Rong-Feng},
        title = "{Light Curves of Partial Tidal Disruption Events}",
      journal = {\apj},
         year = 2021,
        month = jun,
       volume = {914},
       number = {1},
          eid = {69},
        pages = {69},
          doi = {10.3847/1538-4357/abf9a7},
archivePrefix = {arXiv},
       eprint = {2104.08827},
 primaryClass = {astro-ph.HE},
       adsurl = {https://ui.adsabs.harvard.edu/abs/2021ApJ...914...69C}
}

@ARTICLE{Kormendy2013,
       author = {{Kormendy}, John and {Ho}, Luis C.},
        title = "{Coevolution (Or Not) of Supermassive Black Holes and Host Galaxies}",
      journal = {\araa},
         year = 2013,
        month = aug,
       volume = {51},
       number = {1},
        pages = {511-653},
          doi = {10.1146/annurev-astro-082708-101811},
archivePrefix = {arXiv},
       eprint = {1304.7762},
 primaryClass = {astro-ph.CO},
       adsurl = {https://ui.adsabs.harvard.edu/abs/2013ARA&A..51..511K}
}

@ARTICLE{Wang2023,
       author = {{Wang}, Tinggui and {Liu}, Guilin and {Cai}, Zhenyi and {Geng}, Jinjun and {Fang}, Min and {He}, Haoning and {Jiang}, Ji-an and {Jiang}, Ning and {Kong}, Xu and {Li}, Bin and {Li}, Ye and {Luo}, Wentao and {Pan}, Zhizheng and {Wu}, Xuefeng and {Yang}, Ji and {Yu}, Jiming and {Zheng}, Xianzhong and {Zhu}, Qingfeng and {Cai}, Yi-Fu and {Chen}, Yuanyuan and {Chen}, Zhiwei and {Dai}, Zigao and {Fan}, Lulu and {Fan}, Yizhong and {Fang}, Wenjuan and {He}, Zhicheng and {Hu}, Lei and {Hu}, Maokai and {Jin}, Zhiping and {Jiang}, Zhibo and {Li}, Guoliang and {Li}, Fan and {Li}, Xuzhi and {Liang}, Runduo and {Lin}, Zheyu and {Liu}, Qingzhong and {Liu}, Wenhao and {Liu}, Zhengyan and {Liu}, Wei and {Liu}, Yao and {Lou}, Zheng and {Qu}, Han and {Sheng}, Zhenfeng and {Shi}, Jianchun and {Shu}, Yiping and {Su}, Zhenbo and {Sun}, Tianrui and {Wang}, Hongchi and {Wang}, Huiyuan and {Wang}, Jian and {Wang}, Junxian and {Wei}, Daming and {Wei}, Junjie and {Xue}, Yongquan and {Yan}, Jingzhi and {Yang}, Chao and {Yuan}, Ye and {Yuan}, Yefei and {Zhang}, Hongxin and {Zhang}, Miaomiao and {Zhao}, Haibin and {Zhao}, Wen},
        title = "{Science with the 2.5-meter Wide Field Survey Telescope (WFST)}",
      journal = {Science China Physics, Mechanics, and Astronomy},
         year = 2023,
        month = oct,
       volume = {66},
       number = {10},
          eid = {109512},
        pages = {109512},
          doi = {10.1007/s11433-023-2197-5},
archivePrefix = {arXiv},
       eprint = {2306.07590},
 primaryClass = {astro-ph.IM},
       adsurl = {https://ui.adsabs.harvard.edu/abs/2023SCPMA..6609512W}
}

@ARTICLE{Auchettl2017,
       author = {{Auchettl}, Katie and {Guillochon}, James and {Ramirez-Ruiz}, Enrico},
        title = "{New Physical Insights about Tidal Disruption Events from a Comprehensive Observational Inventory at X-Ray Wavelengths}",
      journal = {\apj},
         year = 2017,
        month = apr,
       volume = {838},
       number = {2},
          eid = {149},
        pages = {149},
          doi = {10.3847/1538-4357/aa633b},
archivePrefix = {arXiv},
       eprint = {1611.02291},
 primaryClass = {astro-ph.HE},
       adsurl = {https://ui.adsabs.harvard.edu/abs/2017ApJ...838..149A}
}

@ARTICLE{Hammerstein2023,
       author = {{Hammerstein}, Erica and {van Velzen}, Sjoert and {Gezari}, Suvi and {Cenko}, S. Bradley and {Yao}, Yuhan and {Ward}, Charlotte and {Frederick}, Sara and {Villanueva}, Natalia and {Somalwar}, Jean J. and {Graham}, Matthew J. and {Kulkarni}, Shrinivas R. and {Stern}, Daniel and {Andreoni}, Igor and {Bellm}, Eric C. and {Dekany}, Richard and {Dhawan}, Suhail and {Drake}, Andrew J. and {Fremling}, Christoffer and {Gatkine}, Pradip and {Groom}, Steven L. and {Ho}, Anna Y.~Q. and {Kasliwal}, Mansi M. and {Karambelkar}, Viraj and {Kool}, Erik C. and {Masci}, Frank J. and {Medford}, Michael S. and {Perley}, Daniel A. and {Purdum}, Josiah and {van Roestel}, Jan and {Sharma}, Yashvi and {Sollerman}, Jesper and {Taggart}, Kirsty and {Yan}, Lin},
        title = "{The Final Season Reimagined: 30 Tidal Disruption Events from the ZTF-I Survey}",
      journal = {\apj},
         year = 2023,
        month = jan,
       volume = {942},
       number = {1},
          eid = {9},
        pages = {9},
          doi = {10.3847/1538-4357/aca283},
archivePrefix = {arXiv},
       eprint = {2203.01461},
 primaryClass = {astro-ph.HE},
       adsurl = {https://ui.adsabs.harvard.edu/abs/2023ApJ...942....9H}
}

@ARTICLE{Piran2015,
       author = {{Piran}, Tsvi and {Svirski}, Gilad and {Krolik}, Julian and {Cheng}, Roseanne M. and {Shiokawa}, Hotaka},
        title = "{‧Disk Formation Versus Disk Accretion{\textemdash}What Powers Tidal Disruption Events?}",
      journal = {\apj},
         year = 2015,
        month = jun,
       volume = {806},
       number = {2},
          eid = {164},
        pages = {164},
          doi = {10.1088/0004-637X/806/2/164},
archivePrefix = {arXiv},
       eprint = {1502.05792},
 primaryClass = {astro-ph.HE},
       adsurl = {https://ui.adsabs.harvard.edu/abs/2015ApJ...806..164P}
}

@ARTICLE{Law-Smith2020,
       author = {{Law-Smith}, Jamie A.~P. and {Coulter}, David A. and {Guillochon}, James and {Mockler}, Brenna and {Ramirez-Ruiz}, Enrico},
        title = "{Stellar Tidal Disruption Events with Abundances and Realistic Structures (STARS): Library of Fallback Rates}",
      journal = {\apj},
         year = 2020,
        month = dec,
       volume = {905},
       number = {2},
          eid = {141},
        pages = {141},
          doi = {10.3847/1538-4357/abc489},
archivePrefix = {arXiv},
       eprint = {2007.10996},
 primaryClass = {astro-ph.HE},
       adsurl = {https://ui.adsabs.harvard.edu/abs/2020ApJ...905..141L}
}

@ARTICLE{Nixon2022,
       author = {{Nixon}, C.~J. and {Coughlin}, Eric R.},
        title = "{Stellar Revival and Repeated Flares in Deeply Plunging Tidal Disruption Events}",
      journal = {\apjl},
         year = 2022,
        month = mar,
       volume = {927},
       number = {2},
          eid = {L25},
        pages = {L25},
          doi = {10.3847/2041-8213/ac5118},
archivePrefix = {arXiv},
       eprint = {2202.00014},
 primaryClass = {astro-ph.HE},
       adsurl = {https://ui.adsabs.harvard.edu/abs/2022ApJ...927L..25N}
}

@ARTICLE{Ryu2020,
       author = {{Ryu}, Taeho and {Krolik}, Julian and {Piran}, Tsvi and {Noble}, Scott C.},
        title = "{Tidal Disruptions of Main-sequence Stars. III. Stellar Mass Dependence of the Character of Partial Disruptions}",
      journal = {\apj},
         year = 2020,
        month = dec,
       volume = {904},
       number = {2},
          eid = {100},
        pages = {100},
          doi = {10.3847/1538-4357/abb3ce},
archivePrefix = {arXiv},
       eprint = {2001.03503},
 primaryClass = {astro-ph.HE},
       adsurl = {https://ui.adsabs.harvard.edu/abs/2020ApJ...904..100R}
}

@ARTICLE{Price2012,
       author = {{Price}, Daniel J.},
        title = "{Smoothed particle hydrodynamics and magnetohydrodynamics}",
      journal = {Journal of Computational Physics},
         year = 2012,
        month = feb,
       volume = {231},
       number = {3},
        pages = {759-794},
          doi = {10.1016/j.jcp.2010.12.011},
archivePrefix = {arXiv},
       eprint = {1012.1885},
 primaryClass = {astro-ph.IM},
       adsurl = {https://ui.adsabs.harvard.edu/abs/2012JCoPh.231..759P}
}

@ARTICLE{VanVelzen2021,
       author = {{van Velzen}, Sjoert and {Gezari}, Suvi and {Hammerstein}, Erica and {Roth}, Nathaniel and {Frederick}, Sara and {Ward}, Charlotte and {Hung}, Tiara and {Cenko}, S. Bradley and {Stein}, Robert and {Perley}, Daniel A. and {Taggart}, Kirsty and {Foley}, Ryan J. and {Sollerman}, Jesper and {Blagorodnova}, Nadejda and {Andreoni}, Igor and {Bellm}, Eric C. and {Brinnel}, Valery and {De}, Kishalay and {Dekany}, Richard and {Feeney}, Michael and {Fremling}, Christoffer and {Giomi}, Matteo and {Golkhou}, V. Zach and {Graham}, Matthew J. and {Ho}, Anna. Y.~Q. and {Kasliwal}, Mansi M. and {Kilpatrick}, Charles D. and {Kulkarni}, Shrinivas R. and {Kupfer}, Thomas and {Laher}, Russ R. and {Mahabal}, Ashish and {Masci}, Frank J. and {Miller}, Adam A. and {Nordin}, Jakob and {Riddle}, Reed and {Rusholme}, Ben and {van Santen}, Jakob and {Sharma}, Yashvi and {Shupe}, David L. and {Soumagnac}, Maayane T.},
        title = "{Seventeen Tidal Disruption Events from the First Half of ZTF Survey Observations: Entering a New Era of Population Studies}",
      journal = {\apj},
         year = 2021,
        month = feb,
       volume = {908},
       number = {1},
          eid = {4},
        pages = {4},
          doi = {10.3847/1538-4357/abc258},
archivePrefix = {arXiv},
       eprint = {2001.01409},
 primaryClass = {astro-ph.HE},
       adsurl = {https://ui.adsabs.harvard.edu/abs/2021ApJ...908....4V}
}

@ARTICLE{Hinkle2024,
       author = {{Hinkle}, Jason T. and {Auchettl}, Katie and {Hoogendam}, Willem B. and {Payne}, Anna V. and {Holoien}, Thomas W. -S. and {Shappee}, Benjamin J. and {Tucker}, Michael A. and {Kochanek}, Christopher S. and {Stanek}, K.~Z. and {Vallely}, Patrick J. and {Angus}, Charlotte R. and {Ashall}, Chris and {de Jaeger}, Thomas and {Desai}, Dhvanil D. and {Do}, Aaron and {Fausnaugh}, Michael M. and {Huber}, Mark E. and {Rickards Vaught}, Ryan J. and {Shi}, Jennifer},
        title = "{On the Double: Two Luminous Flares from the Nearby Tidal Disruption Event ASASSN-22ci (AT2022dbl) and Connections to Repeating TDE Candidates}",
      journal = {arXiv e-prints},
         year = 2024,
        month = dec,
          eid = {arXiv:2412.15326},
        pages = {arXiv:2412.15326},
          doi = {10.48550/arXiv.2412.15326},
archivePrefix = {arXiv},
       eprint = {2412.15326},
 primaryClass = {astro-ph.HE},
       adsurl = {https://ui.adsabs.harvard.edu/abs/2024arXiv241215326H}
}

@ARTICLE{Alexander2001,
       author = {{Alexander}, Tal and {Livio}, Mario},
        title = "{Tidal Scattering of Stars on Supermassive Black Holes in Galactic Centers}",
      journal = {\apjl},
         year = 2001,
        month = oct,
       volume = {560},
       number = {2},
        pages = {L143-L146},
          doi = {10.1086/324324},
archivePrefix = {arXiv},
       eprint = {astro-ph/0109237},
 primaryClass = {astro-ph},
       adsurl = {https://ui.adsabs.harvard.edu/abs/2001ApJ...560L.143A}
}

@ARTICLE{Manukian2013,
       author = {{Manukian}, Haik and {Guillochon}, James and {Ramirez-Ruiz}, Enrico and {O'Leary}, Ryan M.},
        title = "{Turbovelocity Stars: Kicks Resulting from the Tidal Disruption of Solitary Stars}",
      journal = {\apjl},
         year = 2013,
        month = jul,
       volume = {771},
       number = {2},
          eid = {L28},
        pages = {L28},
          doi = {10.1088/2041-8205/771/2/L28},
archivePrefix = {arXiv},
       eprint = {1305.4634},
 primaryClass = {astro-ph.HE},
       adsurl = {https://ui.adsabs.harvard.edu/abs/2013ApJ...771L..28M}
}

@ARTICLE{Neustadt2020,
       author = {{Neustadt}, J.~M.~M. and {Holoien}, T.~W. -S. and {Kochanek}, C.~S. and {Auchettl}, K. and {Brown}, J.~S. and {Shappee}, B.~J. and {Pogge}, R.~W. and {Dong}, Subo and {Stanek}, K.~Z. and {Tucker}, M.~A. and {Bose}, S. and {Chen}, Ping and {Ricci}, C. and {Vallely}, P.~J. and {Prieto}, J.~L. and {Thompson}, T.~A. and {Coulter}, D.~A. and {Drout}, M.~R. and {Foley}, R.~J. and {Kilpatrick}, C.~D. and {Piro}, A.~L. and {Rojas-Bravo}, C. and {Buckley}, D.~A.~H. and {Gromadzki}, M. and {Dimitriadis}, G. and {Siebert}, M.~R. and {Do}, A. and {Huber}, M.~E. and {Payne}, A.~V.},
        title = "{To TDE or not to TDE: the luminous transient ASASSN-18jd with TDE-like and AGN-like qualities}",
      journal = {\mnras},
         year = 2020,
        month = may,
       volume = {494},
       number = {2},
        pages = {2538-2560},
          doi = {10.1093/mnras/staa859},
archivePrefix = {arXiv},
       eprint = {1910.01142},
 primaryClass = {astro-ph.HE},
       adsurl = {https://ui.adsabs.harvard.edu/abs/2020MNRAS.494.2538N}
}

@ARTICLE{Mockler2019,
       author = {{Mockler}, Brenna and {Guillochon}, James and {Ramirez-Ruiz}, Enrico},
        title = "{Weighing Black Holes Using Tidal Disruption Events}",
      journal = {\apj},
         year = 2019,
        month = feb,
       volume = {872},
       number = {2},
          eid = {151},
        pages = {151},
          doi = {10.3847/1538-4357/ab010f},
archivePrefix = {arXiv},
       eprint = {1801.08221},
 primaryClass = {astro-ph.HE},
       adsurl = {https://ui.adsabs.harvard.edu/abs/2019ApJ...872..151M}
}

@ARTICLE{Harris2020,
  author  = {Harris, Charles R. and Millman, K. Jarrod and van der Walt, Stéfan J and Gommers, Ralf and Virtanen, Pauli and Cournapeau, David and Wieser, Eric and Taylor, Julian and Berg, Sebastian and Smith, Nathaniel J. and Kern, Robert and Picus, Matti and Hoyer, Stephan and van Kerkwijk, Marten H. and Brett, Matthew and Haldane, Allan and Fernández del Río, Jaime and Wiebe, Mark and Peterson, Pearu and Gérard-Marchant, Pierre and Sheppard, Kevin and Reddy, Tyler and Weckesser, Warren and Abbasi, Hameer and Gohlke, Christoph and Oliphant, Travis E.},
  title   = {Array programming with {NumPy}},
  journal = {Nature},
  year    = {2020},
  volume  = {585},
  pages   = {357–362},
  doi     = {10.1038/s41586-020-2649-2}
}

@Article{Hunter2007,
  Author    = {Hunter, J. D.},
  Title     = {Matplotlib: A 2D graphics environment},
  Journal   = {Computing in Science \& Engineering},
  Volume    = {9},
  Number    = {3},
  Pages     = {90--95},
  publisher = {IEEE COMPUTER SOC},
  doi       = {10.1109/MCSE.2007.55},
  year      = 2007
}

@ARTICLE{Price2007,
       author = {{Price}, Daniel J.},
        title = "{splash: An Interactive Visualisation Tool for Smoothed Particle Hydrodynamics Simulations}",
      journal = {\pasa},
         year = 2007,
        month = oct,
       volume = {24},
       number = {3},
        pages = {159-173},
          doi = {10.1071/AS07022},
archivePrefix = {arXiv},
       eprint = {0709.0832},
 primaryClass = {astro-ph},
       adsurl = {https://ui.adsabs.harvard.edu/abs/2007PASA...24..159P}
}

@ARTICLE{Virtanen2020,
  author  = {Virtanen, Pauli and Gommers, Ralf and Oliphant, Travis E. and
            Haberland, Matt and Reddy, Tyler and Cournapeau, David and
            Burovski, Evgeni and Peterson, Pearu and Weckesser, Warren and
            Bright, Jonathan and {van der Walt}, St{\'e}fan J. and
            Brett, Matthew and Wilson, Joshua and Millman, K. Jarrod and
            Mayorov, Nikolay and Nelson, Andrew R. J. and Jones, Eric and
            Kern, Robert and Larson, Eric and Carey, C J and
            Polat, {\.I}lhan and Feng, Yu and Moore, Eric W. and
            {VanderPlas}, Jake and Laxalde, Denis and Perktold, Josef and
            Cimrman, Robert and Henriksen, Ian and Quintero, E. A. and
            Harris, Charles R. and Archibald, Anne M. and
            Ribeiro, Ant{\^o}nio H. and Pedregosa, Fabian and
            {van Mulbregt}, Paul and {SciPy 1.0 Contributors}},
  title   = {{{SciPy} 1.0: Fundamental Algorithms for Scientific
            Computing in Python}},
  journal = {Nature Methods},
  year    = {2020},
  volume  = {17},
  pages   = {261--272},
  adsurl  = {https://rdcu.be/b08Wh},
  doi     = {10.1038/s41592-019-0686-2},
}

@ARTICLE{Zhong2014,
       author = {{Zhong}, Shiyan and {Berczik}, Peter and {Spurzem}, Rainer},
        title = "{Super Massive Black Hole in Galactic Nuclei with Tidal Disruption of Stars}",
      journal = {\apj},
         year = 2014,
        month = sep,
       volume = {792},
       number = {2},
          eid = {137},
        pages = {137},
          doi = {10.1088/0004-637X/792/2/137},
archivePrefix = {arXiv},
       eprint = {1407.3537},
 primaryClass = {astro-ph.GA},
       adsurl = {https://ui.adsabs.harvard.edu/abs/2014ApJ...792..137Z}
}

@ARTICLE{Kubli2025,
       author = {{Kubli}, Noah and {Franchini}, Alessia and {Coughlin}, Eric R. and {Nixon}, C.~J. and {Keller}, Sebastian and {Capelo}, Pedro R. and {Mayer}, Lucio},
        title = "{Tidal disruption events with SPH-EXA: resolving the return of the stream}",
      journal = {arXiv e-prints},
         year = 2025,
        month = oct,
          eid = {arXiv:2510.26663},
        pages = {arXiv:2510.26663},
          doi = {10.48550/arXiv.2510.26663},
archivePrefix = {arXiv},
       eprint = {2510.26663},
 primaryClass = {astro-ph.HE},
       adsurl = {https://ui.adsabs.harvard.edu/abs/2025arXiv251026663K}
}

@ARTICLE{Hu2025,
       author = {{Fitz Hu}, Fangyi and {Mandel}, Ilya and {Nealon}, Rebecca and {Price}, Daniel J.},
        title = "{Converged simulations of the nozzle shock in tidal disruption events}",
      journal = {arXiv e-prints},
         year = 2025,
        month = oct,
          eid = {arXiv:2510.04790},
        pages = {arXiv:2510.04790},
          doi = {10.48550/arXiv.2510.04790},
archivePrefix = {arXiv},
       eprint = {2510.04790},
 primaryClass = {astro-ph.HE},
       adsurl = {https://ui.adsabs.harvard.edu/abs/2025arXiv251004790F}
}

@ARTICLE{Lacy1982,
       author = {{Lacy}, J.~H. and {Townes}, C.~H. and {Hollenbach}, D.~J.},
        title = "{The nature of the central parsec of the Galaxy}",
      journal = {\apj},
         year = 1982,
        month = nov,
       volume = {262},
        pages = {120-134},
          doi = {10.1086/160402},
       adsurl = {https://ui.adsabs.harvard.edu/abs/1982ApJ...262..120L}
}

@ARTICLE{Coughlin2019,
       author = {{Coughlin}, Eric R. and {Nixon}, C.~J.},
        title = "{Partial Stellar Disruption by a Supermassive Black Hole: Is the Light Curve Really Proportional to t $^{-9/4}$?}",
      journal = {\apjl},
         year = 2019,
        month = sep,
       volume = {883},
       number = {1},
          eid = {L17},
        pages = {L17},
          doi = {10.3847/2041-8213/ab412d},
archivePrefix = {arXiv},
       eprint = {1907.03034},
 primaryClass = {astro-ph.GA},
       adsurl = {https://ui.adsabs.harvard.edu/abs/2019ApJ...883L..17C}
}

@ARTICLE{Miller2025,
       author = {{Miller}, Adam A. and {Abrams}, Natasha S. and {Aldering}, Greg and {Anand}, Shreya and {Angus}, Charlotte R. and {Arcavi}, Iair and {Baltay}, Charles and {Bauer}, Franz E. and {Brethauer}, Daniel and {Bloom}, Joshua S. and {Bommireddy}, Hemanth and {Catelan}, M{\'a}rcio and {Chornock}, Ryan and {Clark}, Peter and {Collett}, Thomas E. and {Dimitriadis}, Georgios and {Faris}, Sara and {F{\"o}rster}, Francisco and {Franckowiak}, Anna and {Frohmaier}, Christopher and {Galbany}, Llu{\'\i}s and {Galleguillos}, Renato B. and {Goobar}, Ariel and {Graur}, Or and {Guti{\'e}rrez}, Claudia P. and {Hall}, Saarah and {Hammerstein}, Erica and {Herner}, Kenneth R. and {Hook}, Isobel M. and {Huston}, Macy J. and {Johansson}, Joel and {Kilpatrick}, Charles D. and {Kim}, Alex G. and {Knop}, Robert A. and {Kowalski}, Marek P. and {Kwok}, Lindsey A. and {LeBaron}, Natalie and {Lin}, Kenneth W. and {Liu}, Chang and {Lu}, Jessica R. and {Lu}, Wenbin and {Lunnan}, Ragnhild and {Maguire}, Kate and {Makrygianni}, Lydia and {Margutti}, Raffaella and {Maoz}, Dan and {Veres}, Patrik Mil{\'a}n and {Moore}, Thomas and {Nayana}, A.~J. and {Nicholl}, Matt and {Nordin}, Jakob and {Oates}, S.~R. and {Pignata}, Giuliano and {Polin}, Abigail and {Poznanski}, Dovi and {Prieto}, Jose L. and {Rabinowitz}, David L. and {Rehemtulla}, Nabeel and {Rigault}, Mickael and {Ryczanowski}, Dan and {Sarin}, Nikhil and {Schulze}, Steve and {Shah}, Ved G. and {Sheng}, Xinyue and {Shilling}, Samuel P.~R. and {Simmons}, Brooke D. and {Singh}, Avinash and {Smith}, Graham P. and {Smith}, Mathew and {Sollerman}, Jesper and {Soumagnac}, Maayane T. and {Stubbs}, Christopher W. and {Sullivan}, Mark and {Suresh}, Aswin and {Trakhtenbrot}, Benny and {Ward}, Charlotte and {Wiston}, Eli and {Xiong}, Helen and {Yao}, Yuhan and {Nugent}, Peter E.},
        title = "{The La Silla Schmidt Southern Survey}",
      journal = {\pasp},
         year = 2025,
        month = sep,
       volume = {137},
       number = {9},
          eid = {094204},
        pages = {094204},
          doi = {10.1088/1538-3873/ae02c5},
archivePrefix = {arXiv},
       eprint = {2503.14579},
 primaryClass = {astro-ph.IM},
       adsurl = {https://ui.adsabs.harvard.edu/abs/2025PASP..137i4204M}
}

@ARTICLE{Tonry2018,
       author = {{Tonry}, J.~L. and {Denneau}, L. and {Heinze}, A.~N. and {Stalder}, B. and {Smith}, K.~W. and {Smartt}, S.~J. and {Stubbs}, C.~W. and {Weiland}, H.~J. and {Rest}, A.},
        title = "{ATLAS: A High-cadence All-sky Survey System}",
      journal = {\pasp},
         year = 2018,
        month = jun,
       volume = {130},
       number = {988},
        pages = {064505},
          doi = {10.1088/1538-3873/aabadf},
archivePrefix = {arXiv},
       eprint = {1802.00879},
 primaryClass = {astro-ph.IM},
       adsurl = {https://ui.adsabs.harvard.edu/abs/2018PASP..130f4505T}
}

@ARTICLE{Kochanek2017,
       author = {{Kochanek}, C.~S. and {Shappee}, B.~J. and {Stanek}, K.~Z. and {Holoien}, T.~W.-S. and {Thompson}, Todd A. and {Prieto}, J.~L. and {Dong}, Subo and {Shields}, J.~V. and {Will}, D. and {Britt}, C. and {Perzanowski}, D. and {Pojma{\'n}ski}, G.},
        title = "{The All-Sky Automated Survey for Supernovae (ASAS-SN) Light Curve Server v1.0}",
      journal = {\pasp},
         year = 2017,
        month = oct,
       volume = {129},
       number = {980},
        pages = {104502},
          doi = {10.1088/1538-3873/aa80d9},
archivePrefix = {arXiv},
       eprint = {1706.07060},
 primaryClass = {astro-ph.SR},
       adsurl = {https://ui.adsabs.harvard.edu/abs/2017PASP..129j4502K}
}

@ARTICLE{Dai2015,
       author = {{Dai}, Lixin and {McKinney}, Jonathan C. and {Miller}, M. Coleman},
        title = "{Soft X-Ray Temperature Tidal Disruption Events from Stars on Deep Plunging Orbits}",
      journal = {\apjl},
         year = 2015,
        month = oct,
       volume = {812},
       number = {2},
          eid = {L39},
        pages = {L39},
          doi = {10.1088/2041-8205/812/2/L39},
archivePrefix = {arXiv},
       eprint = {1507.04333},
 primaryClass = {astro-ph.HE},
       adsurl = {https://ui.adsabs.harvard.edu/abs/2015ApJ...812L..39D}
}

@ARTICLE{Evans1989,
       author = {{Evans}, Charles R. and {Kochanek}, Christopher S.},
        title = "{The Tidal Disruption of a Star by a Massive Black Hole}",
      journal = {\apjl},
         year = 1989,
        month = nov,
       volume = {346},
        pages = {L13},
          doi = {10.1086/185567},
       adsurl = {https://ui.adsabs.harvard.edu/abs/1989ApJ...346L..13E}
}

@ARTICLE{Loeb1997,
       author = {{Loeb}, Abraham and {Ulmer}, Andrew},
        title = "{Optical Appearance of the Debris of a Star Disrupted by a Massive Black Hole}",
      journal = {\apj},
         year = 1997,
        month = nov,
       volume = {489},
       number = {2},
        pages = {573-578},
          doi = {10.1086/304814},
archivePrefix = {arXiv},
       eprint = {astro-ph/9703079},
 primaryClass = {astro-ph},
       adsurl = {https://ui.adsabs.harvard.edu/abs/1997ApJ...489..573L}
}

@ARTICLE{Bennerot2017,
       author = {{Bonnerot}, Cl{\'e}ment and {Rossi}, Elena M. and {Lodato}, Giuseppe},
        title = "{Long-term stream evolution in tidal disruption events}",
      journal = {\mnras},
         year = 2017,
        month = jan,
       volume = {464},
       number = {3},
        pages = {2816-2830},
          doi = {10.1093/mnras/stw2547},
archivePrefix = {arXiv},
       eprint = {1608.00970},
 primaryClass = {astro-ph.HE},
       adsurl = {https://ui.adsabs.harvard.edu/abs/2017MNRAS.464.2816B}
}

@ARTICLE{Hinkle2020,
       author = {{Hinkle}, Jason T. and {Holoien}, Thomas W.-S. and {Shappee}, Benjamin. J. and {Auchettl}, Katie and {Kochanek}, Christopher S. and {Stanek}, K.~Z. and {Payne}, Anna V. and {Thompson}, Todd A.},
        title = "{Examining a Peak-luminosity/Decline-rate Relationship for Tidal Disruption Events}",
      journal = {\apjl},
         year = 2020,
        month = may,
       volume = {894},
       number = {1},
          eid = {L10},
        pages = {L10},
          doi = {10.3847/2041-8213/ab89a2},
archivePrefix = {arXiv},
       eprint = {2001.08215},
 primaryClass = {astro-ph.HE},
       adsurl = {https://ui.adsabs.harvard.edu/abs/2020ApJ...894L..10H}
}

@ARTICLE{Cendes2022,
       author = {{Cendes}, Y. and {Berger}, E. and {Alexander}, K.~D. and {Gomez}, S. and {Hajela}, A. and {Chornock}, R. and {Laskar}, T. and {Margutti}, R. and {Metzger}, B. and {Bietenholz}, M.~F. and {Brethauer}, D. and {Wieringa}, M.~H.},
        title = "{A Mildly Relativistic Outflow Launched Two Years after Disruption in Tidal Disruption Event AT2018hyz}",
      journal = {\apj},
         year = 2022,
        month = oct,
       volume = {938},
       number = {1},
          eid = {28},
        pages = {28},
          doi = {10.3847/1538-4357/ac88d0},
archivePrefix = {arXiv},
       eprint = {2206.14297},
 primaryClass = {astro-ph.HE},
       adsurl = {https://ui.adsabs.harvard.edu/abs/2022ApJ...938...28C}
}

@ARTICLE{Jiang2016,
       author = {{Jiang}, Yan-Fei and {Guillochon}, James and {Loeb}, Abraham},
        title = "{Prompt Radiation and Mass Outflows from the Stream-Stream Collisions of Tidal Disruption Events}",
      journal = {\apj},
         year = 2016,
        month = oct,
       volume = {830},
       number = {2},
          eid = {125},
        pages = {125},
          doi = {10.3847/0004-637X/830/2/125},
archivePrefix = {arXiv},
       eprint = {1603.07733},
 primaryClass = {astro-ph.HE},
       adsurl = {https://ui.adsabs.harvard.edu/abs/2016ApJ...830..125J}
}

@ARTICLE{Metzger2022,
       author = {{Metzger}, Brian D.},
        title = "{Cooling Envelope Model for Tidal Disruption Events}",
      journal = {\apjl},
         year = 2022,
        month = sep,
       volume = {937},
       number = {1},
          eid = {L12},
        pages = {L12},
          doi = {10.3847/2041-8213/ac90ba},
archivePrefix = {arXiv},
       eprint = {2207.07136},
 primaryClass = {astro-ph.HE},
       adsurl = {https://ui.adsabs.harvard.edu/abs/2022ApJ...937L..12M}
}

@ARTICLE{Matthee2024,
       author = {{Matthee}, Jorryt and {Naidu}, Rohan P. and {Brammer}, Gabriel and {Chisholm}, John and {Eilers}, Anna-Christina and {Goulding}, Andy and {Greene}, Jenny and {Kashino}, Daichi and {Labbe}, Ivo and {Lilly}, Simon J. and {Mackenzie}, Ruari and {Oesch}, Pascal A. and {Weibel}, Andrea and {Wuyts}, Stijn and {Xiao}, Mengyuan and {Bordoloi}, Rongmon and {Bouwens}, Rychard and {van Dokkum}, Pieter and {Illingworth}, Garth and {Kramarenko}, Ivan and {Maseda}, Michael V. and {Mason}, Charlotte and {Meyer}, Romain A. and {Nelson}, Erica J. and {Reddy}, Naveen A. and {Shivaei}, Irene and {Simcoe}, Robert A. and {Yue}, Minghao},
        title = "{Little Red Dots: An Abundant Population of Faint Active Galactic Nuclei at z {\ensuremath{\sim}} 5 Revealed by the EIGER and FRESCO JWST Surveys}",
      journal = {\apj},
         year = 2024,
        month = mar,
       volume = {963},
       number = {2},
          eid = {129},
        pages = {129},
          doi = {10.3847/1538-4357/ad2345},
archivePrefix = {arXiv},
       eprint = {2306.05448},
 primaryClass = {astro-ph.GA},
       adsurl = {https://ui.adsabs.harvard.edu/abs/2024ApJ...963..129M}
}

@ARTICLE{Mockler2022,
       author = {{Mockler}, Brenna and {Twum}, Angela A. and {Auchettl}, Katie and {Dodd}, Sierra and {French}, K.~D. and {Law-Smith}, Jamie A.~P. and {Ramirez-Ruiz}, Enrico},
        title = "{Evidence for the Preferential Disruption of Moderately Massive Stars by Supermassive Black Holes}",
      journal = {\apj},
         year = 2022,
        month = jan,
       volume = {924},
       number = {2},
          eid = {70},
        pages = {70},
          doi = {10.3847/1538-4357/ac35d5},
archivePrefix = {arXiv},
       eprint = {2110.03013},
 primaryClass = {astro-ph.HE},
       adsurl = {https://ui.adsabs.harvard.edu/abs/2022ApJ...924...70M}
}

@ARTICLE{Golightly2019,
       author = {{Golightly}, Elen C.~A. and {Coughlin}, Eric R. and {Nixon}, C.~J.},
        title = "{Tidal Disruption Events: The Role of Stellar Spin}",
      journal = {\apj},
         year = 2019,
        month = feb,
       volume = {872},
       number = {2},
          eid = {163},
        pages = {163},
          doi = {10.3847/1538-4357/aafd2f},
archivePrefix = {arXiv},
       eprint = {1901.03717},
 primaryClass = {astro-ph.HE},
       adsurl = {https://ui.adsabs.harvard.edu/abs/2019ApJ...872..163G}
}

@ARTICLE{LiptaiPrice2019,
       author = {{Liptai}, David and {Price}, Daniel J. and {Mandel}, Ilya and {Lodato}, Giuseppe},
        title = "{Disc formation from tidal disruption of stars on eccentric orbits by Kerr black holes using GRSPH}",
      journal = {arXiv e-prints},
         year = 2019,
        month = oct,
          eid = {arXiv:1910.10154},
        pages = {arXiv:1910.10154},
          doi = {10.48550/arXiv.1910.10154},
archivePrefix = {arXiv},
       eprint = {1910.10154},
 primaryClass = {astro-ph.HE},
       adsurl = {https://ui.adsabs.harvard.edu/abs/2019arXiv191010154L}
}

@ARTICLE{Bandopadhyay2026,
       author = {{Bandopadhyay}, Ananya and {Coughlin}, Eric R. and {Nixon}, C.~J.},
        title = "{The Maximum Gravity Model for Partial Tidal Disruption Events: Mass Loss, Peak Fallback Rate, and Dependence on Stellar Properties}",
      journal = {\apj},
         year = 2026,
        month = feb,
       volume = {998},
       number = {1},
          eid = {81},
        pages = {81},
          doi = {10.3847/1538-4357/ae31e3},
archivePrefix = {arXiv},
       eprint = {2601.02476},
 primaryClass = {astro-ph.HE},
       adsurl = {https://ui.adsabs.harvard.edu/abs/2026ApJ...998...81B}
}

@ARTICLE{Bonnerot2020,
       author = {{Bonnerot}, Cl{\'e}ment and {Lu}, Wenbin},
        title = "{Simulating disc formation in tidal disruption events}",
      journal = {\mnras},
         year = 2020,
        month = jun,
       volume = {495},
       number = {1},
        pages = {1374-1391},
          doi = {10.1093/mnras/staa1246},
archivePrefix = {arXiv},
       eprint = {1906.05865},
 primaryClass = {astro-ph.HE},
       adsurl = {https://ui.adsabs.harvard.edu/abs/2020MNRAS.495.1374B}
}

@ARTICLE{Hayasaki2013,
       author = {{Hayasaki}, Kimitake and {Stone}, Nicholas and {Loeb}, Abraham},
        title = "{Finite, intense accretion bursts from tidal disruption of stars on bound orbits}",
      journal = {\mnras},
         year = 2013,
        month = sep,
       volume = {434},
       number = {2},
        pages = {909-924},
          doi = {10.1093/mnras/stt871},
archivePrefix = {arXiv},
       eprint = {1210.1333},
 primaryClass = {astro-ph.HE},
       adsurl = {https://ui.adsabs.harvard.edu/abs/2013MNRAS.434..909H}
}

@ARTICLE{Mockler2025,
       author = {{Mockler}, Brenna and {Hammerstein}, Erica and {Coughlin}, Eric R. and {Nicholl}, Matt},
        title = "{Tidal Disruption Events}",
      journal = {arXiv e-prints},
         year = 2025,
        month = nov,
          eid = {arXiv:2511.14911},
        pages = {arXiv:2511.14911},
          doi = {10.48550/arXiv.2511.14911},
archivePrefix = {arXiv},
       eprint = {2511.14911},
 primaryClass = {astro-ph.HE},
       adsurl = {https://ui.adsabs.harvard.edu/abs/2025arXiv251114911M}
}

@ARTICLE{VanVelzen2018,
       author = {{van Velzen}, S.},
        title = "{On the Mass and Luminosity Functions of Tidal Disruption Flares: Rate Suppression due to Black Hole Event Horizons}",
      journal = {\apj},
         year = 2018,
        month = jan,
       volume = {852},
       number = {2},
          eid = {72},
        pages = {72},
          doi = {10.3847/1538-4357/aa998e},
archivePrefix = {arXiv},
       eprint = {1707.03458},
 primaryClass = {astro-ph.HE},
       adsurl = {https://ui.adsabs.harvard.edu/abs/2018ApJ...852...72V}
}

@ARTICLE{Dai2018,
       author = {{Dai}, Lixin and {McKinney}, Jonathan C. and {Roth}, Nathaniel and {Ramirez-Ruiz}, Enrico and {Miller}, M. Coleman},
        title = "{A Unified Model for Tidal Disruption Events}",
      journal = {\apjl},
         year = 2018,
        month = jun,
       volume = {859},
       number = {2},
          eid = {L20},
        pages = {L20},
          doi = {10.3847/2041-8213/aab429},
archivePrefix = {arXiv},
       eprint = {1803.03265},
 primaryClass = {astro-ph.HE},
       adsurl = {https://ui.adsabs.harvard.edu/abs/2018ApJ...859L..20D}
}

@ARTICLE{Charalampopoulos2023,
       author = {{Charalampopoulos}, P. and {Pursiainen}, M. and {Leloudas}, G. and {Arcavi}, I. and {Newsome}, M. and {Schulze}, S. and {Burke}, J. and {Nicholl}, M.},
        title = "{AT 2020wey and the class of faint and fast tidal disruption events}",
      journal = {\aap},
         year = 2023,
        month = may,
       volume = {673},
          eid = {A95},
        pages = {A95},
          doi = {10.1051/0004-6361/202245065},
archivePrefix = {arXiv},
       eprint = {2209.12913},
 primaryClass = {astro-ph.HE},
       adsurl = {https://ui.adsabs.harvard.edu/abs/2023A&A...673A..95C}
}

@ARTICLE{Makrygianni2025,
       author = {{Makrygianni}, Lydia and {Arcavi}, Iair and {Newsome}, Megan and {Bandopadhyay}, Ananya and {Coughlin}, Eric R. and {Linial}, Itai and {Mockler}, Brenna and {Quataert}, Eliot and {Nixon}, Chris and {Godson}, Benjamin and {Pursiainen}, Miika and {Leloudas}, Giorgos and {French}, K. Decker and {Zitrin}, Adi and {Faris}, Sara and {Lam}, Marco C. and {Horesh}, Assaf and {Sfaradi}, Itai and {Fausnaugh}, Michael and {Nakar}, Ehud and {Ackley}, Kendall and {Andrews}, Moira and {Charalampopoulos}, Panos and {Davies}, Benjamin D.~R. and {Dgany}, Yael and {Dyer}, Martin J. and {Farah}, Joseph and {Fender}, Rob and {Green}, David A. and {Howell}, D. Andrew and {Killestein}, Thomas and {Koivisto}, Niilo and {Lyman}, Joseph and {McCully}, Curtis and {Mitchell}, Morgan A. and {Padilla Gonzalez}, Estefania and {Rhodes}, Lauren and {Sahu}, Anwesha and {Terreran}, Giacomo and {Warwick}, Ben},
        title = "{The Double Tidal Disruption Event AT 2022dbl Implies that at Least Some ``Standard'' Optical Tidal Disruption Events Are Partial Disruptions}",
      journal = {\apjl},
         year = 2025,
        month = jul,
       volume = {987},
       number = {1},
          eid = {L20},
        pages = {L20},
          doi = {10.3847/2041-8213/ade155},
archivePrefix = {arXiv},
       eprint = {2505.16867},
 primaryClass = {astro-ph.HE},
       adsurl = {https://ui.adsabs.harvard.edu/abs/2025ApJ...987L..20M}
}

@ARTICLE{VanVelzen2020,
       author = {{van Velzen}, Sjoert and {Holoien}, Thomas W.-S. and {Onori}, Francesca and {Hung}, Tiara and {Arcavi}, Iair},
        title = "{Optical-Ultraviolet Tidal Disruption Events}",
      journal = {\ssr},
         year = 2020,
        month = oct,
       volume = {216},
       number = {8},
          eid = {124},
        pages = {124},
          doi = {10.1007/s11214-020-00753-z},
archivePrefix = {arXiv},
       eprint = {2008.05461},
 primaryClass = {astro-ph.HE},
       adsurl = {https://ui.adsabs.harvard.edu/abs/2020SSRv..216..124V}
}

@ARTICLE{Bricman2023,
       author = {{Bu{\v{c}}ar Bricman}, K. and {van Velzen}, S. and {Nicholl}, M. and {Gomboc}, A.},
        title = "{Rubin Observatory's Survey Strategy Performance for Tidal Disruption Events}",
      journal = {\apjs},
         year = 2023,
        month = sep,
       volume = {268},
       number = {1},
          eid = {13},
        pages = {13},
          doi = {10.3847/1538-4365/ace1e7},
archivePrefix = {arXiv},
       eprint = {2307.01829},
 primaryClass = {astro-ph.HE},
       adsurl = {https://ui.adsabs.harvard.edu/abs/2023ApJS..268...13B}
}

@ARTICLE{Nicholl2022,
       author = {{Nicholl}, Matt and {Lanning}, Daniel and {Ramsden}, Paige and {Mockler}, Brenna and {Lawrence}, Andy and {Short}, Phil and {Ridley}, Evan J.},
        title = "{Systematic light-curve modelling of TDEs: statistical differences between the spectroscopic classes}",
      journal = {\mnras},
         year = 2022,
        month = oct,
       volume = {515},
       number = {4},
        pages = {5604-5616},
          doi = {10.1093/mnras/stac2206},
archivePrefix = {arXiv},
       eprint = {2201.02649},
 primaryClass = {astro-ph.HE},
       adsurl = {https://ui.adsabs.harvard.edu/abs/2022MNRAS.515.5604N}
}

@ARTICLE{Bricman2020,
       author = {{Bricman}, Katja and {Gomboc}, Andreja},
        title = "{The Prospects of Observing Tidal Disruption Events with the Large Synoptic Survey Telescope}",
      journal = {\apj},
         year = 2020,
        month = feb,
       volume = {890},
       number = {1},
          eid = {73},
        pages = {73},
          doi = {10.3847/1538-4357/ab6989},
archivePrefix = {arXiv},
       eprint = {1906.08235},
 primaryClass = {astro-ph.HE},
       adsurl = {https://ui.adsabs.harvard.edu/abs/2020ApJ...890...73B}
}

@ARTICLE{Angus2026,
       author = {{Angus}, C.~R. and {Smith}, A.~J. and {Magill}, D. and {Ramsden}, P. and {Sarin}, N. and {Nicholl}, M. and {Mockler}, B. and {Hammerstein}, E. and {Stein}, R. and {Yao}, Y. and {de Boer}, T. and {Chambers}, K.~C. and {Huber}, M.~E. and {Lin}, C.-C. and {Lowe}, T.~B. and {Magnier}, E.~A. and {Smartt}, S.~J. and {Wainscoat}, R.~J.},
        title = "{Can tidal disruption event models reliably measure black hole masses?}",
      journal = {arXiv e-prints},
         year = 2026,
        month = jan,
          eid = {arXiv:2601.04406},
        pages = {arXiv:2601.04406},
          doi = {10.48550/arXiv.2601.04406},
archivePrefix = {arXiv},
       eprint = {2601.04406},
 primaryClass = {astro-ph.HE},
       adsurl = {https://ui.adsabs.harvard.edu/abs/2026arXiv260104406A}
}

@ARTICLE{vanVelzen2011,
       author = {{van Velzen}, Sjoert and {Farrar}, Glennys R. and {Gezari}, Suvi and {Morrell}, Nidia and {Zaritsky}, Dennis and {{\"O}stman}, Linda and {Smith}, Mathew and {Gelfand}, Joseph and {Drake}, Andrew J.},
        title = "{Optical Discovery of Probable Stellar Tidal Disruption Flares}",
      journal = {\apj},
         year = 2011,
        month = nov,
       volume = {741},
       number = {2},
          eid = {73},
        pages = {73},
          doi = {10.1088/0004-637X/741/2/73},
archivePrefix = {arXiv},
       eprint = {1009.1627},
 primaryClass = {astro-ph.CO},
       adsurl = {https://ui.adsabs.harvard.edu/abs/2011ApJ...741...73V}
}

@ARTICLE{Pasham20204,
       author = {{Pasham}, Dheeraj and {Coughlin}, E.~R. and {Guolo}, M. and {Wevers}, T. and {Nixon}, C.~J. and {Hinkle}, Jason T. and {Bandopadhyay}, A.},
        title = "{A Potential Second Shutoff from AT2018fyk: An Updated Orbital Ephemeris of the Surviving Star under the Repeating Partial Tidal Disruption Event Paradigm}",
      journal = {\apjl},
         year = 2024,
        month = aug,
       volume = {971},
       number = {2},
          eid = {L31},
        pages = {L31},
          doi = {10.3847/2041-8213/ad57b3},
archivePrefix = {arXiv},
       eprint = {2406.18124},
 primaryClass = {astro-ph.HE},
       adsurl = {https://ui.adsabs.harvard.edu/abs/2024ApJ...971L..31P}
}

@ARTICLE{Wevers2023,
       author = {{Wevers}, T. and {Coughlin}, E.~R. and {Pasham}, D.~R. and {Guolo}, M. and {Sun}, Y. and {Wen}, S. and {Jonker}, P.~G. and {Zabludoff}, A. and {Malyali}, A. and {Arcodia}, R. and {Liu}, Z. and {Merloni}, A. and {Rau}, A. and {Grotova}, I. and {Short}, P. and {Cao}, Z.},
        title = "{Live to Die Another Day: The Rebrightening of AT 2018fyk as a Repeating Partial Tidal Disruption Event}",
      journal = {\apjl},
         year = 2023,
        month = jan,
       volume = {942},
       number = {2},
          eid = {L33},
        pages = {L33},
          doi = {10.3847/2041-8213/ac9f36},
archivePrefix = {arXiv},
       eprint = {2209.07538},
 primaryClass = {astro-ph.HE},
       adsurl = {https://ui.adsabs.harvard.edu/abs/2023ApJ...942L..33W}
}

@ARTICLE{Wen2024,
       author = {{Wen}, S. and {Jonker}, P.~G. and {Levan}, A.~J. and {Li}, D. and {Stone}, N.~C. and {Zabludoff}, A.~I. and {Cao}, Z. and {Wevers}, T. and {Pasham}, D.~R. and {Lewin}, C. and {Kara}, E.},
        title = "{AT2018fyk: Candidate Tidal Disruption Event by a (Super)Massive Black Hole Binary}",
      journal = {\apj},
         year = 2024,
        month = aug,
       volume = {970},
       number = {2},
          eid = {116},
        pages = {116},
          doi = {10.3847/1538-4357/ad4da3},
archivePrefix = {arXiv},
       eprint = {2405.00894},
 primaryClass = {astro-ph.HE},
       adsurl = {https://ui.adsabs.harvard.edu/abs/2024ApJ...970..116W}
}

@ARTICLE{Andalman2022,
       author = {{Andalman}, Zachary L. and {Liska}, Matthew T.~P. and {Tchekhovskoy}, Alexander and {Coughlin}, Eric R. and {Stone}, Nicholas},
        title = "{Tidal disruption discs formed and fed by stream-stream and stream-disc interactions in global GRHD simulations}",
      journal = {\mnras},
         year = 2022,
        month = feb,
       volume = {510},
       number = {2},
        pages = {1627-1648},
          doi = {10.1093/mnras/stab3444},
archivePrefix = {arXiv},
       eprint = {2008.04922},
 primaryClass = {astro-ph.HE},
       adsurl = {https://ui.adsabs.harvard.edu/abs/2022MNRAS.510.1627A}
}

@ARTICLE{Li2005,
       author = {{Li}, Li-Xin and {Zimmerman}, Erik R. and {Narayan}, Ramesh and {McClintock}, Jeffrey E.},
        title = "{Multitemperature Blackbody Spectrum of a Thin Accretion Disk around a Kerr Black Hole: Model Computations and Comparison with Observations}",
      journal = {\apjs},
         year = 2005,
        month = apr,
       volume = {157},
       number = {2},
        pages = {335-370},
          doi = {10.1086/428089},
archivePrefix = {arXiv},
       eprint = {astro-ph/0411583},
 primaryClass = {astro-ph},
       adsurl = {https://ui.adsabs.harvard.edu/abs/2005ApJS..157..335L}
}

@ARTICLE{Berger2026,
       author = {{Berger}, Vera and {Kara}, Erin and {Chakraborty}, Joheen and {Masterson}, Megan and {Burdge}, Kevin},
        title = "{Disk-to-corona State Transition and Extreme X-Ray Variability in the Tidal Disruption Event AT2019teq}",
      journal = {\apj},
         year = 2026,
        month = mar,
       volume = {999},
       number = {2},
          eid = {265},
        pages = {265},
          doi = {10.3847/1538-4357/ae3006},
archivePrefix = {arXiv},
       eprint = {2601.04311},
 primaryClass = {astro-ph.HE},
       adsurl = {https://ui.adsabs.harvard.edu/abs/2026ApJ...999..265B}
}

@ARTICLE{Chakrobatory2026,
       author = {{Chakraborty}, Joheen and {Masterson}, Megan and {Mummery}, Andrew and {Kara}, Erin and {Panagiotou}, Christos and {Arcodia}, Riccardo and {Berger}, Vera},
        title = "{X-Ray Spectral-timing Properties of Tidal Disruption Events}",
      journal = {\apj},
         year = 2026,
        month = mar,
       volume = {1000},
       number = {1},
          eid = {95},
        pages = {95},
          doi = {10.3847/1538-4357/ae4876},
archivePrefix = {arXiv},
       eprint = {2602.16868},
 primaryClass = {astro-ph.HE},
       adsurl = {https://ui.adsabs.harvard.edu/abs/2026ApJ..1000...95C}
}

@ARTICLE{Yao2026,
       author = {{Yao}, Yuhan and {Chornock}, Ryan and {Mummery}, Andrew and {Margutti}, Raffaella and {Gilfanov}, Marat and {Guolo}, Muryel and {Coughlin}, Eric R. and {Lu}, Wenbin and {Chakraborty}, Joheen and {Pasham}, Dheeraj R. and {Alexander}, Kate D. and {Aspegren}, Olivia and {Angus}, Charlotte R. and {Guo}, Xinze and {Hall}, Xander J. and {Hammerstein}, Erica and {Hinds}, K.-Ryan and {Ho}, Anna Y.~Q. and {Huang}, Xiaoshan and {Kammoun}, Elias and {LeBaron}, Natalie and {Lucchini}, Matteo and {McGrath}, Zo{\"e} and {Nicholl}, Matt and {Perley}, Daniel A. and {Rich}, R. Michael and {Schroeder}, Genevieve and {Sheng}, Xinyue and {Sollerman}, Jesper and {Somalwar}, Jean and {Wise}, Jacob R. and {Coughlin}, Michael W. and {Drake}, Andrew and {Graham}, Matthew J. and {Helou}, George and {Jaimes}, Joahan C. and {Kasliwal}, Mansi M. and {Mahabal}, Ashish A. and {Medvedev}, Pavel and {Purdum}, Josiah and {Rusholme}, Ben and {Sunyaev}, Rashid},
        title = "{AT2024lhc and AT2024kmq in the landscape of featureless tidal disruption events}",
      journal = {arXiv e-prints},
         year = 2026,
        month = feb,
          eid = {arXiv:2602.21624},
        pages = {arXiv:2602.21624},
          doi = {10.48550/arXiv.2602.21624},
archivePrefix = {arXiv},
       eprint = {2602.21624},
 primaryClass = {astro-ph.HE},
       adsurl = {https://ui.adsabs.harvard.edu/abs/2026arXiv260221624Y}
}

@ARTICLE{Balbus1999,
       author = {{Balbus}, Steven A. and {Papaloizou}, John C.~B.},
        title = "{On the Dynamical Foundations of {\ensuremath{\alpha}} Disks}",
      journal = {\apj},
         year = 1999,
        month = aug,
       volume = {521},
       number = {2},
        pages = {650-658},
          doi = {10.1086/307594},
archivePrefix = {arXiv},
       eprint = {astro-ph/9903035},
 primaryClass = {astro-ph},
       adsurl = {https://ui.adsabs.harvard.edu/abs/1999ApJ...521..650B}
}


\end{document}